\documentclass[useAMS,usenatbib]{mn2e}
\usepackage{epsfig,graphics}

\def\Re{\mbox{$R_{\rm eff}$}}

\def\Msun{\mbox{$M_\odot$}}

\def\Yst{\mbox{$\Upsilon_{*}$}}

\def\mst{\mbox{$M_{\star}$}}

\def\lsim{\mathrel{\rlap{\lower3.5pt\hbox{\hskip0.5pt$\sim$}}
    \raise0.5pt\hbox{$<$}}}                % less than or approx. symbol
\def\gsim{~\rlap{$>$}{\lower 1.0ex\hbox{$\sim$}}}

\def\Msun{\mbox{$M_\odot$}}

\def\lsim{\mathrel{\rlap{\lower3.5pt\hbox{\hskip0.5pt$\sim$}}
    \raise0.5pt\hbox{$<$}}}
\def\gsim{~\rlap{$>$}{\lower 1.0ex\hbox{$\sim$}}}

\def\sig{\mbox{$\sigma$}}
\def\sigc{\mbox{$\sigma_{0}$}}
\def\Re{\mbox{$R_{\rm eff}$}}

\def\mst{\mbox{$M_{*}$}}

\def\gZ{\mbox{$\nabla_{\rm Z}$}}
\def\gage{\mbox{$\nabla_{\rm age}$}}

\def\gug{\mbox{$\nabla_{\rm u-g}$}}
\def\ggr{\mbox{$\nabla_{\rm g-r}$}}
\def\ggi{\mbox{$\nabla_{\rm g-i}$}}
\def\ggz{\mbox{$\nabla_{\rm g-z}$}}

\title[Color gradients]{Color and stellar population gradients in galaxies. Correlation with mass.}

\author[Tortora et al.]{\noindent
C.~Tortora$^{1,2}$\thanks{E-mail: ctortora@na.astro.it},
N.R.~Napolitano$^{3}$, V.F.~Cardone$^{4}$, M.~Capaccioli$^{2}$,
P.~Jetzer$^{1}$ \and R.~Molinaro$^{3,5}$
\\~\\
$^1$ Universit$\ddot{a}$t Z$\ddot{u}$rich, Institut f$\ddot{u}$r
Theoretische Physik, Winterthurerstrasse 190,
CH-8057, Z$\ddot{u}$rich, Switzerland\\
$^2$ Dipartimento di Scienze Fisiche, Universit\`{a} di Napoli
Federico II, Compl. Univ. Monte
S. Angelo, 80126 - Napoli, Italy\\
$^3$ INAF -- Osservatorio Astronomico di
Capodimonte, Salita Moiariello, 16, 80131 - Napoli, Italy\\
$^4$ Dipartimento di Fisica Generale "A. Avogadro", Universit\`a
di Torino and Istituto Nazionale di Fisica Nucleare - \\ Sezione
di
Torino, Via Pietro Giuria 1, 10125 - Torino, Italy\\
$^5$ Dipartimento di Fisica, Politecnico di Torino, Corso Duca
degli Abruzzi 24, 10129 - Torino, Italy}

\begin{document}
\date{Accepted  Received }
\pagerange{\pageref{firstpage}--\pageref{lastpage}} \pubyear{xxxx}
\maketitle

\label{firstpage}

\begin{abstract}
We analyze the color gradients (CGs) of $\sim 50\,000$ nearby
Sloan Digital Sky Survey (SDSS) galaxies estimated by their
photometrical parameters (S$\rm \acute{e}$rsic index, total
magnitude, and effective radius). From synthetic spectral models
based on a simplified star formation recipe, we derive the mean
spectral properties, and explain the observed radial trends of the
color as gradients of the stellar population age and metallicity.
Color gradients have been correlated with color, luminosity, size,
velocity dispersion, and stellar mass. Distinct behaviours are
found for early- and late-type galaxies (ETGs and LTGs), pointing
to slightly different physical processes at work in different
morphological types and at different mass scales.

In particular, the most massive ETGs ($\mst \gsim 10^{11} \,
\Msun$) have shallow (even flat) CGs in correspondence of shallow
(negative) metallicity gradients. In the stellar mass range
$10^{10.3-10.5} \lsim \mst \lsim 10^{11} \, \Msun$, the
metallicity gradients reach their minimum of $\sim -0.5 \, \rm
dex^{-1}$. At $\mst \sim 10^{10.3-10.5} \, \Msun$, color and
metallicity gradient slopes suddenly change. They turn out to
anti-correlate with the mass, becoming highly positive at the very
low masses, the transition from negative to positive occurring at
$\mst \sim 10^{9-9.5} \, \Msun$. These correlations are mirrored
by similar trends of CGs with the effective radius and the
velocity dispersion. We have also found that age gradients
anti-correlate with metallicity gradients, as predicted by
hierarchical cosmological simulations for ETGs. On the other side,
LTGs have gradients which systematically decrease with mass (and
are always more negative than in ETGs), consistently with the
expectation from gas infall and supernovae feedback scenarios.

Metallicity is found to be the main driver of the trend of color
gradients, especially for LTGs, but age gradients are not
negligible and seem to play a significant role too. Owing to the
large dataset, we have been able to highlight that older galaxies
have systematically shallower age and metallicity gradients than
younger ones.

The emerging picture is qualitatively consistent with the
predictions from hydrodynamical and chemo-dynamical simulations.
In particular, our results for high-mass galaxies are in perfect
agreement with predictions based on the merging scenario, while
the evolution of LTGs and younger and less massive ETGs seems to
be mainly driven by infall and SN feedback.
\end{abstract}


\section{Introduction}\label{sec:intro}

Galaxy formation is a complicated matter which has not yet come to
a complete and coherent understanding. The standard cosmological
paradigm, the so-called $\Lambda$CDM, predicts that dark matter
(DM) haloes evolve hierarchically since early epochs, with smaller
units merging into massive structures (\citealt{Kauff+93},
\citealt{deLucia06}, \citealt{Ruszkowski+09}), which find positive
evidences in the strong size evolution in massive galaxies since
$z \sim 3$ (\citealt{Glazebrook+04}, \citealt{Daddi+05},
\citealt{Trujillo+06}). This scenario is at variance with the
evidences that today's high mass galaxies formed most of their
stars in earlier epochs and over a shorter time interval than
low-mass ones. This ``downsizing'' scheme
(\citealt{Cowie96,Gallazzi05,Jimenez+05,Nelan+05,Thomas2005,Treu2005b,
Bundy+06,Cimatti+06,Pannella+06,Panter+07,Fontanot+09}) seems at
odd with the $\Lambda$CDM hierarchical growth and seems rather to
support a ``monolithic--like'' formation scheme (see e.g. Chiosi
\& Carraro 2002) which broadly predicts an inside-out formation of
stars after an early dissipative collapse.

However, it is increasingly clearer that the downsizing is
compatible with the hierarchical model if the feedback processes
in the mass accretion history of the $\Lambda$CDM dark haloes are
taken into account (see e.g., \citealt{Neistein+06},
\citealt{Cattaneo+08}, Conroy \& Wechsler 2008,
\citealt{Cattaneo+10}). On the one hand, the gas cooling and the
shock heating (\citealt{dek_birn06}, \citealt{Cattaneo+08}) are
the main drivers of the star formation (SF) activity during the
hierarchical growth of the DM haloes; on the other side supernovae
(SNe), active galactic nuclei (AGN), merging, harassment,
strangulation etc., generally inhibit the stellar formation
(\citealt{Dekel86,Recchi01,Pipino02,Scannapieco06,DiMatteo05,
deLucia06,Cattaneo+06,Kaviraj07,Schawinski07,Cattaneo+08,AS08,
Khalatyan08,Romeo+08,Tortora2009AGN,weinmann09}).

The different processes might rule the star formation at the
global galaxy scale, or act at sub-galactic scales (e.g. the
nuclear regions vs outskirts) such that they are expected to
introduce a gradient of the main stellar properties with the
radius that shall leave observational signatures in galaxy
colours.

This paper is motivated by the fact that color gradients (CGs) are
efficient markers of the stellar properties variations within
galaxies, in particular as they mirror the gradients of star ages
and metallicities (\citealt{1972ApJ...176...21S}), although it is
not yet fully clear whether metallicity is the main driver of the
CGs in normal spheroidal (\citealt{Saglia+00, TO2000,
LaBarbera2002, LaBarbera2003, Ko04, Spolaor09, Rawle+09}) and
late-type systems (\citealt{MacArthur+04}, \citealt{Taylor+05}),
or age plays also an significant role (\citealt{Saglia+00,
LaBarbera2003, MacArthur+04, Spolaor09, Rawle+09}).

CGs are primarily a tool to discriminate the two broad formation
scenarios (monolithic vs hierarchical), but more importantly they
provide a deeper insight on the different mechanisms ruling the
galaxy evolution. As these mechanisms depend on the galaxy mass
scale, the widest mass (and luminosity) observational baseline is
needed to remark the relative effectiveness of the different
physical processes and their correlation with the observed
population gradients.

The observational picture is so complex to justify a schematic
review of the evidences accumulated so far. The facts are as
follows.

1) On average, nearby elliptical and spiral galaxies are bluer
outwards (\citealt{FIH89}, \citealt{Peletier+90a,Peletier+90b},
\citealt{BP94}, \citealt{TO2003}), while dwarfs show mainly redder
outskirts (\citealt{Vader+88}, \citealt{KD89},
\citealt{Chaboyer94}, \citealt{Tully+96}, \citealt{Jansen+00}).
Recently, deeper investigations of later-type galaxies have been
done and have highlighted the presence of non-monotonic colour and
stellar population gradients in these systems
(\citealt{MacArthur+04}, \citealt{BTP08}, \citealt{MacArthur+09},
\citealt{MS+09}, \citealt{SB+09}).

2) Spectral line indices (mainly measured in early-type galaxies,
ETGs) are somehow a more efficient tracer of the stellar
population properties and generally found to change with the
radius like CGs (\citealt{KoAr99}, \citealt{Kuntschner06},
\citealt{MacArthur+09}, \citealt{Rawle+09}).

3) Interpreting CGs in terms of metallicity gradients, in
high-mass galaxies typically $d \log Z / d \log R \sim -0.3$,
which is shallower than predicted by simulations of dissipative
``monolithic'' collapse (where $-0.5$ or a steeper value is
expected, see e.g. \citealt{Larson74, Larson75},
\citealt{Carlberg84}, \citealt{AY87}, \citealt{Ko04}, see also
below).

4) There are contradictory evidences of a correlations of CGs with
galaxy mass and luminosity. Earlier studies have claimed a weak
correlation with the physical properties of galaxies (e.g., mass,
luminosity, etc., \citealt{Peletier+90a}, \citealt{Davies+93},
\citealt{KoAr99}, \citealt{TO2003}), at variance with the typical
monolithic collapse predictions. Recently, a stronger correlation
with mass has started to emerge (e.g. \citealt{Forbes+05}),
pointing to a metallicity gradient decreasing with the mass for
low-mass galaxies (\citealt{Spolaor09}, \citealt{Rawle+09}),
accordingly with the monolithic scenario. Instead, high-mass
galaxies show too shallow gradients to match the predictions of
the monolithic collapse, but compatible with galaxy merging.

This wealth of observational evidences must be confronted with the
model predictions coming from the different galaxy formation
scenarios.

1) Steep gradients are expected when stars form during strong
dissipative (monolithic) collapses in deep potential wells of
galaxy cores where the gas is more efficiently retained, with a
consequent longer star formation activity and a longer chemical
enrichment in the inner than in outer regions (negative
metallicity gradients). On top of that, the delayed onset of winds
from supernovae, causing a further metal supply in the central
regions, would contribute to reinforce the steepness of these
gradients (see, e.g. \citealt{Pipino08}). As these processes are
regulated by the galaxy potential depth, which is somehow related
to the galaxy mass (and luminosity) \footnote{It remains to see
whether there might be a correlation with the the DM fraction of
the systems which is also a function of the galaxy stellar mass
(see e.g., \citealt{Nap05}, \citealt{Cappellari06},
\citealt{Conroy08}, \citealt{Tortora2009}).}, high-mass galaxies
are expected to have metallicity gradients which are steeper than
lower-mass ones. The latter seem to have almost no gradients
(\citealt{Gibson97}, \citealt{CC02}, \citealt{KG03}), although the
results here are based on limited galaxy samples.

2) Within the hierarchical picture, merging induce a meshing of
stellar populations, and lead to a more uniform metallicity
distribution and to shallower CGs (\citealt{White80},
\citealt{Ko04}). Strong jets from powerful AGNs can also quench
the star formation on the global galaxy scale and flatten colors
gradients in the host systems (\citealt{Tortora2009AGN}).

3) Metallicity and color gradients seem also to depend on the
efficiency of the dissipative processes in dark haloes
(\citealt{Hopkins+09a}), with just a weak dependence on the
remnant mass (\citealt{BS99}, \citealt{Ko04},
\citealt{Hopkins+09b}), in a way more similar of the observed
gradients. However, while merging is crucial to smear out the
color radial variation in high-mass galaxies (\citealt{Ko04}), it
seems unimportant for low-mass systems (\citealt{deLucia06},
\citealt{Cattaneo+08}). Instead, the energy from stellar winds and
supernovae and the effect of dissipative collapse might induce
even positive steep gradients (\citealt{Mori+97}).

~\\

CGs are the most direct observables to investigate the effect of
the physical processes (such as merging, AGN, SNe, stellar
feedback), which drive the galaxy evolution as a function of the
main galaxy parameters: luminosity, mass and central velocity
dispersion. In this paper we analyze optical CGs in about
$50\,000$ SDSS local galaxies, spanning a wide range of
luminosities and masses. CGs are obtained in an indirect way from
the photometrical parameters of the individual galaxies (effective
radius, S$\rm \acute{e}$rsic index, total magnitude). Using
synthetic spectral models, we compute age and metallicity
gradients for early- and late-type systems, and analyze the trends
with mass. The sample allows us to investigate the distribution of
the CGs over an unprecedented baseline of galaxy sizes,
luminosities, velocity dispersions, and stellar masses. The
observed trends are interpreted by means of quite different
physical phenomena at the various mass scales. This approach,
while exposed to the uncertainties on the structural parameters
(e.g. S$\rm \acute{e}$rsic index, $n$, and effective radius, \Re)
has the advantage of dealing with large statistics; furthermore it
is exportable to higher redshift galaxy surveys which are
generally limited to rest-frame visual bands. In this respect we
pay particular care in the check of the consistency of our results
with independent analyses. We will mainly concentrate on ETGs, for
which a wide collection of results (having an homogenized gradient
definition) is available from literature. On the other hand, due
to the uncertainties of the structural parameters of late-type
galaxies (LTGs), we will take the results on this galaxy sample
with the right caution, being well aware that our findings might
be only a benchmark for more accurate analyses.

In \S \ref{sec:data} we present the data sample and spectral
models. In \S \ref{sec:results} we show the main results of our
analysis, discussing CGs as a function of structural parameters,
stellar masses, and stellar properties derived from the fitting
procedure. In the same Section, age and metallicity gradients are
shown, and discussed within the galaxy formation scenarios in \S
\ref{sec:disc}. We finally draw some conclusions in \S
\ref{sec:conclusions}.

In the following, we use a cosmological model with $(\Omega_{m},
\, \Omega_{\Lambda}, \, h) = (0.3, \, 0.7, \, 0.7)$, where $h =
H_{0}/100 \, \textrm{km} \, \textrm{s}^{-1} \, \textrm{Mpc}^{-1}$
(\citealt{WMAP, WMAP2}), corresponding to a Universe age of
$t_{\rm univ}=13.5 \, \rm Gyr$.

\begin{figure*}
\psfig{file= 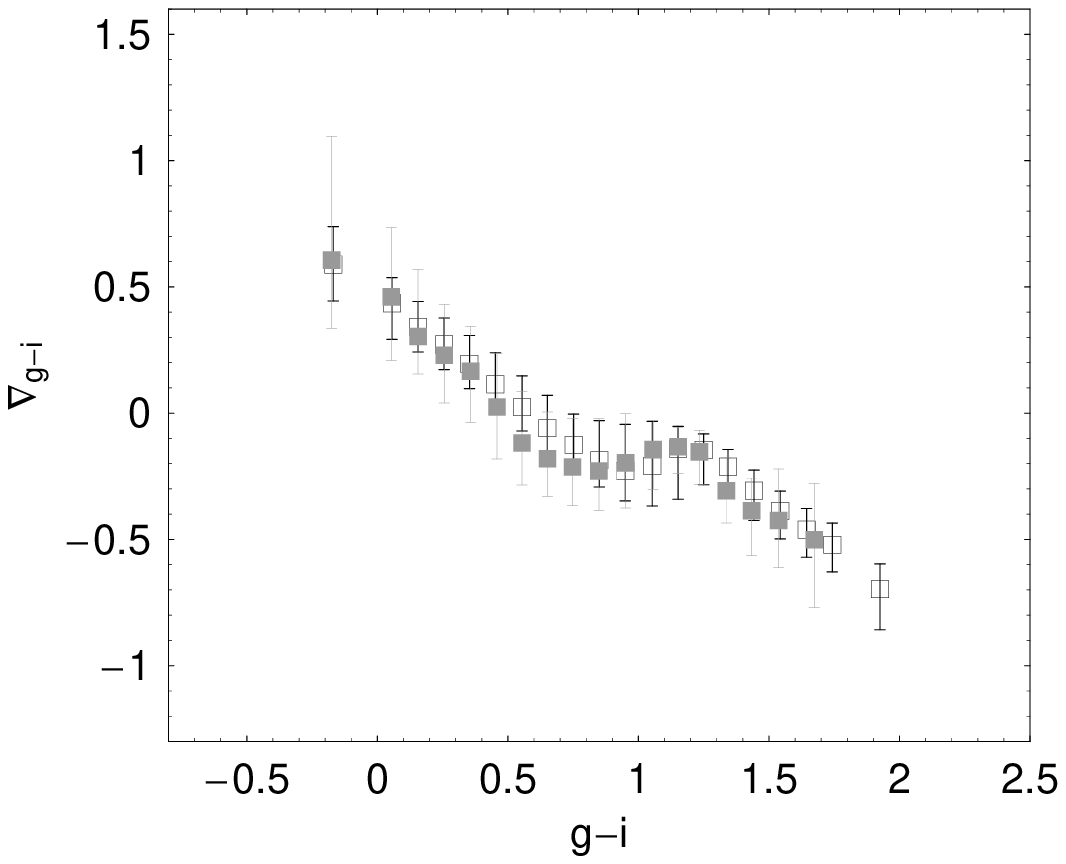, width=0.43\textwidth}
\psfig{file= 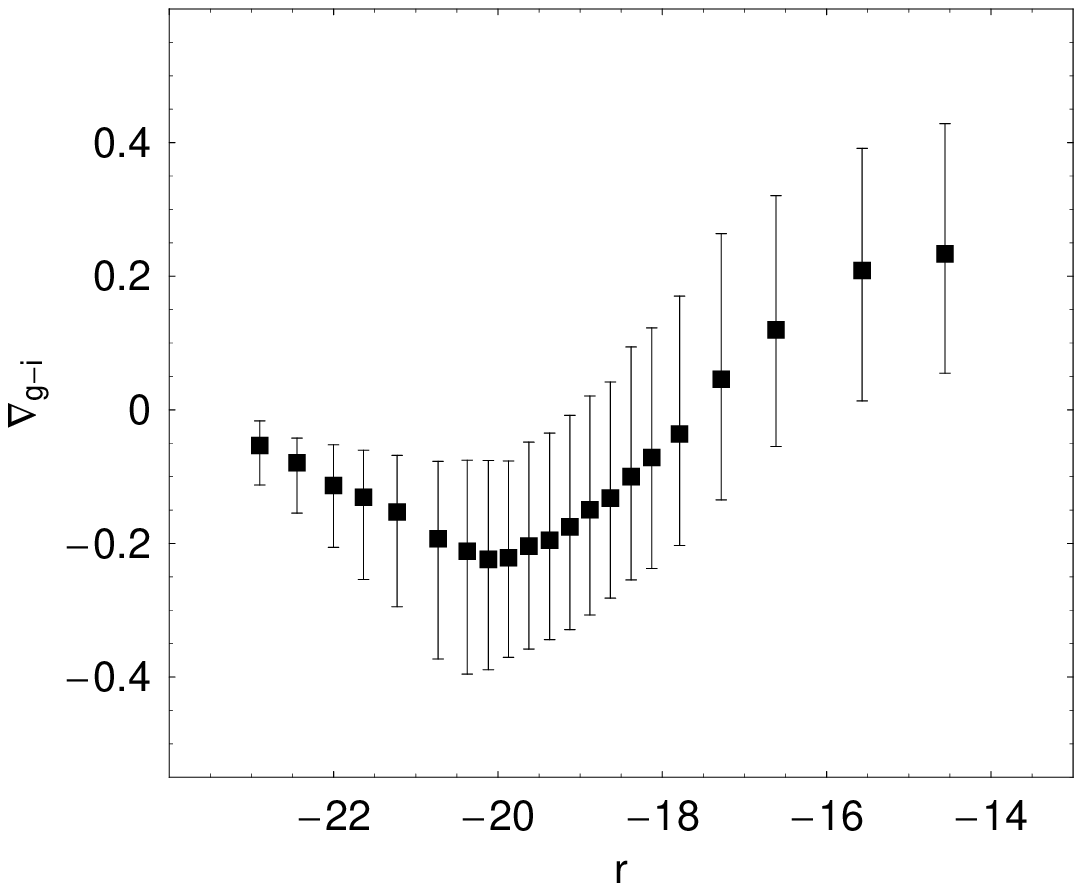, width=0.43\textwidth}\\
\psfig{file= 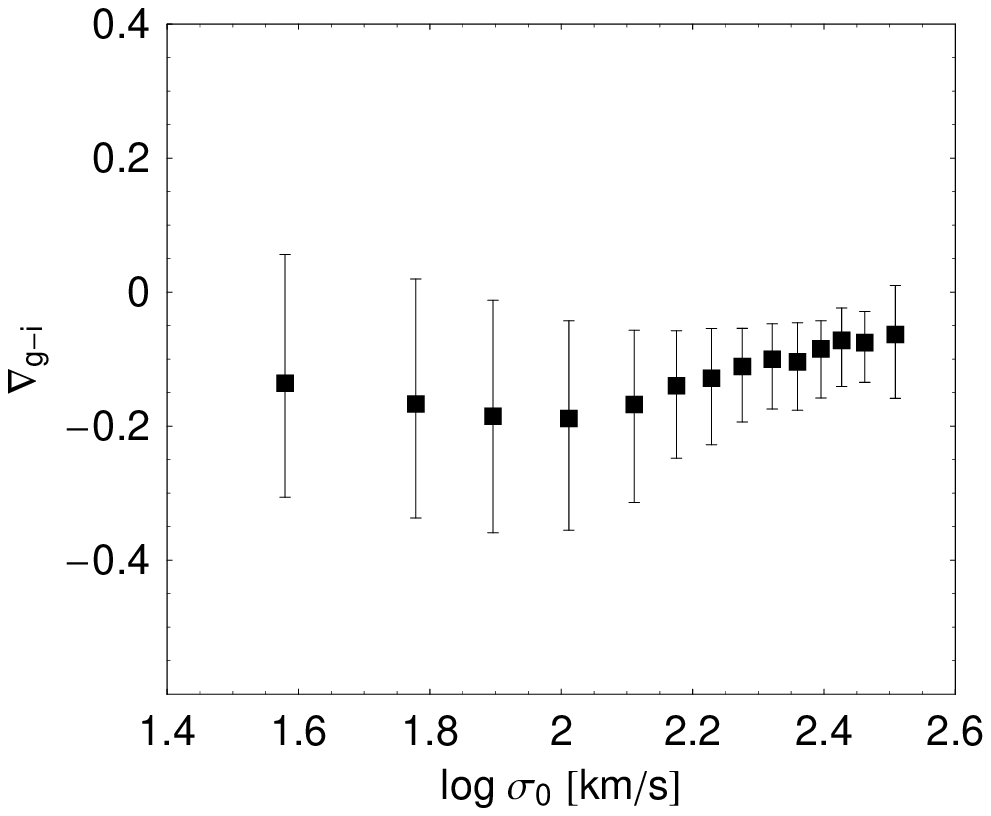, width=0.43\textwidth} \psfig{file=
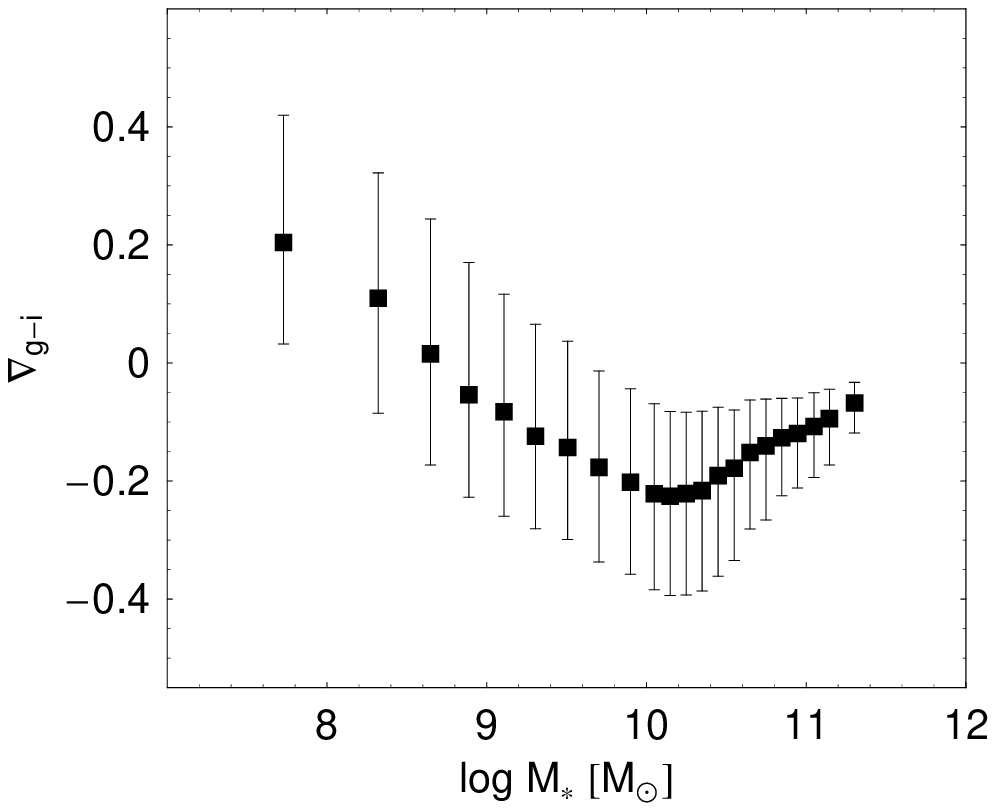, width=0.43\textwidth} \psfig{file= 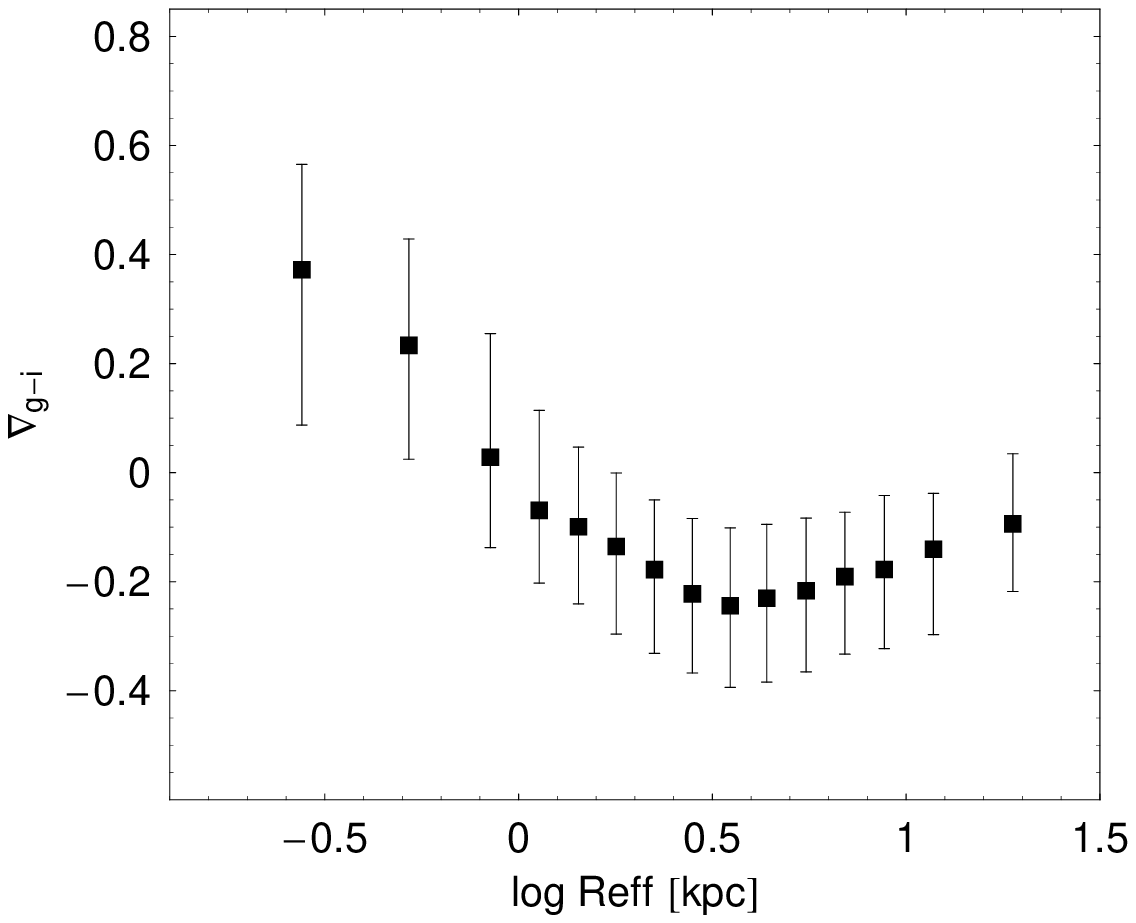,
width=0.43\textwidth}
\psfig{file= 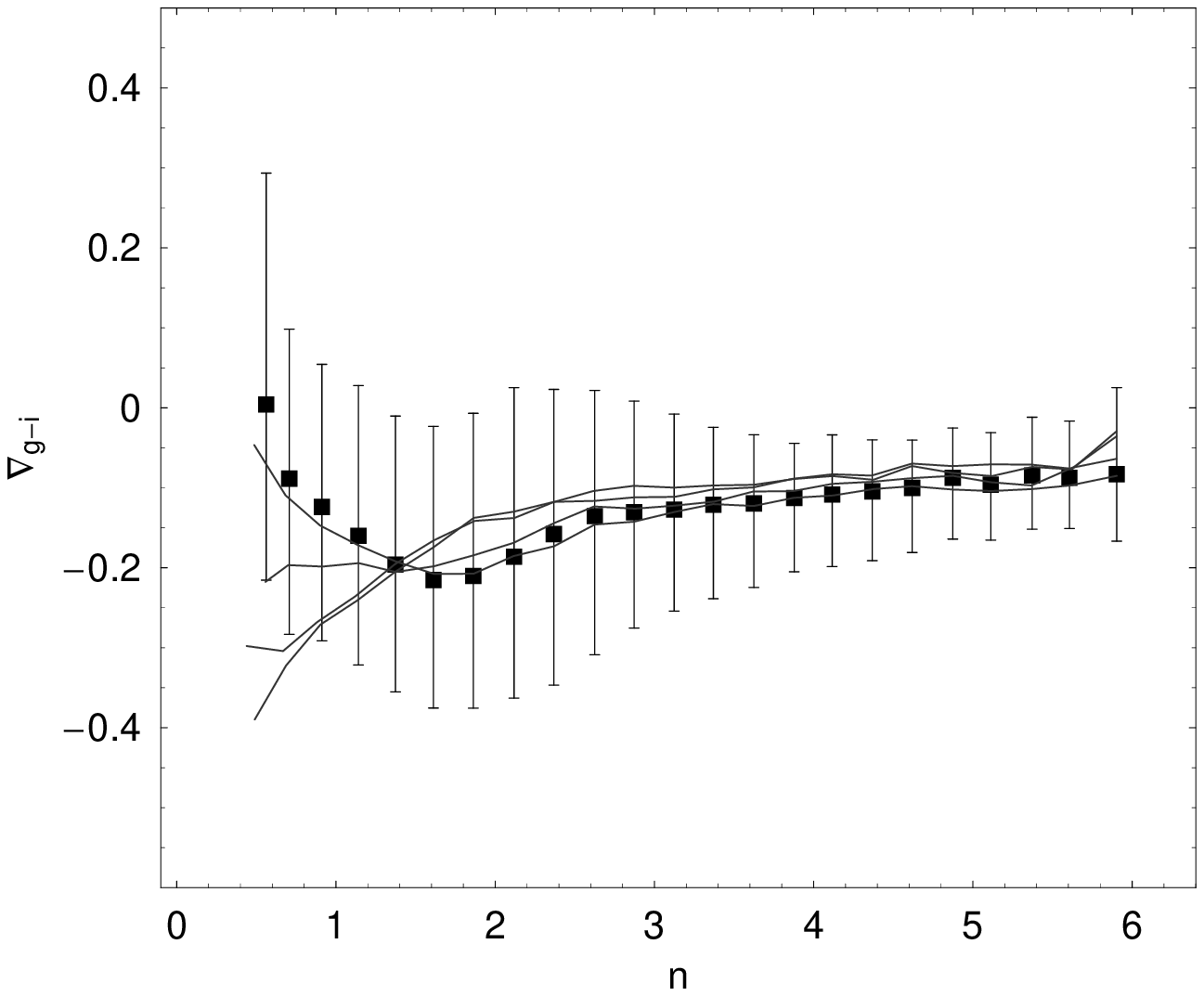, width=0.43\textwidth}\\
\caption{Gradients of the $g-i$ colour, as a function of various
observed and derived quantities. Medians of binned CGs are plotted
together with the $25-75\%$ quantiles, shown as error bars. {\it
Top-left.} CG as a function of color, black open and gray squares
are for colors at $R_{1}$ and $R_{2}$, respectively. {\it
Top-right.} CG as a function of $r$-band magnitude. {\it
Middle-left.} CG as a function of logarithm of $\sigc$, which is
defined as the velocity dispersion within a circular aperture of
radius $0.1 \Re$ (using $r$-band $\Re$) from the SDSS velocity
dispersion $\sigma_{ap}$, using the relation in
\citet{Jorgensen+95,Jorgensen+96}. {\it Middle-right.} Gradient as
a function of total stellar mass $\mst$ (assuming a Chabrier IMF),
which is the output of our stellar population analysis fitting the
total colors. {\it Bottom-left.} CG as a function of logarithmic
of $r$-band \Re\ (the results does not depend on the band). {\it
Bottom-right.} CG as a function of i-band S$\rm \acute{e}$rsic
index $n$; the lines are the the median trends for the S$\rm
\acute{e}$rsic index in the other bands.}\label{fig: fig0}
\end{figure*}

\section{Data}\label{sec:data}

\subsection{Galaxy sample}\label{sec:sample}
Our database consists of $50\,000$ low redshift ($0.0033 \leq z
\leq 0.05$) galaxies in the NYU Value-Added Galaxy Catalog
extracted from SDSS DR4 (\citealt[hereafter
B05]{Blanton05})\footnote{The catalog is available at: {\tt
http://sdss.physics.nyu.edu/vagc/lowz.html}.}. B05 have performed
a set of quality checks contemplating the analysis of large and
complex galaxies, incorrectly deblended in the SDSS pipeline, and
of the star/galaxy separation, with a number of eyeball
inspections on objects in the catalogue. The catalog includes
Petrosian magnitudes and structural parameters in $ugriz$ bands
such as the concentrations $C = R_{90}/R_{50}$, where $R_{90}$ and
$R_{50}$ are the Petrosian radii enclosing $90 \%$ and $50\%$ of
total luminosity. Effective radius, $R_{\rm eff}$, and the S$\rm
\acute{e}$rsic indices $n$ are taken from B05, where, a 1D
seeing-convolved S$\rm \acute{e}$rsic profile is fitted to
galaxies. The adopted redshift range sets the completeness level
at a Petrosian magnitude $r \simeq 18$ and at a surface brightness
within $R_{50}$ $\mu_{50} \simeq 24.5 \, \rm mag \, arcsec^{-2}$.
This luminosity limited sample still suffers from an
incompleteness at low surface brightness levels, e.g. in the
regime of dwarf galaxies with $\mst \lsim 10^{8.3-8.5} \, \rm
\Msun$ (\citealt{Blanton05a}, \citealt{BGD08}, \citealt{LW09}).
Moreover, the incompleteness, almost negligible for more massive
galaxies, comes back again for the most massive ones ($\mst \gsim
\, 10^{11}\, \rm \Msun$) as the sample is biased against large
objects at lower redshifts in consequence of a bias in the SDSS
photometric pipeline (\citealt{Blanton05a}). We address the
systematics of the SDSS sample more in detail in App.
\ref{app:appA4}.

Petrosian magnitudes are reduced to $z=0$ (K-term;
\citealt{Blanton03a}), corrected for galactic extinction using the
\cite{SFD98} dust map, converted to our adopted cosmology
(\citealt{Blanton03b}), and further corrected for the missing flux
of Petrosian magnitudes. The latter correction is almost
unchanged using the relation in \cite{Graham+05}.

The sample spans a wide range of luminosities, masses, and colors,
including galaxies lying both in the red-sequence (RS) and in the
blue cloud (BC). In particular, we have sorted out ETGs by
adopting the following two criteria:
\begin{itemize}
\item the S$\rm \acute{e}$rsic index satisfies the condition $2.5 \leq n \leq 5.5$;
\item the concentration index $C > 2.6$
(\citealt{Shimasaku+01}, \citealt{Padmanabhan+04},
\citealt{Cardone2008})\footnote{In order to retain the low mass
red systems, no lower cut has been applied to the velocity
dispersion. However, very low values ($\sigc < 70 \, \rm km/s$)
may be prone to large systematics, due to the signal-to-noise
ratio and instrumental resolution of the SDSS spectra.}.
\end{itemize}
The final ETG sample consists of $10\,508$ galaxies. Of the
remaining entries of our database, $27\,813$ are LTGs, defined as
objects with $C \leq 2.6$, $n \leq 2.5$, and $\sigc\leq 150~\rm
km/s$, and $9\,359$ are intermediate morphology galaxies, with $C
\leq 2.6$ ($>2.6$) and $n > 2.5$ ($\leq 2.5$), hereafter named
Intermediate-type galaxies (ITGs)\footnote{Note that $2120$
galaxies are cut away by the conservative upper limit on $n$,
dictated by the fitting S$\rm \acute{e}$rsic code limit of
$n=6$.}. Further details on the adopted morphology criteria are
given in the App. \ref{app:appA5}. In this paper we will consider
ETGs and LTGs only, with a main focus on ETGs.

\begin{figure*}
\psfig{file= 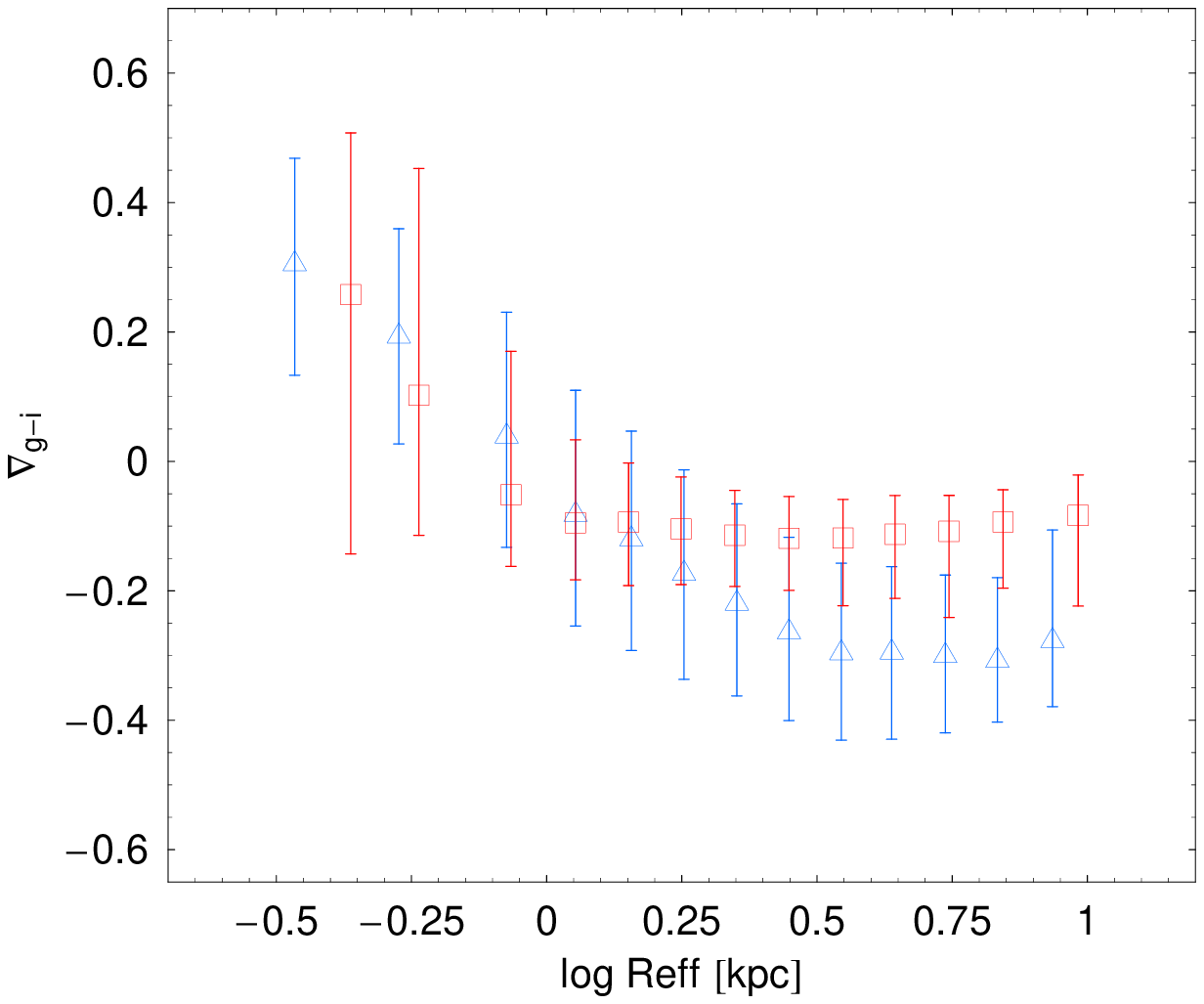, width=0.33\textwidth} \psfig{file=
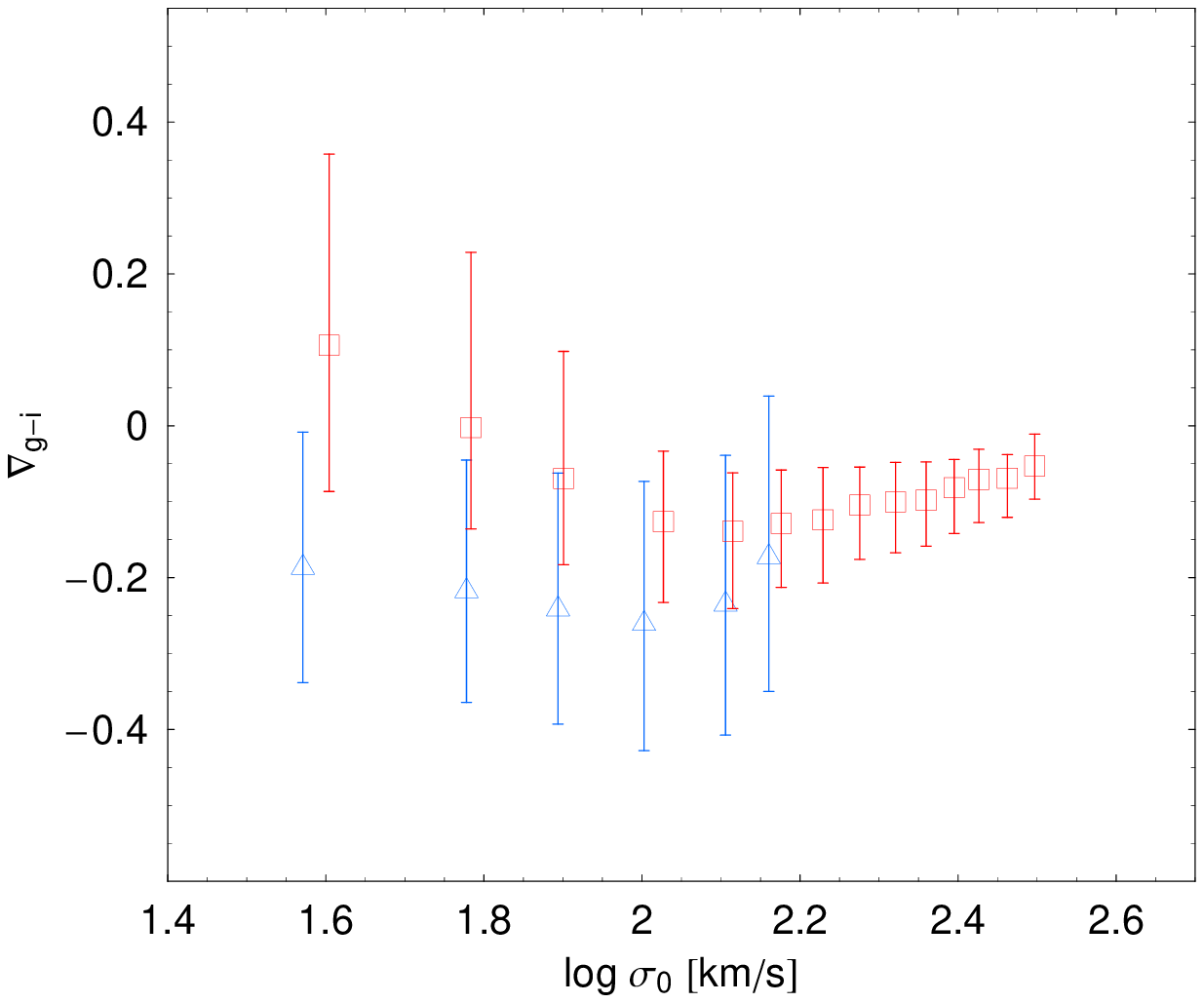, width=0.33\textwidth} \psfig{file= 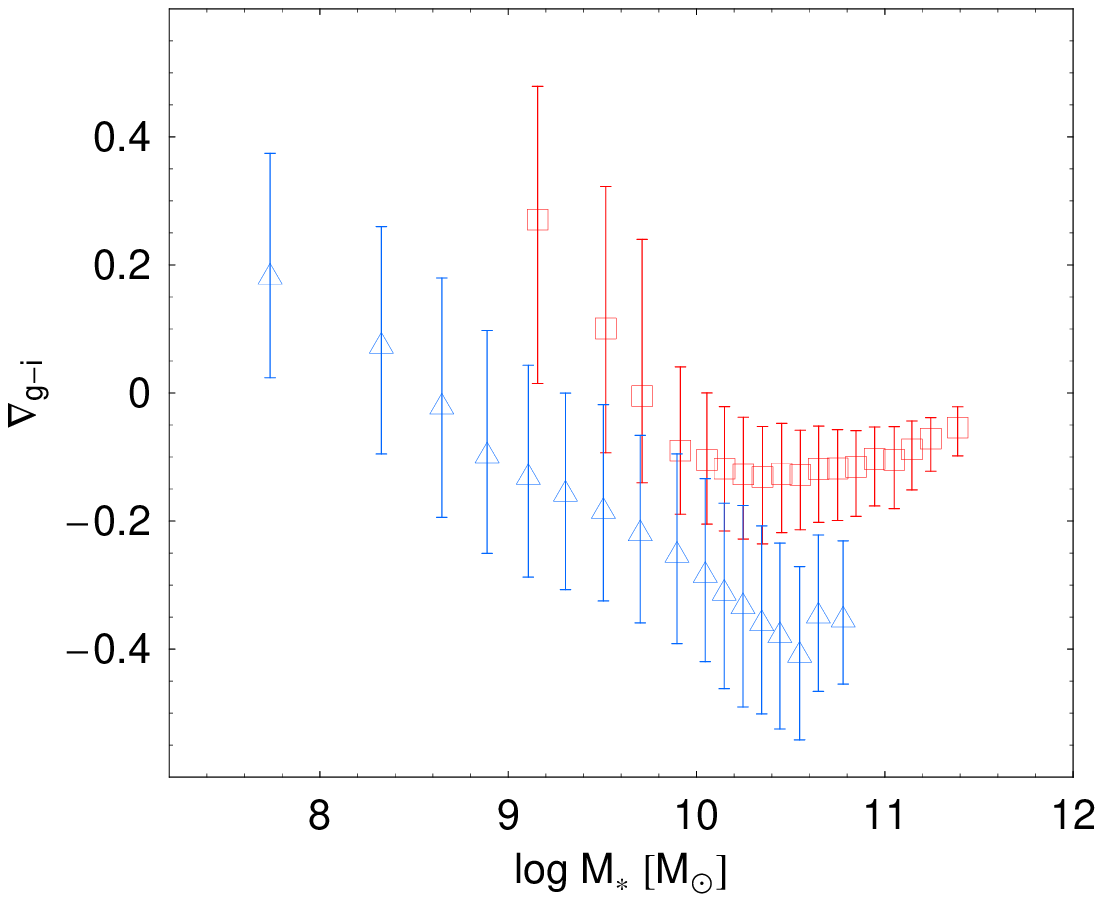,
width=0.33\textwidth} \caption{\ggi\ as a function of effective
radius (left panel), velocity dispersion (middle panel) and
stellar mass (right panel) for ETGs (red symbols) and LTGs (blue
symbols).} \label{fig: fig2}
\end{figure*}

\subsection{Spectral models and gradients}\label{sec:sp_model_gr}

The stellar population analysis of the selected sample is based on
a set of ``single burst'' synthetic stellar spectra using the
prescription of \citet[hereafter BC03]{BC03}. These models are
widely used in literature both for ellipticals and spirals (e.g.,
\citealt{Trager+00}, \citealt{BdeJ01}, \citealt{Ganda+07},
\citealt{Peletier+07}), but we are aware that, in some cases, e.g.
in later-type systems with very protracted star formations
(\citealt{Kennicutt83}), they may fail to reproduce correctly the
galaxy properties (\citealt{MacArthur+09}). A finer analysis would
require a more realistic SF history for LTGs (e.g., an
exponentially decaying or a constant SF), which is beyond the
scope of this paper.

We adopt a \cite{Chabrier01,Chabrier02,Chabrier03} initial mass
function (IMF), noting that the choice of a \cite{Salpeter55} IMF
would affect the total stellar mass only, shifting $\log \mst$
upward by $\sim 0.25$ and leaving all the other stellar population
parameters unchanged (e.g. \citealt{Tortora2009}). Our choice of
the BC03 models is driven by their high versatility and ability to
span the stellar parameter space, e.g. metallicities and ages.
There are other options (\citealt{Rettura06}, \citealt{KG07},
\citealt{Tortora2009} for a comparison of the results from
different prescriptions), but none is found to be bias free
(\citealt{Maraston05}, \citealt{vdW+06}, \citealt{Maraston+09},
\citealt{Conroy09a,Conroy10,Conroy10b}).

In our model procedures we adopt a metallicity ranging within
$0.0001 - 0.05$, and the Universe age, $t_{\rm univ}$ (in the
assumed cosmology), as an upper limit for the galaxy ages.
Synthetic colors have been obtained after convolving the spectra
with the corresponding SDSS filter bandpass functions. To improve
the sensitivity of the results to the small differences in both
$Z$ and ages, we have adopted a ``mesh refinement'' procedure
which interpolates the synthetic models over the initial grid of
equally spaced bins in the logarithm of $Z$ and age. To estimate
the galaxy age, $Z$, and the stellar mass-to-light ratio, \Yst,
we use a $\chi^2$ minimization fit (see \citealt{Tortora2009,
Tortora2009AGN} for details) to the observed colors ($u-g$, $g-r$,
$g-i$, and $g-z$). It is known that the use of just the optical
broad bands is severely prone to the well known age-metallicity
degeneracy. However, following previous indications (e.g.
\citealt{Wu+05}) we have verified in App. B that the use of near
infrared (NIR) photometry reduces the overall uncertainties but
does not modify the estimates of the intrinsic population
parameters based on optical bands only. We shall return on this
issue later on.

We have used the structural parameters given by B05 to derive the
color profile $(X-Y)(R)$ of each galaxy as the differences between
the (logarithmic) surface brightness measurements in the two
bands, $X$ and $Y$. The CG is defined as the angular coefficient
of the relation $X-Y$ vs $\log R/\Re$, $\displaystyle \nabla_{X-Y}
= \frac{\delta (X-Y)}{\delta \log (R/R_{\rm eff})}$, measured in
$\rm mag/dex$ (omitted in the following unless needed for
clarity). The fit of each color profile is performed in the range
$R_{1}= 0.1 \Re \leq R \leq R_{2}= \Re$. We have verified that a
choice of different radial ranges for the fit leaves the global
trends unaffected. In particular, by setting $R_{2}= 2 \Re$ the
gradients changed by $\lsim 0.01-0.02 \, \rm mag$, where $\Re$ is
the $r$-band effective radius (\citealt{Peletier+90b},
\citealt{LaBarbera2005}). Slightly different definitions are also
used elsewhere (see discussion in \citealt{Liu+09}).

By definition, a positive CG, $\nabla_{X-Y}>0$, means that a
galaxy is redder as $R$ increases, while it is bluer outward for a
negative gradient. The fit of synthetic colors is performed on the
colors at $R_{1}$ and $R_{2}$ and on the total integrated colors.
Following \cite{LaBarbera2005} we define the stellar parameters
gradients as $\gage =\log [\rm age_{2}/\rm age_{1}]$ and $\gZ=\log
[\rm Z_{2}/ \rm Z_{1}]$, where $(age_{i}, Z_{i})$ with $i=1,2$ are
the estimated age and metallicity at $R_{1}$ and $R_{2}$,
respectively.

A full test of the reliability of our modelling technique, the
presence of spuriously generated correlations, and the
contribution from internal dust extinction are addressed in App.
\ref{app:app_fit}.

\section{Results}\label{sec:results}
We show how color and stellar population gradients depend on
structural parameters and mass, and compare the results with the
predictions from galaxy formation scenarios. The median trends,
e.g. in bins of mass, velocity dispersion, colors etc., will be
represented along with the 25-75th percentiles in each bin.

The average behaviour of galaxies is analyzed in the
parameter space, and the galaxy properties discussed after splitting
the sample in the two morphological classes, ETGs and LTGs.
We concentrate mainly on the analysis of ETGs and compare our
findings with a wide collection of independent data and
simulations.

\subsection{CGs as a function of structural parameters, luminosity and
mass}\label{sec:results_1}

Fig. \ref{fig: fig0} presents the gradients \ggi\ as a function of
integrated colors, structural parameters, $r$-band magnitude,
central velocity dispersion, and stellar mass. We adopted the
$g-i$ color as a reference having varified that the use of
different colors does not affect our conclusions. As shown in the
top-left panel of the figure, CGs are, on average, positive for
the galaxies with bluer colors, $g-i \lsim 0.5$, while they reach
increasingly negative values for redder galaxies (e.g.
\citealt{PC05, Lee+07, Suh+10}). Since $g-i \sim 0.5$ marks the
transition from the RS to the BC, this result allows us to broadly
associate negative and positive CGs to the RS and BC,
respectively. Furthermore, galaxies with intermediate colors ($g-i
\sim 0.5-0.7$ or equivalently $g-r \sim 0.2-0.6$) have almost null
(or slightly negative) gradients. The same trend holds by
considering central and outer colors (at $R=R_1$ and $R_{2}$). We
will return to this issue later on.

Tight trends are found when plotting the gradients against the
$r$-band total magnitude and stellar mass, while the dependence on
\sigc\ is weaker. In the latter case, gradients look negative
almost everywhere with the less massive galaxies ($\log\sigc<2\,
\rm km/s$) having the smallest $\ggi$ ($\sim -0.2$). Although our
sample does not include very massive galaxies ($\mst \gsim
10^{11.5}\, \rm \Msun$), our results seem to show a stable trend
at the large mass scales, pointing to even shallower gradients. We
note here that the smaller range shown by $\ggi$ as a function of
\sigc\ in the central panel (left column) of Fig. \ref{fig: fig0}
is mainly due  to the mix of the ETGs and LTGs. They have
different trends with small \sigc\ in the low mass regime (see
also \S\ref{sec:morph}), thus producing a dilution of the average
gradients with respect, e.g., the larger median $\ggi$ ($\sim 0
\div 0.2$) found for the low stellar mass (right column).

\begin{figure}
\psfig{file= 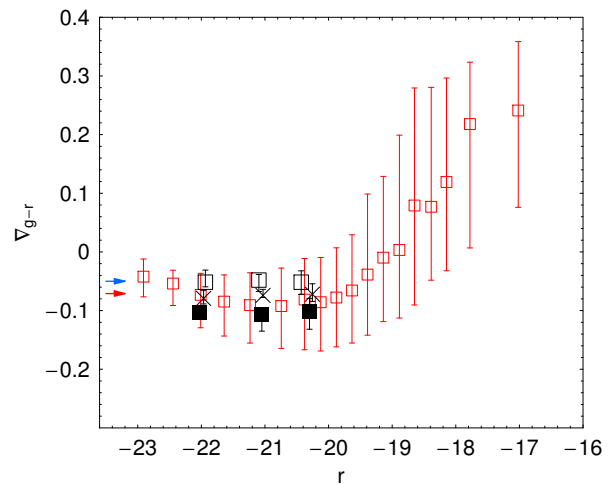, width=0.45\textwidth} \caption{CG \ggr\ as
a function of $r$-band magnitude for ETG sample. Superimposed are
the results from \citet{LaBarbera2005} for high-richness (open
boxes), low-richness (filled boxes), and all (crosses) the
clusters. The blue and red arrows are the median \ggr\ for the
massive galaxies analyzed in \citet{Wu+05} and
\citet{LaBarbera2009} respectively; see the text for details.}
\label{fig: fig3}
\end{figure}

\begin{table}
\centering \caption{Slopes of the correlation between CGs (\gug,
\ggr, \ggi, \ggz), \gage\ and \gZ\ vs $\log \sigc$ and $\log \mst$
for ETGs and LTGs. The errors on slopes are computed as $1\sigma$
uncertainty by a bootstrap method. $^{\large \star}$ For the LTG
sample, splitting galaxies by masses smaller/larger than $\log
\mst \sim 10.3$ would result in slopes of $-0.07\pm0.01$ and
$0.26\pm0.25$, respectively.}\label{tab:slopes_grad_Z_age}
\scriptsize
\begin{tabular}{lccc} \hline \hline
 \rm Correlation & \multicolumn{2}{c}{$\gamma_{\rm ETG}$} & $\gamma_{\rm LTG}$  \\
  \hline
   & $\log \sigc < 2.2$ & $\log \sigc > 2.2$ &   \\
$\gug - \log \sigc$ & $-0.60\pm 0.03$ & $0.40 \pm 0.05$ & $0.02\pm 0.05$\\
$\ggr - \log \sigc$ & $-0.27\pm 0.01$ & $0.19 \pm 0.01$ & $-0.02\pm 0.02$\\
$\ggi - \log \sigc$ & $-0.34\pm 0.02$ & $0.25 \pm 0.04$ & $-0.03\pm 0.02$\\
$\ggz - \log \sigc$ & $-0.40\pm 0.02$ & $0.17 \pm 0.04$ & $-0.09
\pm 0.03$\\
$\gage-\log \sigc$ & $-0.10\pm0.03$ & $0.04\pm0.17$ & $0.14 \pm 0.02$\\
$\gZ-\log \sigc$ & $-0.86\pm0.07$ & $1.02\pm0.24$ & $0.06 \pm 0.10$\\
\hline
   & $\log \mst < 10.3$ & $\log \mst > 10.3$ &   \\
$\gug - \log \mst$ & $-0.43 \pm 0.04$ & $0.12 \pm 0.02$ & $-0.30 \pm 0.01$\\
$\ggr - \log \mst$ & $-0.24 \pm 0.02$ & $0.05 \pm 0.01$ & $-0.13 \pm 0.01$\\
$\ggi - \log \mst$ & $-0.34 \pm 0.02$ & $0.07 \pm 0.01$ & $-0.18 \pm 0.01$\\
$\ggz - \log \mst$ & $-0.36 \pm 0.03$ & $0.02 \pm 0.01$ & $-0.18 \pm 0.01$\\
$\gage-\log \mst^{\large \star}$ & $-0.18\pm0.06$ & $0.24\pm0.03$ & $-0.01\pm 0.01$\\
$\gZ-\log \mst$ & $-0.46\pm0.06$ & $0.10\pm0.04$ & $-0.45\pm 0.01$\\
\hline \hline
\end{tabular}
\end{table}

The correlations with luminosity and mass show that, starting from
the bright/massive end, the CGs become steeper, with the steepest
negative gradients ($\sim -0.2$) corresponding to $r \sim -20$ mag
and $\mst \sim 10^{10.3}\, \rm \Msun$. From there, bluer and
later-type systems begin to dominate, and the gradients invert the
trend with luminosity and mass, becoming positive at $r \sim
-17.5$ mag and $\mst \sim 10^{8.7}\, \rm \Msun$. The lowest mass
systems are those with the highest, positive gradients ($\sim
0.2$). The mass scale at which the CG trend inverts is in
remarkable agreement with the typical mass scale break for
``bright'' and ``ordinary'' galaxies (Capaccioli et al. 1992, see
also \S\ref{sec:morph}) or for  star-forming and passive systems
(\citealt{Kauffmann2003}, \citealt{DA08}).

The above trend holds also for the correlation with \Re\ and, less
strongly, with the S$\rm \acute{e}$rsic index, $n$, as expected in
view of the tight correlation with stellar mass and luminosity
(\citealt{Caon93}, \citealt{PS97}, \citealt{GG03},
\citealt{Shen2003}, \citealt{Trujillo+04}, \citealt{Tortora2009}).
Note that the minimum at $n\sim 2$ for the trend of the gradient
(Fig. \ref{fig: fig0}, bottom right) roughly coincides with the
value used here to set the separation between ETG and LTG systems,
suggesting differences in the CG properties.

\begin{figure*}
\psfig{file= 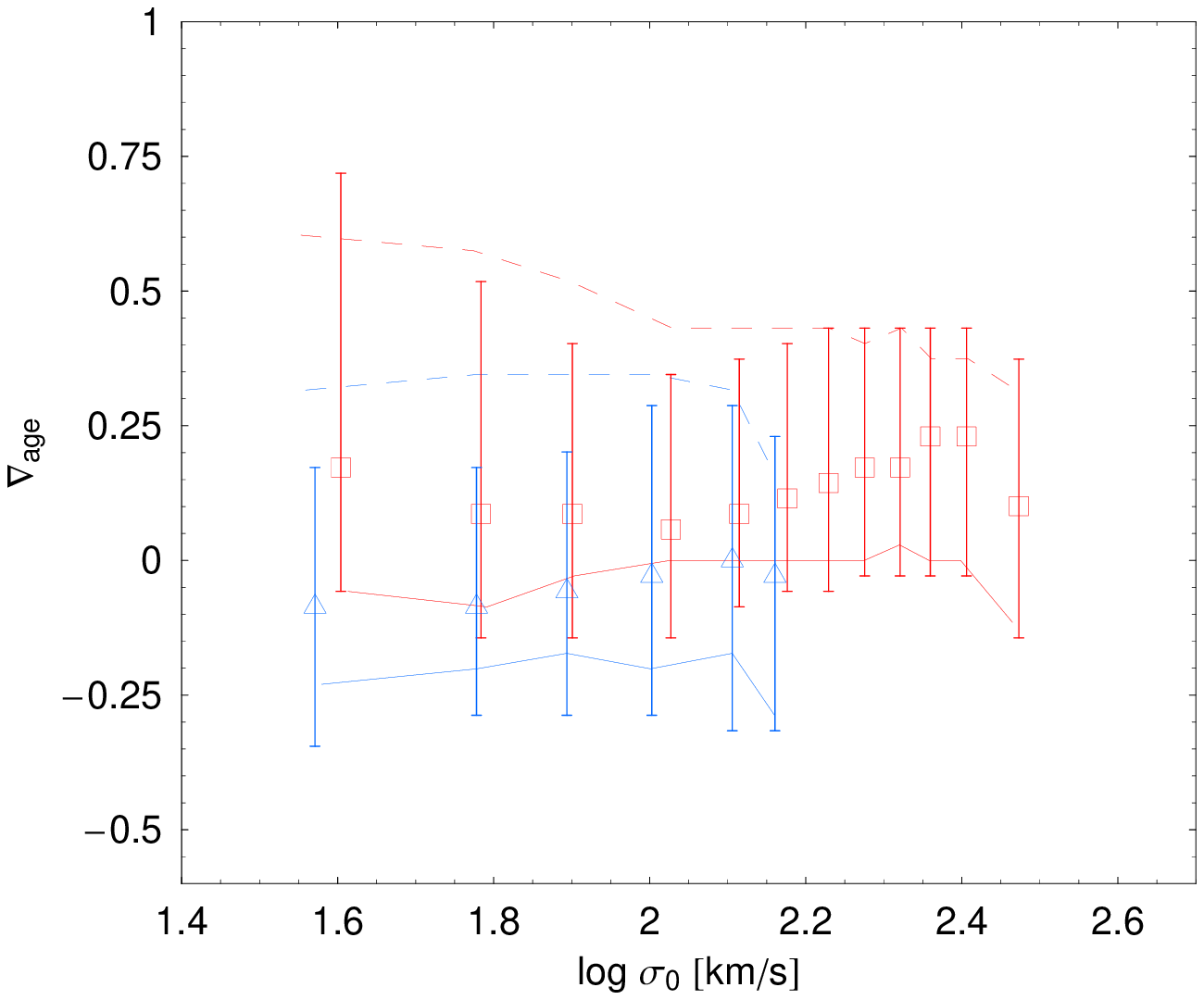, width=0.43\textwidth} \psfig{file=
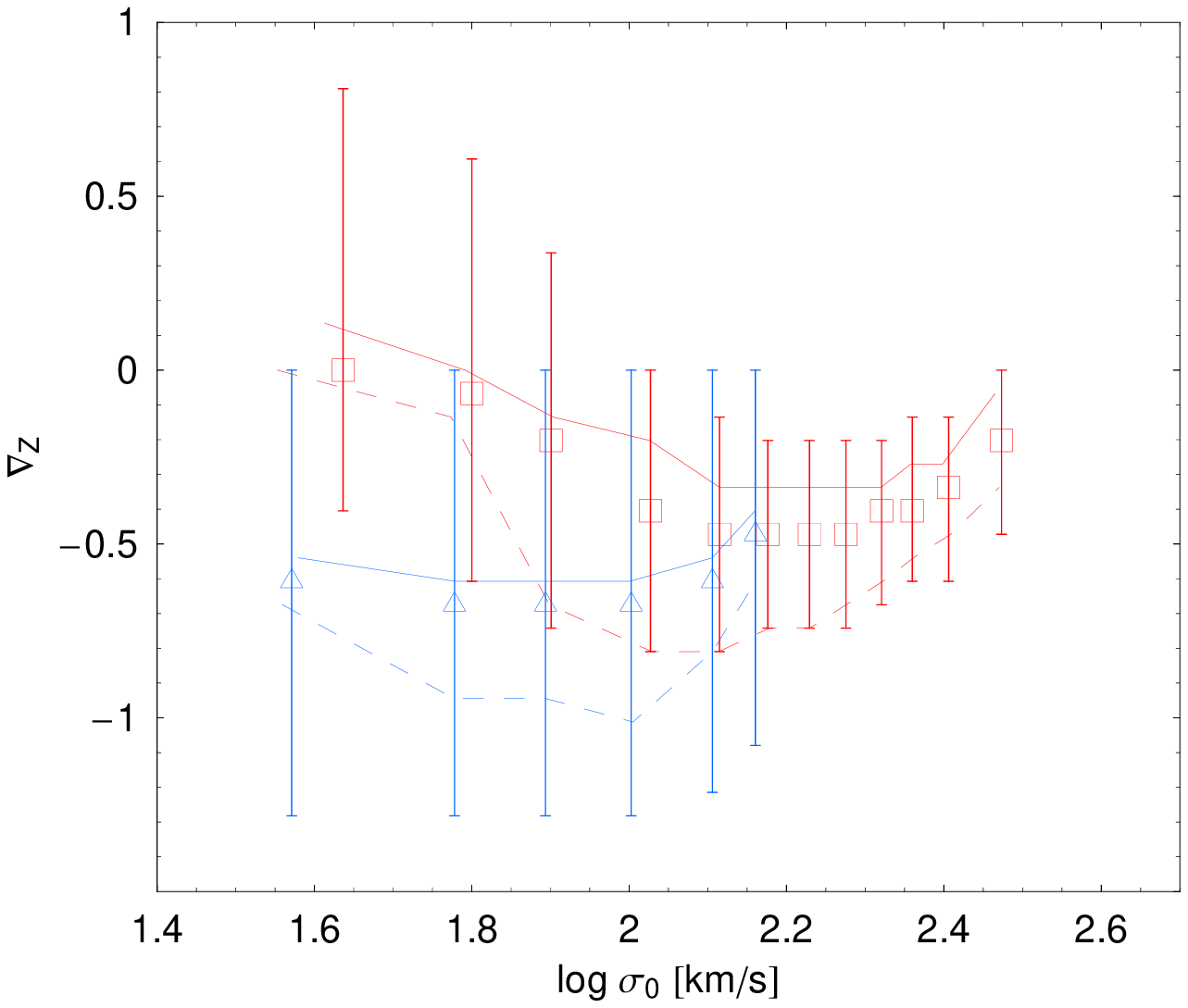, width=0.43\textwidth}\\
\psfig{file= 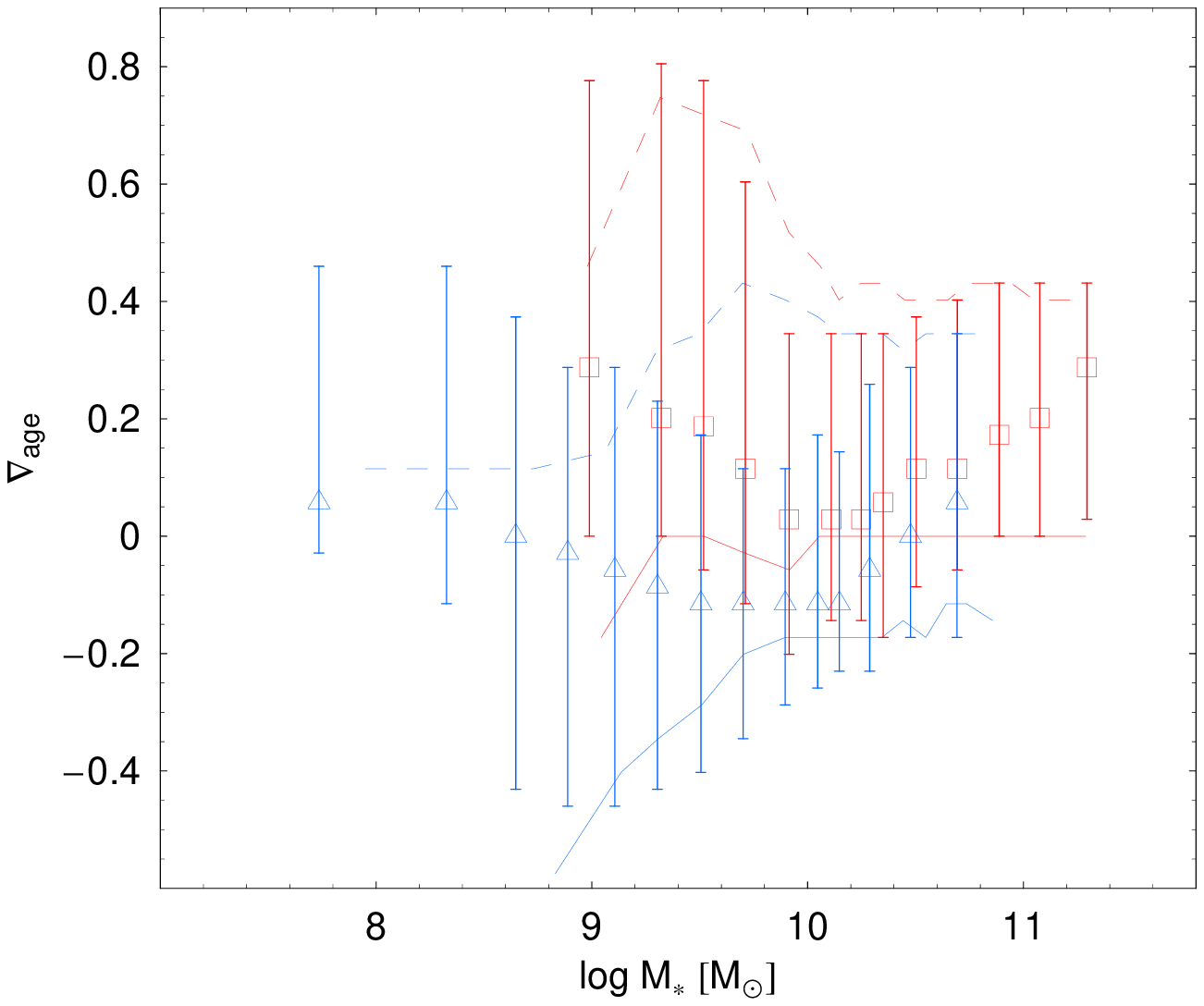, width=0.43\textwidth} \psfig{file=
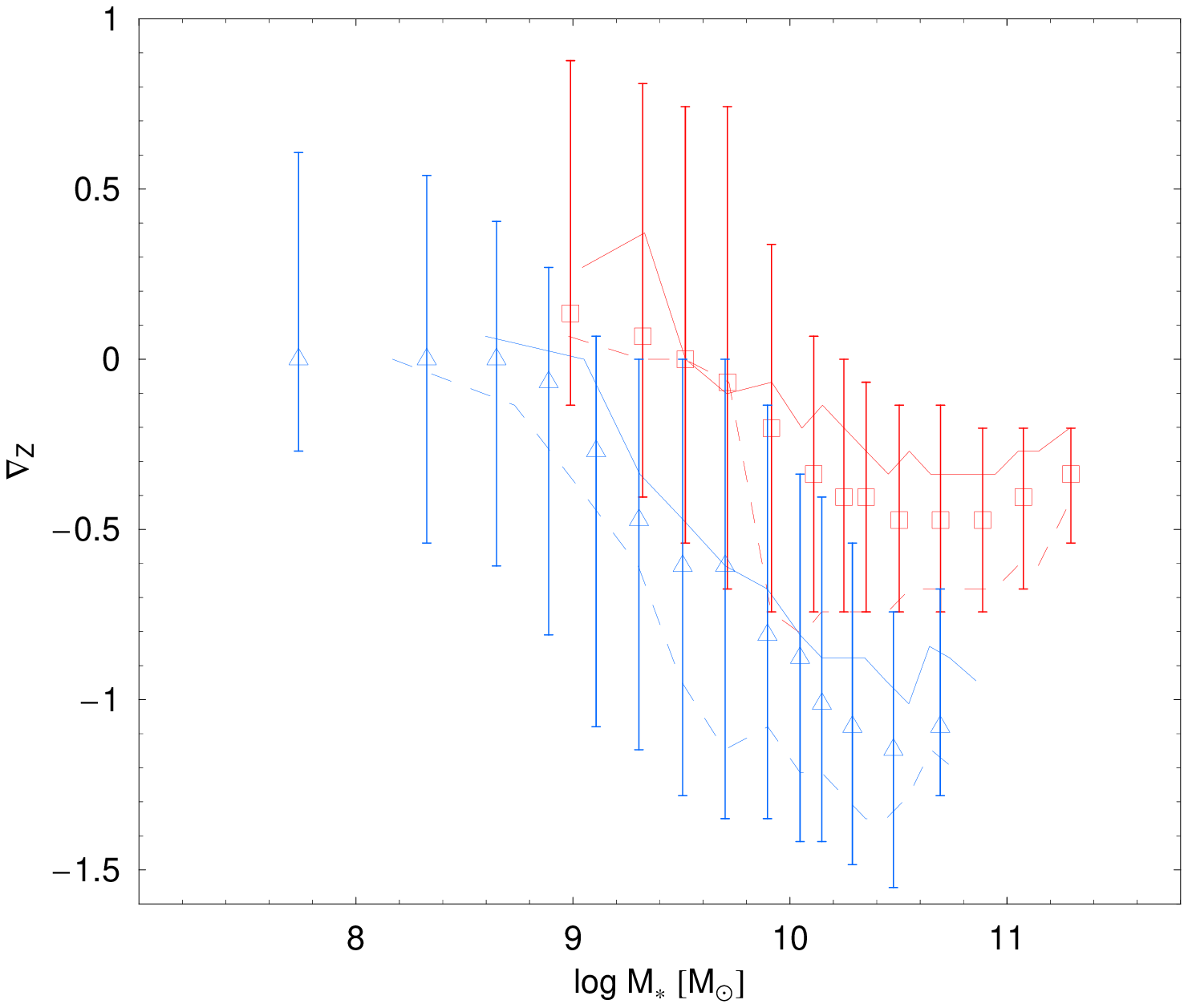, width=0.43\textwidth}\caption{Age and metallicity
gradients as a function of velocity dispersion (top panels) and
stellar mass (bottom panels). Solid and dashed lines are for
galaxies with central stellar populations older and younger than
$\rm 6 \, Gyr$, respectively. Symbols are as in Fig. \ref{fig:
fig2}} \label{fig: fig4}
\end{figure*}

\subsection{Gradients in morphologically selected
galaxies}\label{sec:morph}

Following the dependence of \ggi\ vs $n$ just discussed, ETG and
LTG samples will be now considered separately. In Fig. \ref{fig:
fig2} we plot \ggi\ versus \Re, \sigc , and \mst\ for ETGs (red
symbols) and LTGs (blue symbols). The slopes of the linear
best-fit relations of the CGs versus \sigc\ and \mst\ are shown in
Table \ref{tab:slopes_grad_Z_age}.

First, ETGs are on average brighter and more massive than LTGs
(\citealt{Kauffmann2003}) and have less dispersed CGs. As a
general result, large \Re\ galaxies are found to have lower
gradients than more compact systems (e.g. \citealt{Hopkins+09a}).
However, for LTGs, \ggi\ monotonically decreases with $\Re$ and
becomes flat at $\log \Re \gsim 0.5$, while for ETG it slightly
increases for $\log \Re \gsim 0.3$, and decreases for smaller \Re.
Similar trends are observed for the gradients as a function of the
stellar mass. LTGs show a monotonic decreasing trend (in
qualitative agreement with \citealt{KA90} and \citealt{Liu+09}),
while a U\,-\,shaped function is found for ETGs with the gradients
definitely decreasing with the mass for $\mst \lsim
10^{10.3-10.5}\, \rm \Msun$, and a mildly increasing for larger
mass values.  This mass scale roughly corresponds to a total
luminosity of $r\sim -20.5$ mag, which is compatible with the
typical luminosity (and mass) scale for ETG dichotomy in the
galaxy structural properties (e.g., \citealt{Capaccioli92a},
\citealt{Capaccioli92b}, \citealt{2003AJ....125.2951G}). As
discussed in \citet{Capaccioli92a} this dichotomy may be related
to the formation mechanism processes, and in particular to the
dominance of the merging events for the ``bright'' sample, which
is compatible with the shallower gradients that we observe for
these systems (see, e.g., Fig. \ref{fig: fig2}, right panel).
Nevertheless, for a fixed stellar mass we observe that ETGs
gradients are, on average, larger than LTG ones.

The same two\,-\,fold trend is shown for ETGs gradients as
function of the velocity dispersion, while no trend with the mass
is observed for LTGs gradients.

As a consistency check, in Fig. \ref{fig: fig3} we superpose to
our $g-r$ gradients as a function of the $r-$band luminosity the
results of \cite{LaBarbera2005} obtained by analyzing a sample of
low redshift luminous ($r \leq -20$ mag) cluster galaxies. Their
value ($\ggr \sim -0.075$) is in agreement with our findings over
the range where the two studies overlap, but we were also able to
identify a change of \ggr\ with our larger luminosity baseline.
The results from the massive galaxies in \cite{Wu+05} and
\cite{LaBarbera2009} are also shown to have \ggr\ consistent with
our results.

Unfortunately, we have no information on the environment of the
individual systems in our sample, to check the dependence of our
results with the local density. There are many lines of evidences
that ETGs in clusters (and in very dense environments in general)
have shallower gradients than systems in less dense environments
(\citealt{TO2000}, \citealt{T+00}, \citealt{LaBarbera2005} -- see
also Fig. \ref{fig: fig3}). However, according to the
morphological segregation, LTGs, which are expected to reside in
the field or in the external regions of clusters, have indeed
lower gradients than ETGs which are found to live mainly in higher
density environments.

\subsection{Interpretation in terms of Z and age
gradients}\label{sec:age_met}

Here we test whether the CGs can be explained in terms of stellar
properties such as age and metallicity. As seen in
\S\ref{sec:data}, we have obtained the gradients of metallicity
and age distributions using two characteristic scales ($0.1 \Re$
and $1 \Re$). The \gage\ and \gZ\ are shown in Fig. \ref{fig:
fig4} as a function of the velocity dispersion and the stellar
mass, and the fitted slopes of such relations are quoted in Table
\ref{tab:slopes_grad_Z_age}.

Some features survive the large scatter. For LTGs the \gage\ is
about zero both with \sigc\ and mass, strongly suggesting that CGs
should not depend on age gradients of the galaxy stellar
population\footnote{The scatter of the gradients around zero has
been suggested to be a manifestation of possible differences in
the formation processes in these systems \cite{MacArthur+09}.}.
Actually, \gZ\ is instead strongly dependent on \mst, with the
lowest metallicity gradients ($\sim -1$) at the largest masses,
while the dependence on \sigc\ is very weak as the
$\gZ\sim(-0.5,-0.6)$ for all \sigc.

ETG age gradients seem basically featureless when plotted against
\sigc\, where we find $\gage > 0$ with typical median values of
$\gage \sim 0.2$. On the other hand, the metallicity gradients
show some features with \gZ\ decreasing for $\log \sigc \lsim 2.2
\, \rm km/s$, with a slope $= -0.86$, and increasing for larger
velocity dispersion, with a slope of $1.02$, reaching the
shallowest values ($\sim -0.2$) at $\log \sigc \gsim 2.4 \, \rm
km/s$ and at the very low $\sigc$ (i.e. $\log \sigc<1.8\, \rm
km/s$) where \gZ$\sim0$. This peculiar trend is mirrored by a
similar dependence on the stellar mass\footnote{For ETGs, stellar
mass and \sigc\ are tightly correlated, thus the correlation and
the interpretation discussed here can be made when dissecting the
similar trend as a function of velocity dispersion. On the
contrary, many low mass LTGs have very large velocity dispersion
which contribute to have almost constant metallicity gradients for
$\log \sigc \lsim 2.2 \, \rm km/s$ and increasing at larger \sigc.
But in this case the scatter is very large suggesting that the
central velocity dispersion is not a representative
parameter of such rotational velocity supported systems.}. This
two-fold behaviour is also significant when we plot \gage\ as a
function of the stellar mass: \gage\ decreases at $\log \mst \lsim
10.3-10.5$, and increases in more massive systems.

\begin{figure}
\psfig{file= 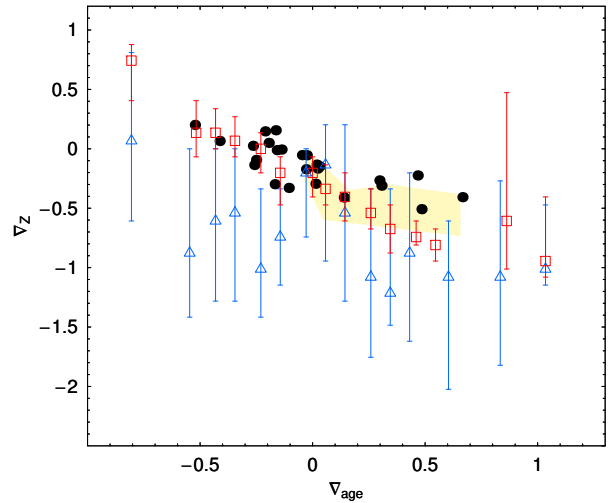, width=0.45\textwidth} \caption{Metallicity
vs age gradients for ETGs and LTGs. The symbols are the same as in
Fig. \ref{fig: fig2} and \ref{fig: fig4}. Metallicity vs age
gradients for ETGs are compared with predictions from gas-rich
mergings in \citet{Hopkins+09a} (yellow shaded region) and cluster
ETGs in \citet{Rawle+09} (black points).} \label{fig: fig5}
\end{figure}

Bright ETGs with $r \lsim -20$ mag ($\log \mst \gsim 10$) have a
median $\gage = 0.144^{+0.004}_{-0.003}$ (with scatter $^{+0.26}
_{-0.17}$) and $\gZ = -0.472^{+0.005}_{-0.004}$ (with scatter
$^{+0.34} _{-0.27}$), where the error bars represent the $1
\sigma$ uncertainty and the scatters the 25-75 percentiles of the
sample distribution. However, we have found that \gage\ and \gZ\
strongly depend on galaxy age: in particular, if we separate the
sample in systems with older and younger than $6 \, \rm Gyr$
central stars, we obtain different age and metallicity gradient
trends as shown in Fig. \ref{fig: fig4}. Older systems have
gradients which are shallower than younger (in particular for
ETGs) and bracket the average trend of the whole samples. A
partial contribution to the scatter is also given by the central
metallicity: the lower-Z systems show generally shallower
gradients and viceversa. However, the effect is significantly
smaller with respect to the central age.

The median values of the older sample is $\gage =
0^{+0.003}_{-0.004}$ (with scatter $^{+0.14} _{-0.14}$) and $\gZ =
-0.337^{+0.006}_{-0.005}$ (with scatter $^{+0.27} _{-0.20}$) while
for the younger one we obtain $\gage = 0.43^{+0.001}_{-0.006}$
(with scatter $^{+0.08} _{-0.20}$) and $\gZ =
-0.67^{+0.007}_{-0.004}$ (with scatter $^{+0.27} _{-0.13}$).
Galaxies with old central ages are fully consistent with the
results of pure metallicity and mixed age+metallicity models in
\cite{LaBarbera2005} and with the best fitted values in
\cite{LaBarbera2009}  (where galaxy ages are in the range of
$7.8-12.6\, \rm Gyr$). The agreement with \cite{Rawle+09} is also
good if we consider that their sample has a median central age of
$\sim 10 \, \rm Gyr$ (calculated within an $\Re/3$ aperture),
comparable to the old objects of our sample: they find an age
gradient of $-0.02 \pm 0.06$ and metallicity gradients of $-0.13
\pm 0.04$. Finally, from the analysis of 36 nearby ETGs with SDSS
and 2MASS photometry, \cite{Wu+05} have derived $\gage = 0.02 \pm
0.04$ (and scatter $0.25$) and $\gZ = -0.25 \pm 0.03$ (and scatter
$0.19$), which turn to be qualitatively consistent again with our
``old'' sample. The inclusion of dust extinction leaves our
general considerations unaffected, as we will show in App.
\ref{app:appA3}. The average trends of the metallicity and age
gradients are almost unchanged, with variations going in the
direction of a better match with other literature data.

The massive ETGs, with young cores ($<6 \, \rm Gyr$) have very
steep metallicity gradients, which are discrepant with respect to
the literature results we have compared with. In App.
\ref{app:appA5} we have checked that this result does not depend
on the contamination of the ETG sample from late-type systems.
These latter, indeed, also show the same dependence of gradients
on central age (see Fig. \ref{fig: fig4}). A clear example of
systems with very steep gradients are the isolated galaxy NGC 821
which have a young central stellar population and a metallicity
gradient of $\gZ = -0.72 \pm 0.04$ or NGC 2865, for which is  $\gZ
=-0.47 \pm 0.05$ (\citealt{Proctor+05}, \citealt{Reda+07}). One
reason why we have been able to pick up this difference between
centrally old and young systems is that most of the samples
analyzed in literature are made up of cluster ellipticals (e.g.
\citealt{Mehlert+03}, \citealt{Rawle+09}), with only few cases of
field galaxies (\citealt{Ogando+05}, \citealt{Reda+07}), while our
sample is composed by galaxies living in various environments. On
average, field galaxies are found to be younger and have a wider
distribution of galaxy age (\citealt{Bernardi+98},
\citealt{Trager+00}, \citealt{Thomas2005}), thus, a large fraction
of our centrally young galaxies would be made of systems in the
field or low density environments with steeper age and metallicity
gradients than those in cluster galaxies (due to the little
interactions with the neighborhood, see Sect. \ref{sec:morph} and
\cite{LaBarbera2005}).

Fig. \ref{fig: fig5} displays a relevant anti-correlation between
age and metallicity gradients: for ETGs the correlation is clear
with a slope of $-0.94 \pm 0.02$, while for LTGs it seems poorer
but still significant, with a slope $-0.48 \pm 0.03$ (see also
Table \ref{tab:slopes_grad_Z_age_2}).  This is an intrinsic
property of the galaxy sample, which is consistent with
predictions from the galaxy merging scenario (e.g.
\citealt{Hopkins+09a}) and with other observational evidences
(e.g. \citealt{Rawle+09}). It is not spuriously induced by the
well known age--metallicity degeneracy (see App. B). In
particular, galaxies with strong negative metallicity gradients
have positive age gradients (coherently with the merging
simulation in the figure), while the occurrence of negative age
gradients is associated to small mass systems with positive
metallicity gradients. The negative correlation found for LTGs is
also qualitatively consistent with the literature on spiral
galaxies (\citealt{MacArthur+04}).

\begin{figure*}
\psfig{file= 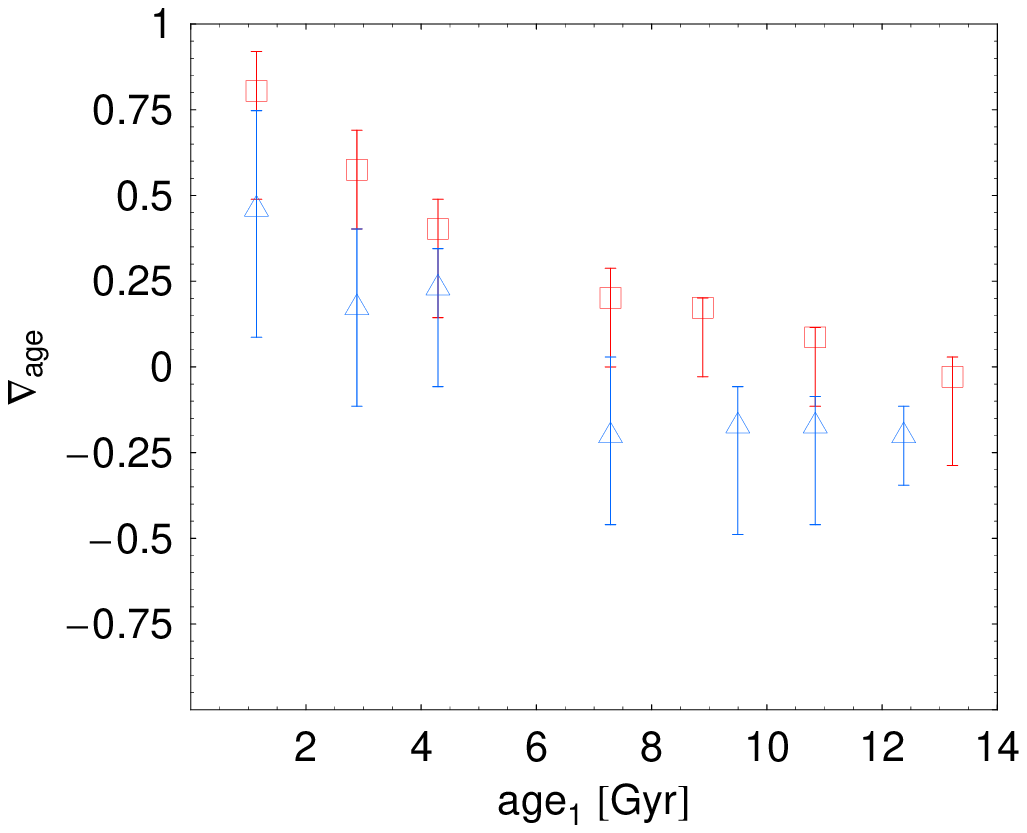, width=0.23\textwidth} \psfig{file=
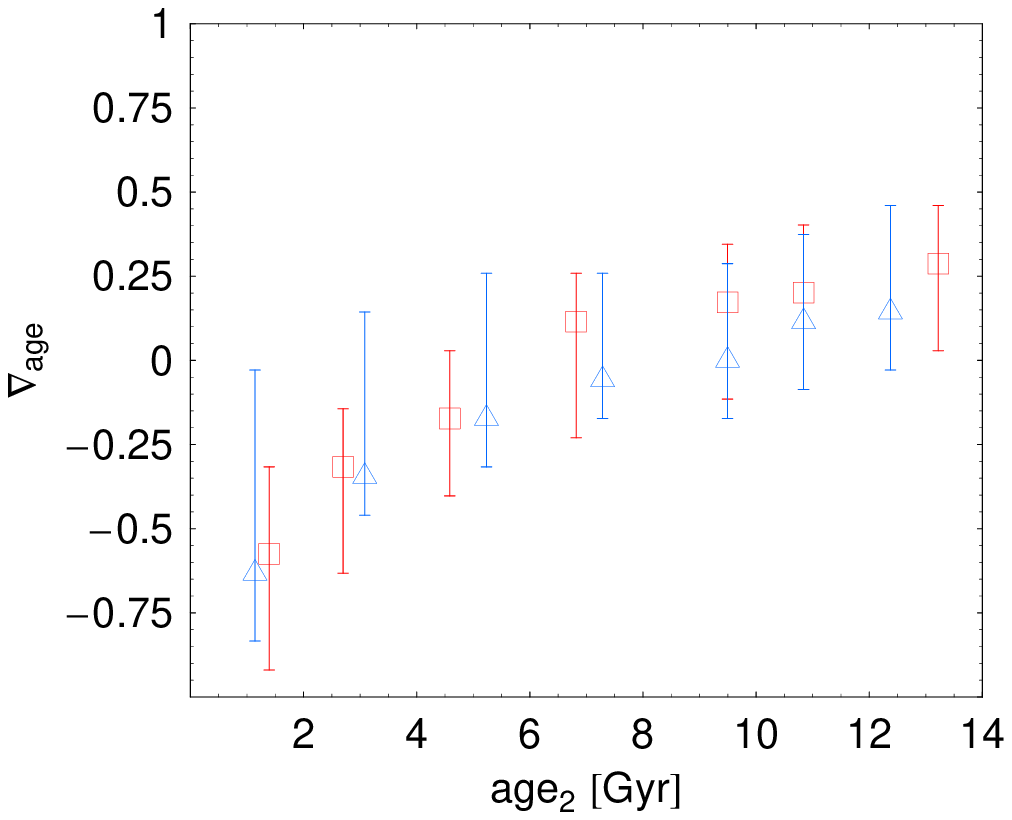, width=0.23\textwidth} \psfig{file= 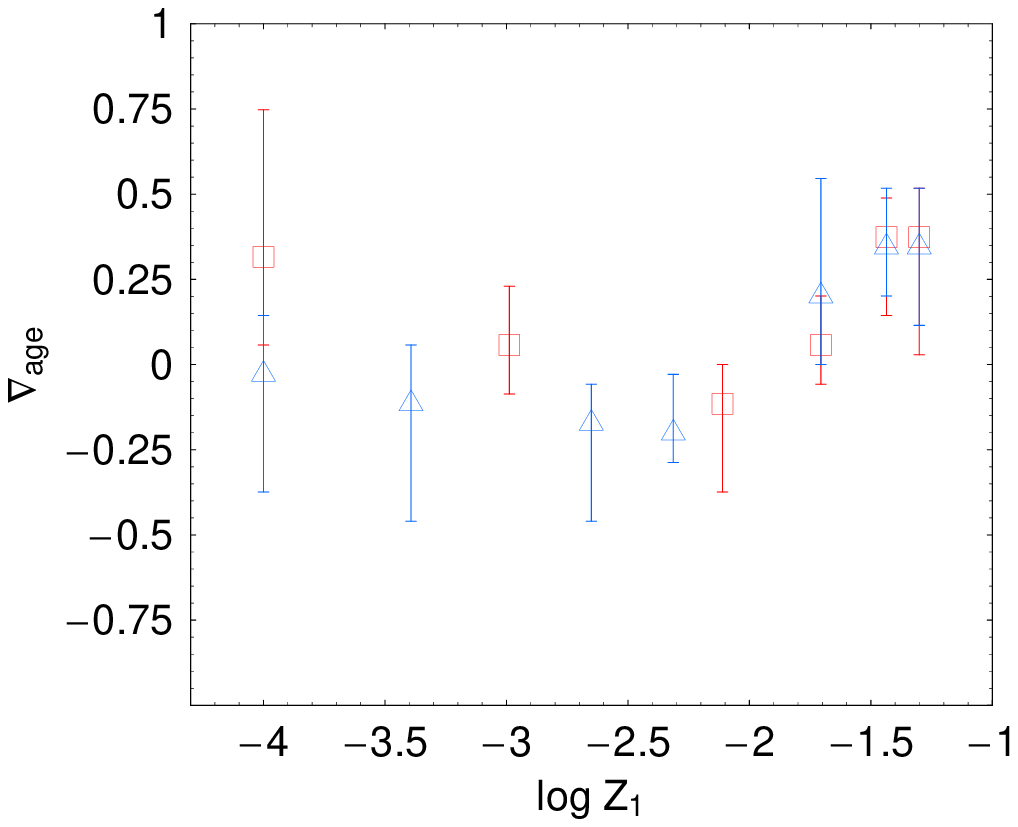,
width=0.23\textwidth} \psfig{file=
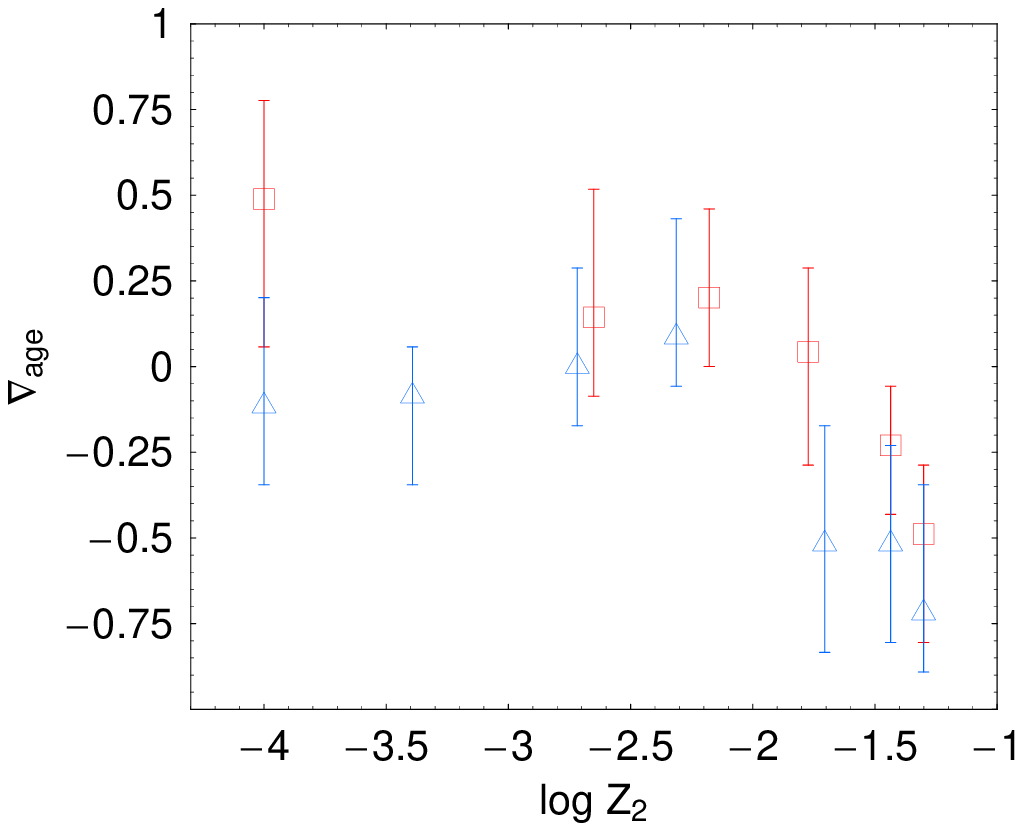, width=0.23\textwidth} \\
\psfig{file= 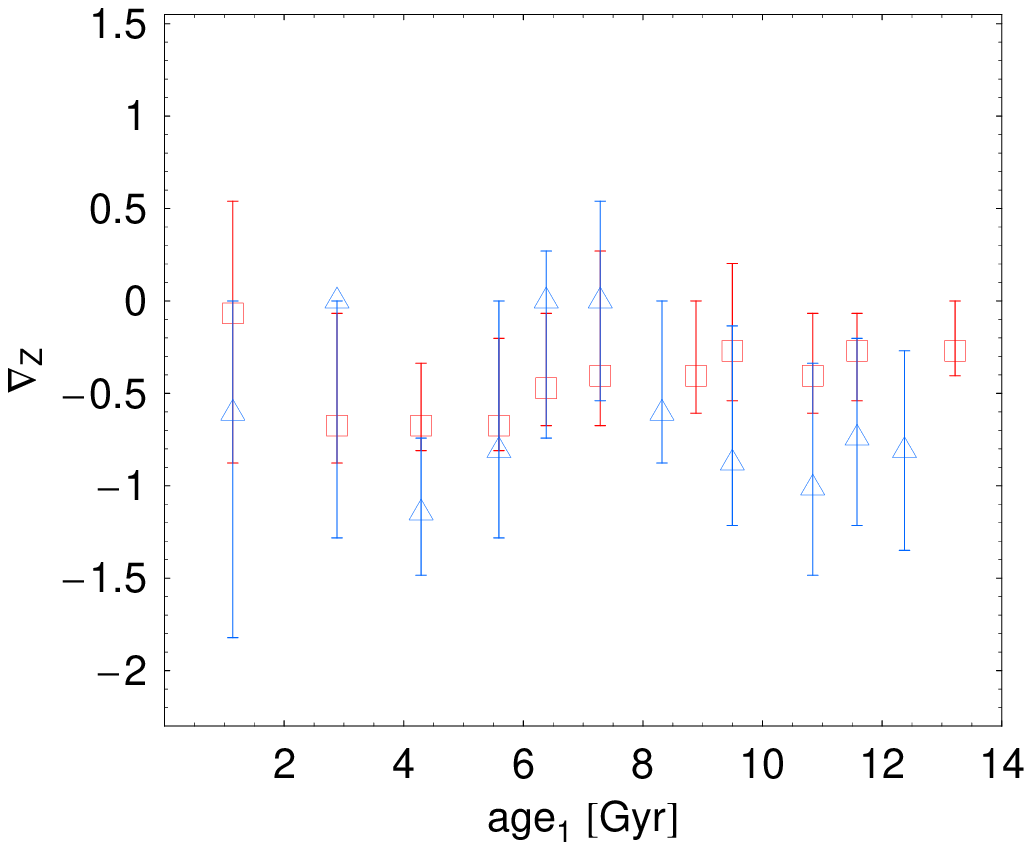, width=0.23\textwidth} \psfig{file=
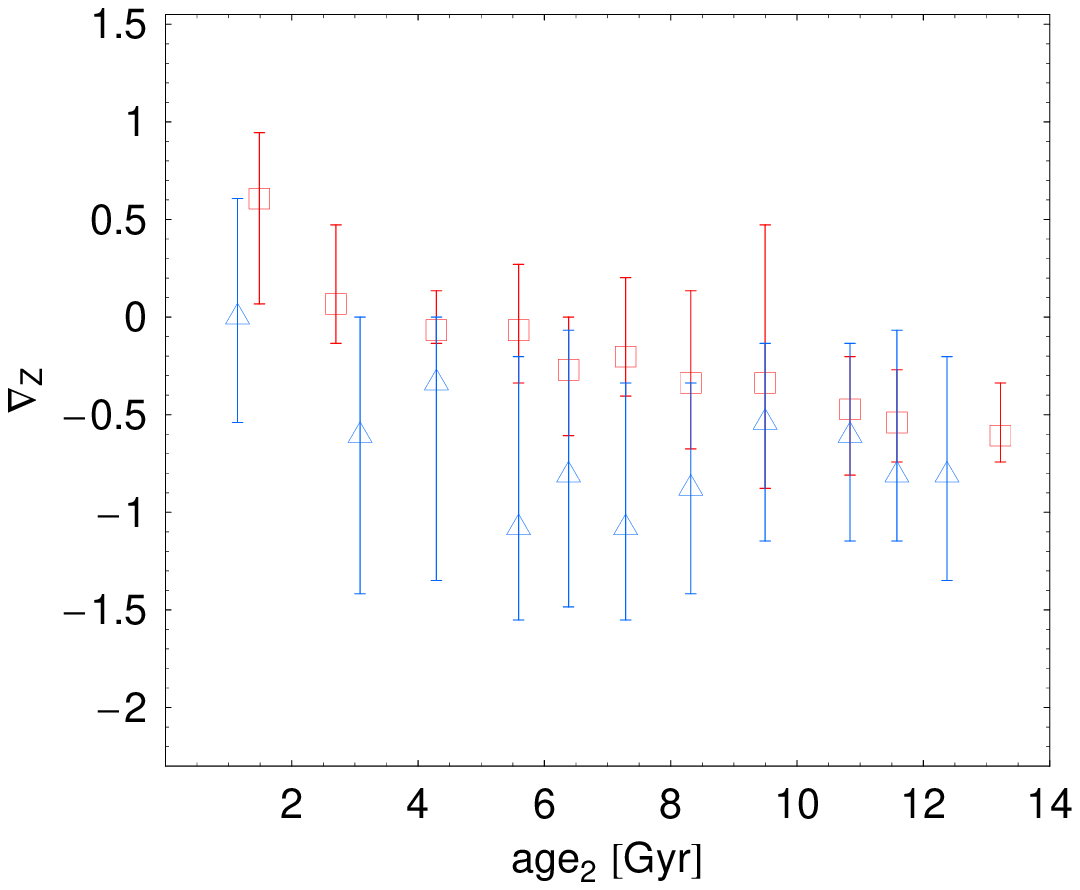, width=0.23\textwidth} \psfig{file= 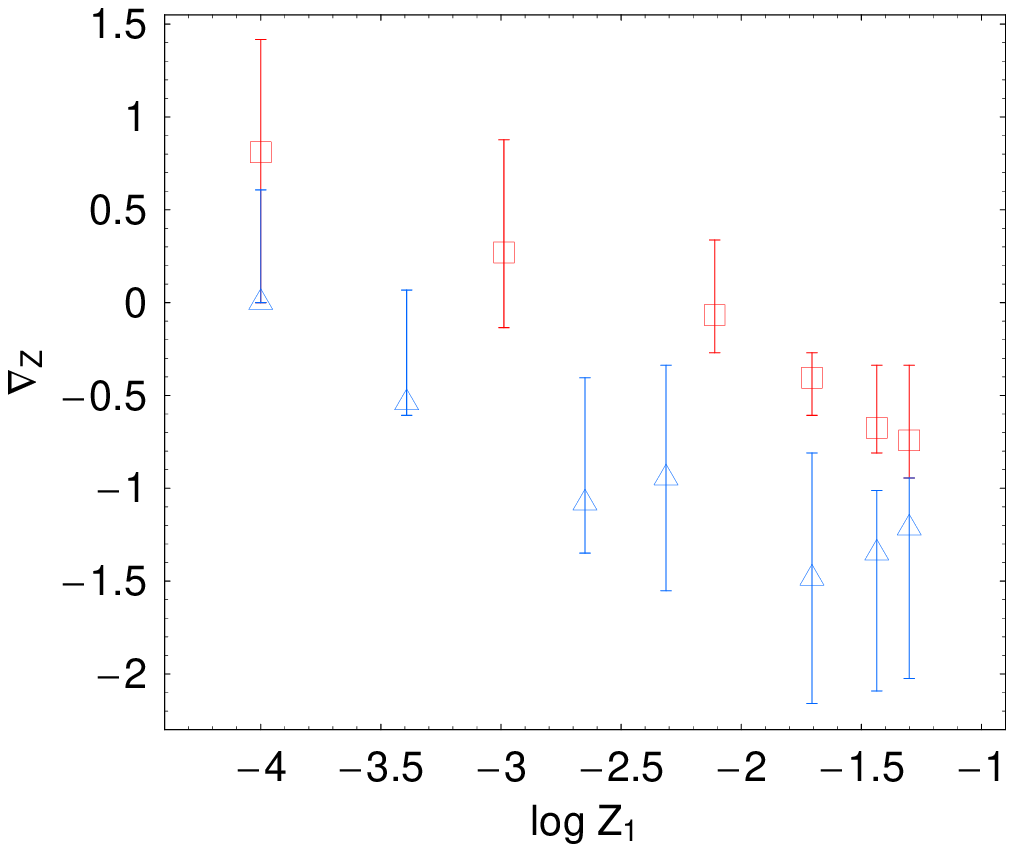,
width=0.23\textwidth} \psfig{file= 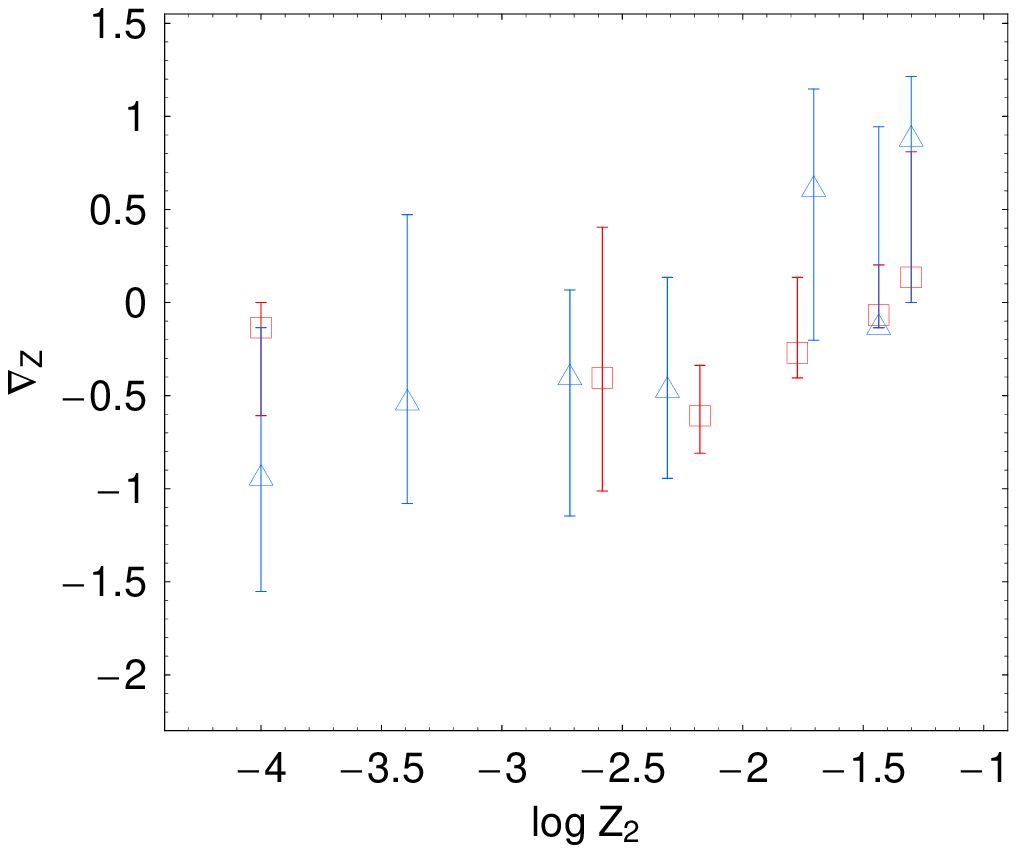,
width=0.23\textwidth}\\ \caption{Correlations of gradients and
fitted stellar parameters. The symbols are the same as in Figs.
\ref{fig: fig2}, \ref{fig: fig4} and \ref{fig: fig5}. {\it Top
panels.} From left, age gradient as a function of $age_{1}$,
$age_{2}$, $Z_{1}$ and $Z_{2}$. {\it Bottom panels.} From left,
metallicity gradient as a function of $age_{1}$, $age_{2}$,
$Z_{1}$ and $Z_{2}$.} \label{fig: fig5bis}
\end{figure*}

In Fig. \ref{fig: fig5bis} we finally show the correlations
between gradients and stellar parameters at different galaxy
radii, with their related slopes listed in Table
\ref{tab:slopes_grad_Z_age_2}. Metallicity and age gradients show
a tight correlation with central metallicity and age,
respectively. For the ETGs, the age gradients are generally
different from zero, positive for small central ages, and null or
negative for old systems. Metallicity gradients are negative for
systems with older stellar populations at \Re\ or with higher
metallicity, while they are about zero for systems with old
central ages or intermediate ages at \Re\ ($4-7$ Gyr). Noticeably,
the very positive \gZ\ and negative \gage\ correspond to very
young populations at \Re , i.e. recent star formation outside the
galaxy cores. These latter systems seem compatible with dwarf
galaxies hosting expanding shells (e.g. \citealt{Mori+97}, see
also \S\ref{sec:disc}).

\begin{table*}
\centering \caption{Slopes of the linear correlations between
stellar population gradients \gZ, \gage, $Z$ and $age$ for ETGs
and LTGs. IN and OUT indicate gradients correlated with inner (at
$R=R_{1}$) and outer (at $R=R_{2}$) metallicity and age. In the
last row the result for correlation between \gZ\ and \gage\ are
reported. The fitting procedure is the same as that in Table
\ref{tab:slopes_grad_Z_age}. $^{(*)}$: here the linear correlation
is a 0-th order approximation.}\label{tab:slopes_grad_Z_age_2}
\begin{tabular}{lcccc} \hline \hline
 \rm Correlation & \multicolumn{2}{c}{$\gamma_{\rm ETG}$} & \multicolumn{2}{c}{$\gamma_{\rm LTG}$}  \\
  \hline
   & $\rm IN$ & $\rm OUT$ & $\rm IN$ & $\rm OUT$  \\
$\gage-\log age$ & $-0.80\pm 0.03$ & $0.93\pm 0.03$ & $-0.78\pm 0.02$ & $0.75\pm 0.03$\\
$\gage-\log Z^{(*)}$ & $0.09\pm 0.01$ & $-0.34\pm 0.01$ & $0.21 \pm 0.01$ & $-0.24\pm 0.02$\\
$\gZ-\log age$ & $0.62\pm 0.05$ & $-1.07\pm 0.04$ & $-0.20\pm 0.06$ & $-0.62\pm 0.04$\\
$\gZ-\log Z$ & $-0.64\pm 0.02$ & $0.06\pm 0.02$ & $-0.49\pm 0.01$ & $0.66\pm 0.02$\\
\hline
$\gZ-\gage$ & \multicolumn{2}{c}{$-0.92\pm 0.02$} &  \multicolumn{2}{c}{$-0.50\pm 0.03$} \\
\hline \hline
\end{tabular}
\end{table*}

Similarly, LTGs have also positive age gradients in presence of
young cores, but in these cases we observe also large central
metallicities (see Fig. \ref{fig: fig5bis}, top row, third panel
from left) and for these systems we also found the strongest
metallicity gradients (bottom raw, third panel of Fig. \ref{fig:
fig5bis}).

Overall, results in Figs. \ref{fig: fig4} and  \ref{fig: fig5bis}
are in substantial agreement with the correlations reported in
\citet{Rawle+09} \footnote{In particular, if we limit our galaxies
sample to luminous systems ($r < -20$ mag) with $Z_{1} \geq 0.008$
and $age_{1} \geq 2 \, \rm Gyr$ and perform the fit with respect
to the central quantities within $\Re / 3$ we find that the slope
of the $\gZ - \log Z_{1}$ relation is $-0.58 \pm 0.04$ in good
agreement with the values in the range $[-0.6,-0.7]$ from
\citet{Rawle+09}. For $\gage - \log age_{1}$ relation we find a
slope of $-0.90 \pm 0.06$, which disagree with the shallower
values reported in \cite{Rawle+09}. A similar disagreement is
found for the $\gZ - \log age_{1}$ and $\gage - \log Z_{1}$
correlations, where we find slopes $0.67 \pm 0.05$ and $0.50 \pm
0.03$ respectively, which are still not perfectly consistent with
their results. These might be related to the small sample adopted,
to the selected environment (i.e. cluster cores), to differences
in population markers (i.e. line strengths rather than colors) and
the use of synthetic spectral models with variable
$\alpha$-enhancement over-abundance.} and with the findings in
hydrodynamical simulations of both elliptical and disc galaxies
(\citealt{Hopkins+09a}, \citealt{SB+09}).

\subsection{Comparison with literature data}\label{sec:compar}

We now proceed to a more detailed comparison of our findings with
a set of literature works which make use of a more sophisticated
analysis, although usually associated to smaller samples. In
particular, we  concentrate on the ETG sample, taking the results
for LTGs with the necessary caution, due to the larger degree of
uncertainties arising from, e.g., 1) the simple assumption on the
star formation recipe, 2) the assumption of no dust gradients (see
also App. \ref{app:appA3}), and 3) the prediction from theoretical
simulations of the formation of such composite and complex systems
which are still in their infancy (\citealt{MacArthur+09},
\citealt{MS+09}, \citealt{SB+09}).

Due to our minimal choice of the wavelength baseline and the
possible biases on the stellar population estimates caused by the
use of the optical band only, the comparison of our results with
literature data is critical to 1) ultimately check the robustness
of our results on the regions of the parameter space overlapping
with independent analysis and 2) assess the gain in the physical
information allowed by our analysis. In Fig. \ref{fig: fig6} we
summarize the metallicity and age gradients for ETGs as a function
of central velocity dispersion, and compare these with the observed
gradients from a collection of literature studies
(\citealt{Mehlert+03}, \citealt{Proctor03}, \citealt{Ogando+05},
\citealt{Reda+07}, \citealt{SB+07}, \citealt{Koleva+09a,
Koleva+09}, \citealt{Spolaor09}, \citealt{Rawle+09}). In the next
section we will expand this comparison to the predictions from
cosmological simulations. When considering massive ETGs, our \gZ\
are systematically steeper than literature data (see top-left of
Fig. \ref{fig: fig6}), although showing on average a lower scatter than
dwarf galaxies.

As seen in Fig. \ref{fig: fig5bis} and also illustrated in Fig.
\ref{fig: fig6} (top-right panel), a significant contribution to
the scatter of the metallicity gradients is linked to the age of
galaxy cores, with older galaxies having, on average, shallower
gradients than younger systems. When considering objects with
$age_{1}> 6 \, \rm Gyr$, the agreement with the other studies
(generally dealing with old systems; see \S\ref{sec:age_met}) is
remarkably good. The same conclusion is reached when considering
the \gage\ (Fig. \ref{fig: fig6}, bottom-right), where the steeper
(positive) gradients are shown by the younger systems. In App.
\ref{app:appA3} we show that shallower age gradients might be
found if the dust extinction is accounted in the stellar models
for the massive ETGs, while metallicity gradients remain almost
unchanged.

The dependence of the gradients on the galaxy age is consistent
with results in the dwarf regime by \cite{Spolaor09} and
\cite{Koleva+09a, Koleva+09}, which find shallower (and tightly
correlated with mass) and steeper gradients in old and young
systems, respectively.

\begin{figure*}
\psfig{file= 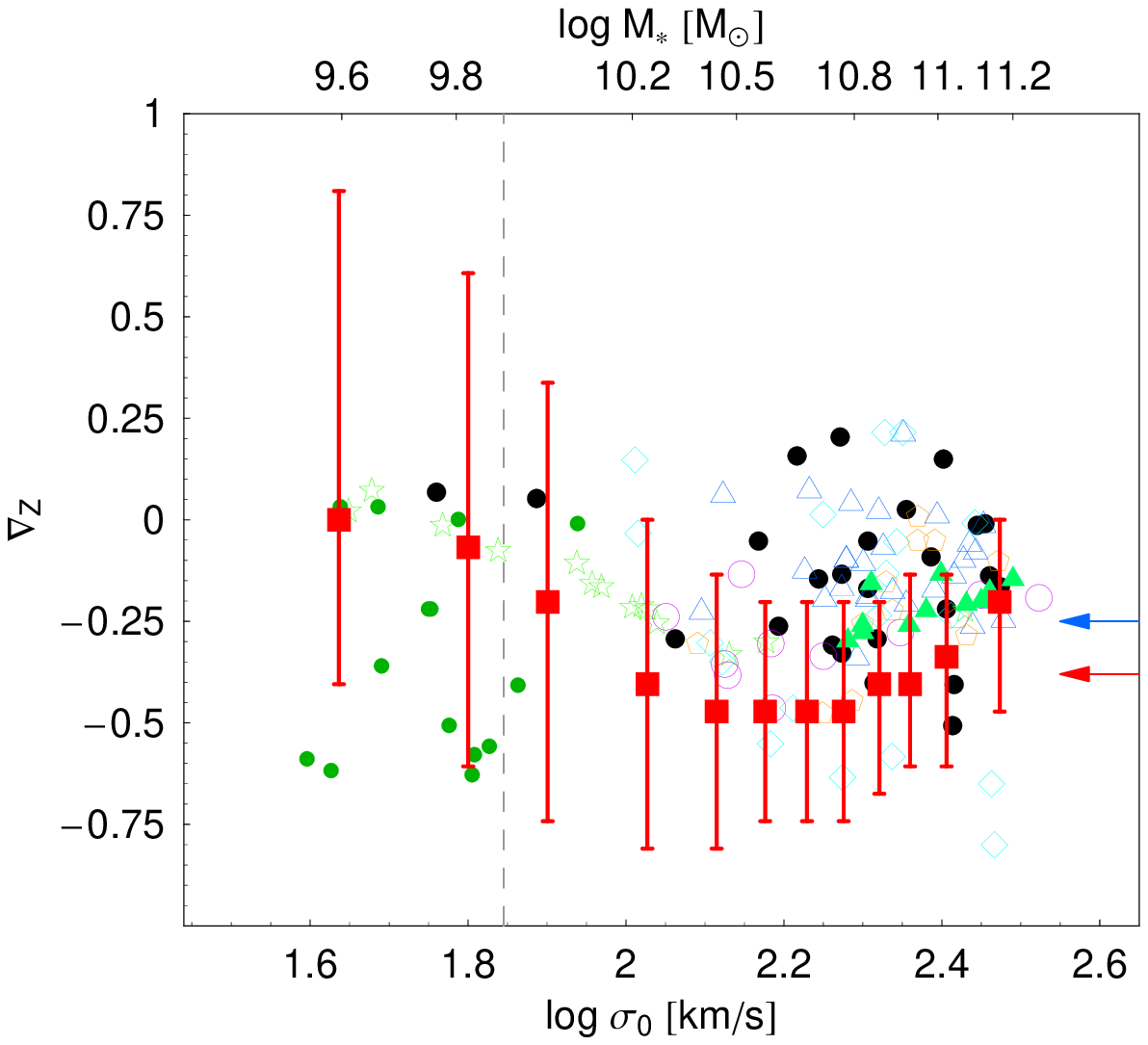, width=0.45\textwidth} \psfig{file=
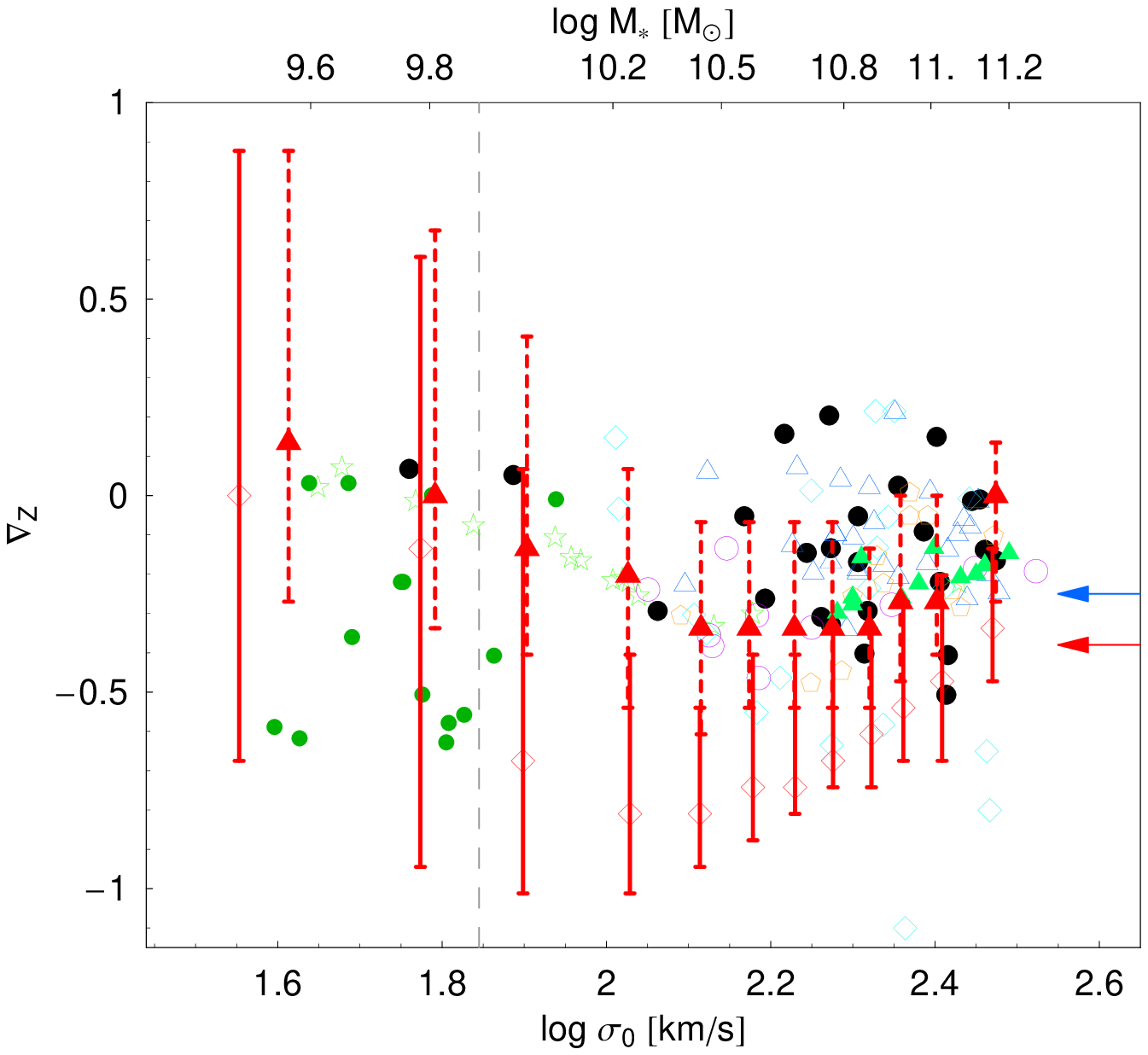, width=0.45\textwidth}\\
\psfig{file= 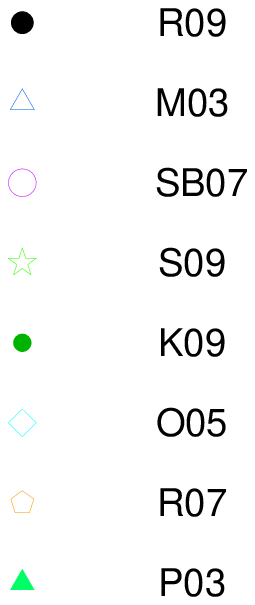, width=0.45\textwidth} \psfig{file=
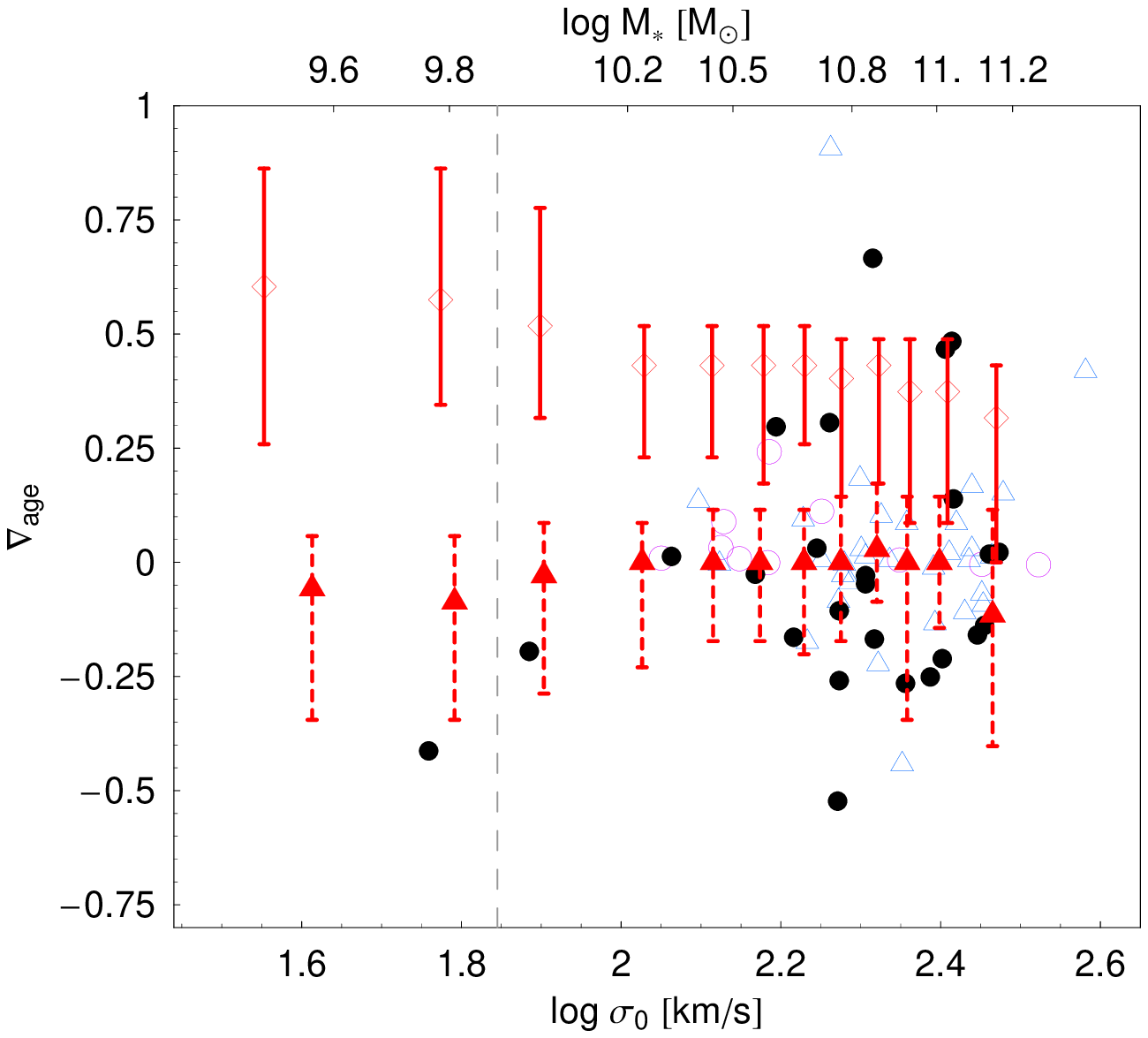, width=0.45\textwidth} \caption{Metallicity gradients
for ETGs as a function of central velocity dispersion compared
with some literature data. Galaxies with velocity dispersion
$\lsim 70 \, \rm km/s$, at the left of the vertical dashed gray
line, are somehow uncertain. {\it Top-left.} Metallicity gradients
are compared with \citet{Rawle+09} (R09) as black points
(including the results for Shapley and A3389 clusters),
\citet{Mehlert+03} (M03) as blue triangles, \citet{SB+07} (SB07)
as purple circles, \citet{Spolaor09} (S09) as green stars,
\citet{Koleva+09a, Koleva+09} (K09) as light green points,
\citet{Ogando+05} (O05) as cyan diamonds, \citet{Reda+07} (R07) as
orange pentagons and \citet{Proctor03} (P03) as green triangles.
The blue and red arrows are for the results in \citet{Wu+05} and
\citet{LaBarbera2009}. {\it Top-right.} Our ETGs divided in young
and old central ages are compared with literature data: diamonds
with continue bars are for ETGs with $0 < age_{1} \leq 6 \, \rm
Gyr$; triangle with dashed bars are for $age_{1}
> 6 \, \rm Gyr$. {\it Bottom-left.} Legend of symbols for literature data.
{\it Bottom-right.} Age gradients divided by age are plotted
versus literature data (M03, SB07 and R09) as above.} \label{fig:
fig6}
\end{figure*}

As a final remark, galaxies with very low velocity dispersion
($\log \sigc \lsim 1.85 \, \rm km/s$, see the vertical dashed gray
line in Figs. \ref{fig: fig6} and \ref{fig: fig9}) could be
affected by some systematics due to the low signal-to-noise ratio
and instrumental resolution of SDSS spectra. Notwithstanding
this caution, our recovered gradients reproduce fairly well  the
literature results for dwarf ETGs (e.g.
\citealt{Spolaor09}). Also, the indication of shallower gradients at
the lowest $\sigc$ seems robust, due to the clear turnover at
$\log \sigc \sim 2.2 \, \rm km/s$ and the inversion in the range
$1.85 \lsim \log \sigc \lsim 2.2 \, \rm km/s$.

\section{Discussion}\label{sec:disc}
Our results seem to support the idea that the metallicity trend versus
the stellar mass for LTGs is mainly driven by the interplay of gas inflow and winds from
supernovae and evolved stars (\citealt{Matteucci94},
\citealt{Ko04}, \citealt{Pipino08}). These processes tend to
increase the central metallicity and prevent the enrichment of the
outer regions. Therefore more massive systems have on average
larger central metallicities which correspond to steeper negative
gradients.

\begin{figure*}
\psfig{file= 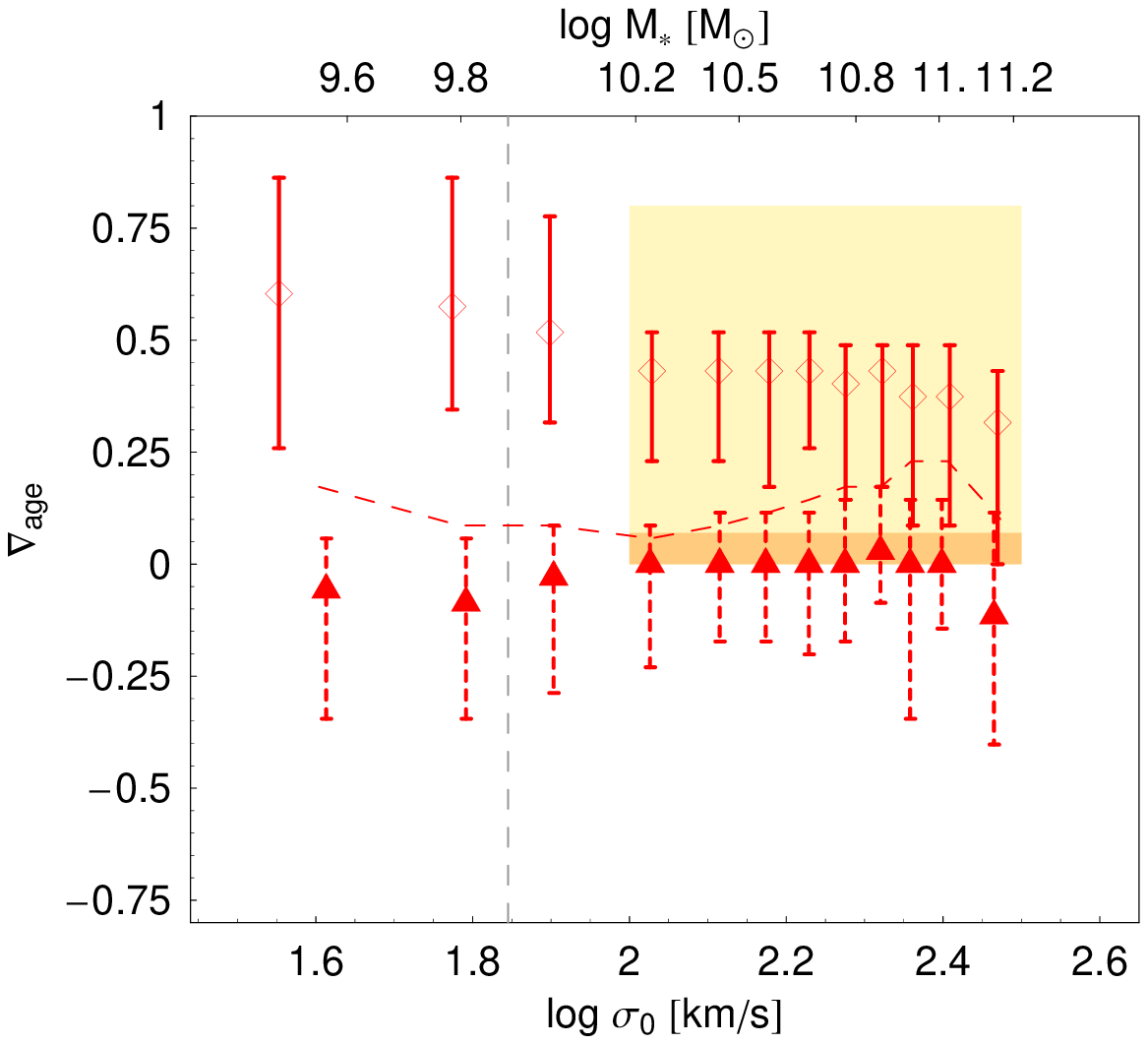, width=0.37\textwidth} \psfig{file=
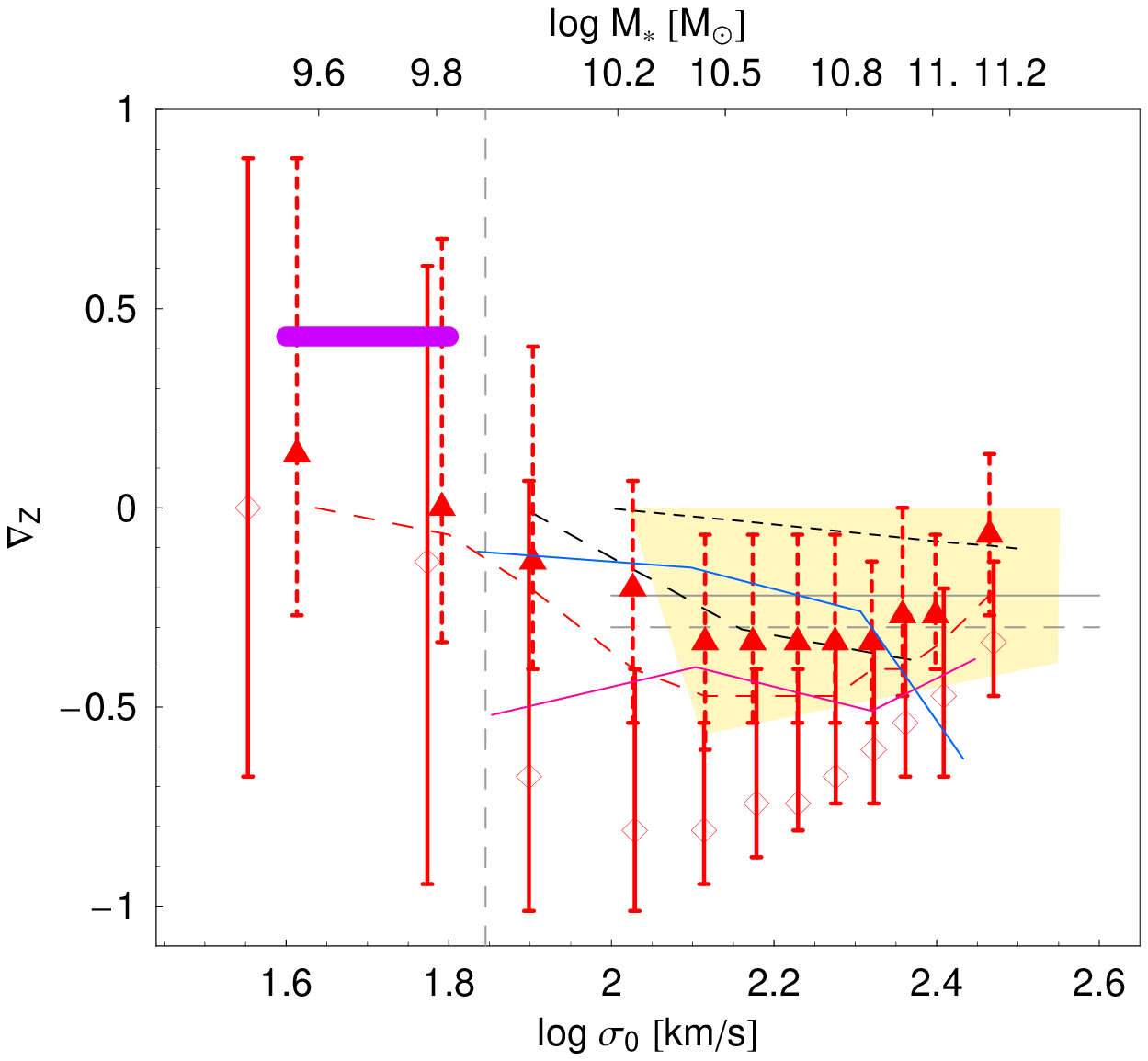, width=0.37\textwidth}  \psfig{file= 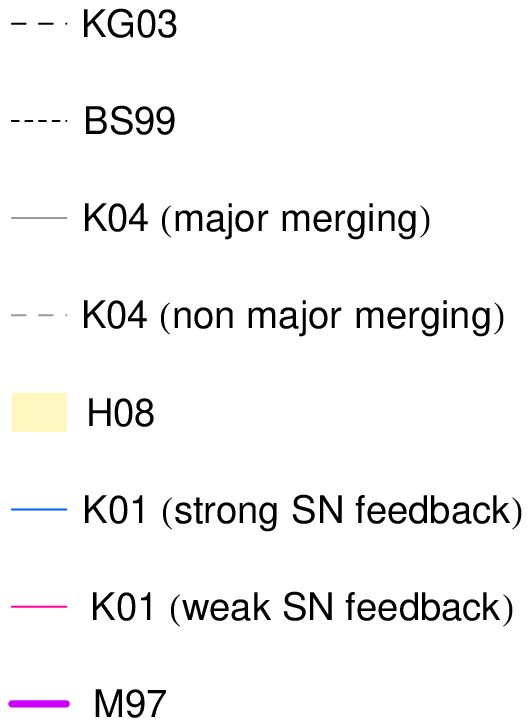,
width=0.2\textwidth} \caption{ {\it Left.} Age gradients versus
simulations. Comparison with gas-rich merging predictions in
\citet{Hopkins+09a}, with the darker (lighter) region showing
galaxies older (younger) than $6 \, \rm Gyr$. {\it Middle.}
Metallicity gradients versus simulations. The predictions from
merging models in \citet{BS99} are shown as short dashed line,
dissipative collapse models in \citet{KG03} as long dashed line,
the remnants of major-mergings between gas-rich disk galaxies in
\citet{Hopkins+09a} as yellow shaded region, the typical gradient
for major (continue gray line) and non-major (dashed gray line)
mergings in \citet{Ko04}. The violet thick line shows the
predictions as in the simulation of \citet{Mori+97}; blue and pink
lines are the result from the chemo-dynamical model in
\citet{Kawata01}, respectively for strong and weak supernovae
feedback models (the B-band magnitude in this work is transformed
in a velocity dispersion \sigc\ using the relation in
\citet{Tortora2009}). The symbols are as in Fig. \ref{fig: fig6}
and the dashed red line is for the whole ETGs sample. {\it Right.}
Symbol legend for simulations.} \label{fig: fig9}
\end{figure*}

Low mass ETGs ($\mst \lsim 10^{10.3-10.5} \, \rm \Msun$, see Table
\ref{tab:slopes_grad_Z_age}) show a similar correlation of the
metallicity gradient with stellar mass (Fig. \ref{fig: fig4}),
which suggests that they might experience the effect of infall/SN
feedback as LTGs. For instance, low mass ETGs in Fig. \ref{fig: fig9} are
consistent with SN feedback (both soft and strong), as predicted in
chemo-dynamical simulations in \citet{Kawata01}. On average,
weaker SN feedback gives gradients in agreement with ours at
intermediate $\log \sigc$ ($\sim 2.2\, \rm km/s$), while the
stronger feedback recipe seems to reproduce better the low $\log \sigc$
side of the correlation. A recipe including the SN feedback with a varying power as a
function of mass (\citealt{Dekel86}, \citealt{dek_birn06}) would
allow a fairly good match of the observed decreasing trend of the gradients for the
less massive galaxies (as in \citealt{KG03}), but it
fails to reproduce our observed gradients for the very massive systems.
Moreover, as seen in Fig. \ref{fig: fig4}, the absolute value
of \gZ\ of ETGs is lower than for the LTGs, probably as a consequence
of the dilution of the gradient due to the higher-density environment
where ETGs generally live in (see also discussion in
\S\ref{sec:morph}). It might be also the effect of other
mechanisms at work such as merging (see,
e.g., \citealt{Ko04}), which however is expected to be less effective in
these mass ranges (\citealt{deLucia06}).

At very low masses ($\mst \lsim 10^{9.5} \, \rm \Msun$ and $\log
\sig_c \lsim 1.8\, \rm km/s$) the \gZ\ turns to even positive
values, which mainly corresponds to systems with negative \gage\ as
in Fig. \ref{fig: fig5} and very low central metallicities as in
Fig. \ref{fig: fig5bis}: these dwarf-like systems are compatible
with the expanding shell model from \citet{Mori+97} (see also Fig.
\ref{fig: fig9}).

At larger masses, the shallow metallicity gradients of ETGs
suggest that these have experienced merging and/or tidal
interactions (at a rate that could increase as a function of the
stellar mass) which have diluted the \gZ. Such events might have
taken place in earlier phases of the galaxy evolution, as
indicated by the presence of null age gradients in old systems. In
Fig. \ref{fig: fig9} we compare the \gage\ to the prediction from
the hierarchical simulations in \citealt{Hopkins+09a}, where a
perfect agreement is found for the old sample (see also the good
agreement with literature data if Fig. \ref{fig: fig6}). From the
same simulations we also see that systems with younger cores are
expected to show positive gradients which are fully consistent
with our low central age systems as in Fig. \ref{fig: fig4} and
also reported in Fig. \ref{fig: fig9}. In the same figure,
metallicity gradients are in qualitative agreement with gradients
of remnants in \cite{Hopkins+09a} and the medians from
\cite{Ko04}, while the merging models in \cite{BS99} reproduce the
mean gradients in galaxies with $\log \sigc \gsim 2.4$, but fail
at lower $\sigc$. Once again, if we consider the older systems
only, they are in a better agreement with the merging simulations:
this is expected since, after the initial gas rich-merging events
that might produce both a larger central metallicity and a
positive age gradient (\citealt{Ko04}, \citealt{MH94}), subsequent
gas poor-merging may dilute the positive age gradient with time as
well as make the metallicity gradients to flatten out
(\citealt{White80}, \citealt{Hopkins+09a}, \citealt{DiMatteo09})
-- see also left column of Fig. \ref{fig: fig5bis}. In this
respect, massive and old ETG systems seem to be fully consistent
with the merging scenario. However, a further mechanism that may
act to produce the shallower (or almost null) color and
metallicity gradients in old massive ETGs, with $\mst \gsim
10^{11} \, \rm \Msun$ and $\log \sigc \gsim 2.4$ (Fig. \ref{fig:
fig4}), might be due to the strong quasar feedback at high
redshift (\citealt{Tortora2009AGN}), while steeper metallicity
gradients at lower masses could be linked to less efficient AGNs.

\section{Conclusions}\label{sec:conclusions}

We have investigated the colour gradients in a sample of $\sim
50\,000$ local galaxies from the SDSS as a function of structural
parameters, luminosity, and stellar mass. CGs have been found to
correlate mainly with luminosity and stellar mass. They have a
negative minimum ($\ggi \sim -0.2$) at $\mst \sim 10^{10.3} \, \rm
\Msun$, and increase with the mass for $\mst \gsim 10^{10.3} \,
\rm \Msun$, the very massive galaxies, $\mst \gsim 10^{11} \, \rm
\Msun$, having the shallower values ($\sim -0.1$). On the other
mass side, the gradients decrease with mass and become positive
for $\mst \lsim 10^{9}  \, \rm \Msun$. These trends are mirrored
by similar behaviours with luminosity and galaxy size, e.g. the
gradients have a minimum at $r \sim -20$ mag and $\log \Re \sim
0.5$, then increase toward the small and the large end of the
parameter distribution, turning to positive values for $\log
\Re\lsim 0$ and $r\gsim -18$ mag. The dependence on velocity
dispersion is very loose. A clear dichotomy is also found when
looking at the dependence on the S$\rm \acute{e}$rsic index $n$
which suggests a distinct behaviour between the ETG and LTG. In
fact, these two families mark clear differences in their trends
with structural parameters (e.g. mass and \sigc).

For LTGs, \ggi\ monotonically decreases with the mass, with more
massive systems having the lowest CGs ($\sim -0.4$) and less
massive systems ($\mst \lsim 10^{8-8.5}\, \rm \Msun$) showing even
positive gradients. ETGs have negative gradients mildly increasing
with mass for $\log\mst \gsim 10^{10.3-10.5}\, \rm \Msun$, which
marks the mass scale for the gradient slope inversion (see Table
\ref{tab:slopes_grad_Z_age}). This result is consistent with
\cite{Spolaor09} and reminiscent of the typical mass scale where
the star formation and structural parameters in galaxies
drastically change (\citealt{Capaccioli92a, Capaccioli92b},
\citealt{Kauffmann2003}, \citealt{Croton06},
\citealt{Cattaneo+08}). A similar trend is observed when plotting
the gradients as a function of effective radius, while a tighter
trend with velocity dispersion is evident for ETGs only (Fig.
\ref{fig: fig4}). This was masked when plotting LTGs and ETGs
together.

We have used galaxy colors at $0.1\Re$ and $1\Re$ in our synthetic
spectral models to determine the variation of age and metallicity
at these radii. The observed trends of the CGs with mass and
\sigc\ are correlated with a similar trends for metallicity and
age gradients. Despite the large scatter of the data, the strong
correlation of metallicity gradients with the central velocity
dispersion of ETGs is clear, with a turnoff point at $\log \sigc
\sim 2.2 \, \rm km/s$  (see Table \ref{tab:slopes_grad_Z_age}).
These results are in very good agreement with a collection of
results from literature (\citealt{Mehlert+03},
\citealt{Proctor03}, \citealt{Ogando+05}, \citealt{Reda+07},
\citealt{SB+07}, \citealt{Koleva+09a, Koleva+09},
\citealt{Spolaor09}, \citealt{Rawle+09}), in particular for the
old galaxies in our sample.

A remarkable result of our analysis is the confirmation
that the galaxy (central) age is one of the main drivers of the
scatter of the age and metallicity gradients, with the older
systems showing generally the shallower gradients with respect the
young ones. In Fig. \ref{fig: fig6}, for instance, we show that in the low
mass regime, at fixed \sigc, older galaxies have on average
shallower gradients, consistently with the results in
\cite{Spolaor09}, while systems with late formation or with a
recent SF episode have a larger spread (and metallicity gradients
down to very low values, $\sim -0.6$), consistently with findings
in \cite{Koleva+09a,Koleva+09}. The measured scatter might be the
consequence of a variety of phenomena affecting the dwarf galaxy
evolution, such as (soft) SN feedback, interaction/merging in the high
density environment, and star-formation by shell expansion (Fig. 8). On the
other mass side, a further factor of the spread of the massive
systems gradients might be the randomness of the initial
conditions of the mergings (\citealt{Rawle+09}).

Previous attempts to quantify the correlations of gradients with
luminosity, mass, or velocity dispersion have often failed, mainly
because of the exiguity of the galaxy sample (e.g.
\citealt{Peletier+90a}, \citealt{KoAr99}, \citealt{TO2003}). Only
recently a clear correlation of the color and metallicity
gradients with mass has been ascertained (see e.g.,
\citealt{Forbes+05}), pointing to different trends for high and
low mass galaxies (\citealt{Spolaor09}, \citealt{Rawle+09}). Our
analysis has confirmed and reinforced these results, as it relies
on one of the largest local galaxy samples including dwarf,
normal, and giant galaxies. Using CGs derived by the structural
parameters in B05, we obtained statistically meaningful trends,
which allow us to fix the link with physical phenomena as well as
to make direct comparisons with predictions from simulations
(\citealt{Mori+97}, \citealt{BS99}, \citealt{Kawata01},
\citealt{KG03}, \citealt{Ko04}, \citealt{Hopkins+09a}).

\begin{figure}
\psfig{file= 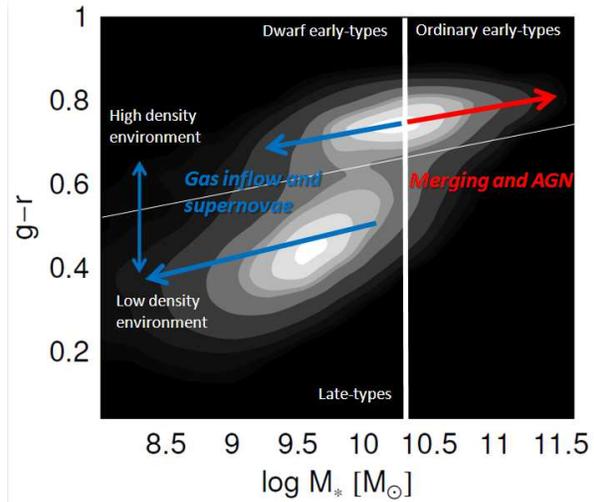, width=0.45\textwidth} \caption{Color-mass
diagram. The contours show the density of data-points with gray
scale going from darker (low density of galaxies) to brighter
regions (high density). The thin white line sets the separation
between RS and BC; the thick vertical one gives the mass scale,
$\mst \sim 10^{10.3}$, which separates galaxies belonging to the
RS in normal and dwarf ETGs, and sets a qualitative upper mass for
LTGs. The arrows give information about the efficiency of the
phenomena which drive the two-fold trend we discuss in the paper.
AGN feedback and merging are important at high mass with an
efficiency increasing with mass, while SN feedback and gas inflow
drive the galaxy evolution in the less massive side, with an
efficiency that is larger at the lowest masses. Galaxies in RS and
BC lie in environments with a different density, which is
manifested as a difference of the gradients in the two samples.}
\label{fig: fin}
\end{figure}

We can finally draw the physical scenario sketched in Fig.
\ref{fig: fin}, resting on the results discussed so far. The
formation of LTGs and less massive ETGs is mainly driven by a
monolithic-like collapse as they have almost null or positive age
gradients and negative metallicity gradients decreasing with the
mass. These are in fact well explained by simple gas inflow and
feedback from SN and evolved stars, and also reproduced in
simulations of dissipative collapse and SN feedback models in Fig.
\ref{fig: fig9} (\citealt{Larson74, Larson75},
\citealt{Carlberg84}, \citealt{AY87}, \citealt{KG03}). The
difference in the magnitude of the gradients between LTGs and
lower mass ETGs mainly resides in the effect of the environment as
the former live in very low density environments, while the latter
stay in massive haloes, together with more massive ETGs. Here they
experience strong gravitational interactions producing shallower
color and metallicity gradients. The efficiency of such phenomena
is strong enough for lower mass ETGs, thus shaping the observed
decreasing trend with mass (see e.g. \citealt{dek_birn06}).

The most massive ETG systems have shallower gradients
(\citealt{BS99}, \citealt{Ko04}, \citealt{Hopkins+09a}),
flattening out with the mass due to the increasing intervention of
(gas-rich and -poor) merging, tidal interactions, and quasar/radio
mode AGN feedback (\citealt{Capaccioli92a}, \citealt{FM2000},
\citealt{deLucia06}, \citealt{dek_birn06}, \citealt{Liu+06},
\citealt{Sijacki+07}, \citealt{Cattaneo+08}). All the simulations
collected in literature fail to reproduce the fine details of the
trends we find, suggesting that a full understanding of physical
processes involved in the galaxy evolution is still missing.

In future analysis we plan to enlarge the wavelength baseline and
also to use line-strength measurements for the stellar model (when
available). We will investigate the systematics induced when other
synthetic prescriptions are assumed and the dependence of the CGs
on the galaxy star formation history. We will derive the $M/L$
gradients, if any, and discuss their impact on the dark matter
content of ETGs and correlate the CGs with the DM fraction of
these systems as probe of the galaxy potential wells. Finally, the
realization of hydrodynamical simulations of jets emitted by AGN
would be useful to shape, together with galaxy merging, the color
and metallicity gradients for massive galaxies.

\section*{Acknowledgments}

We thank the anonymous referee for the useful suggestions which
helped to improve the paper and the robustness of the results. We
also thank F. La Barbera for the fruitful discussion and comments.
CT is supported by the Swiss National Science Foundation and by a
grant from the project Mecenass, funded by the Compagnia di San
Paolo. VFC is supported by Regione Piemonte and Universit\`{a} di
Torino and partially from INFN project PD51.

\appendix

\section{Sample systematics and
contaminants}\label{app:app_sample}

\begin{figure*}
\psfig{file= 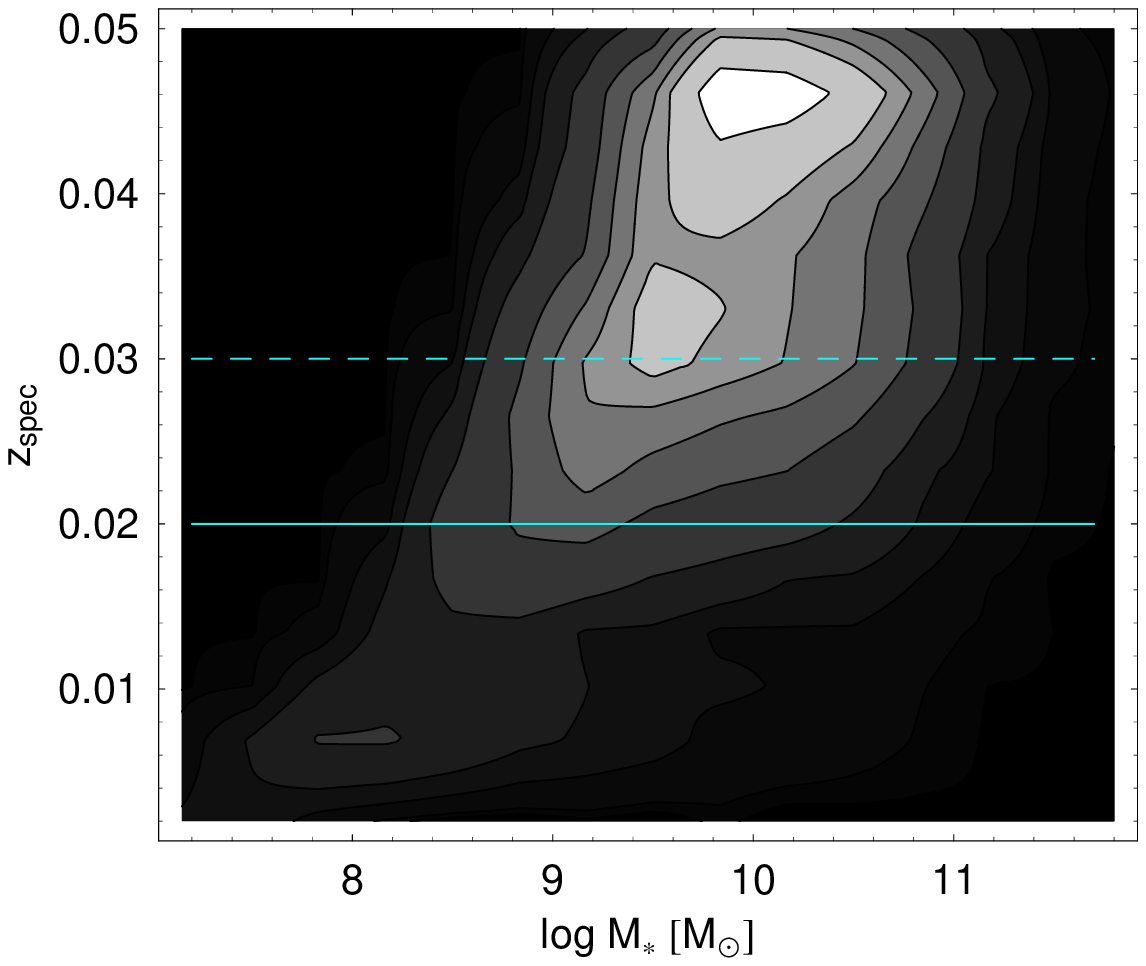, width=0.3\textwidth} \psfig{file=
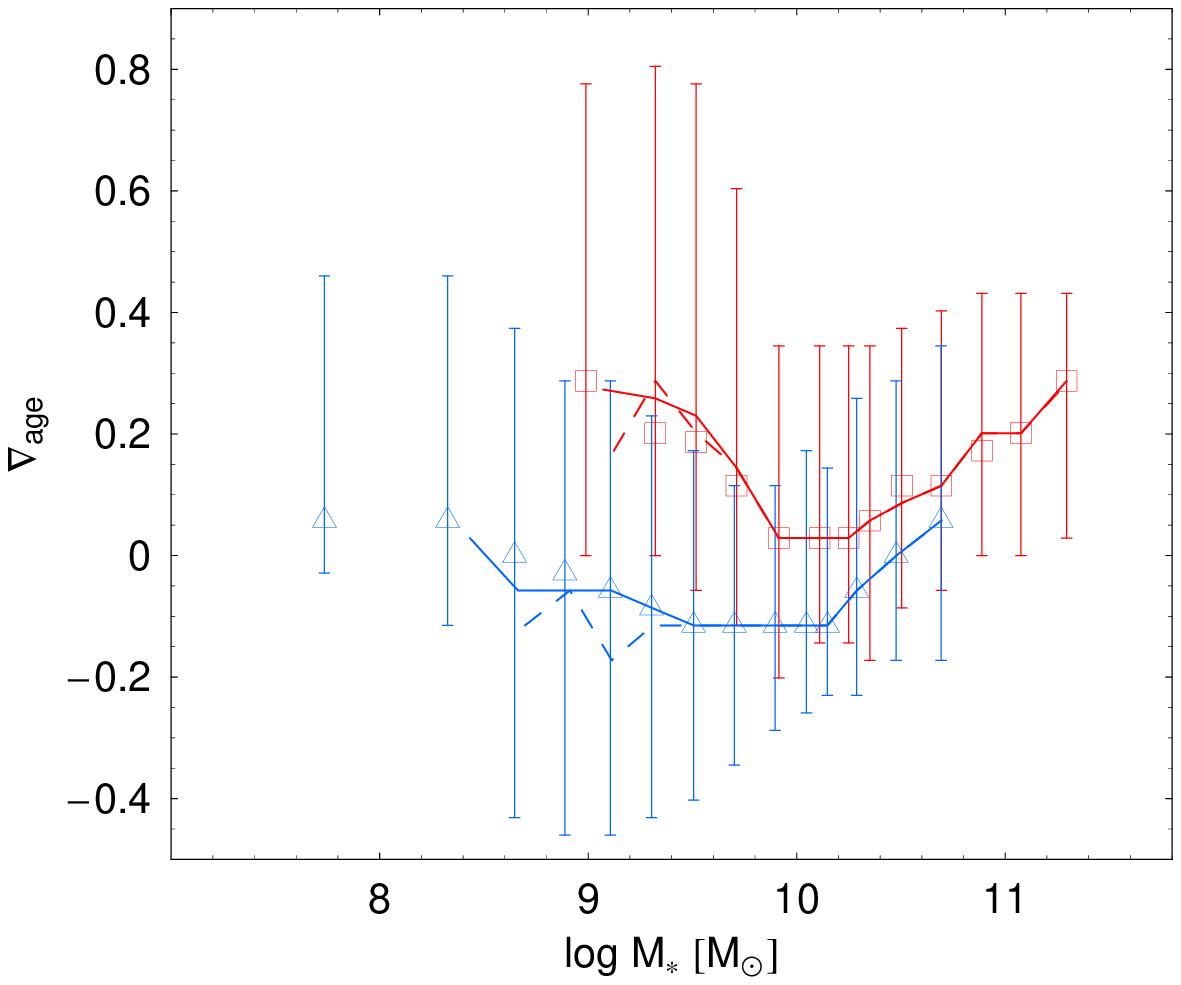, width=0.3\textwidth} \psfig{file= 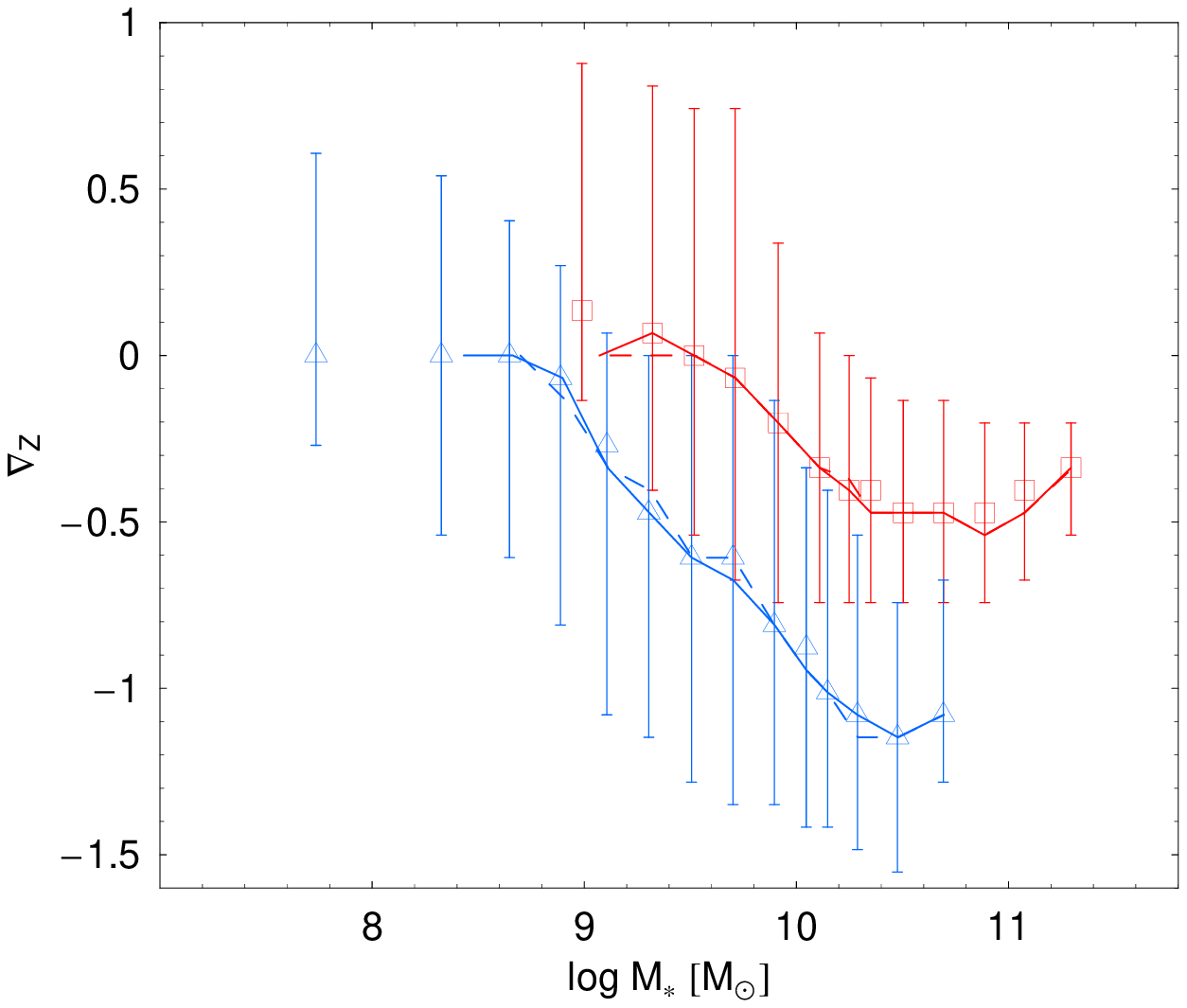,
width=0.3\textwidth}\caption{{\it Left Panel.} Spectroscopic
redshift as a function of stellar mass for the B05 galaxy sample.
The two lines are relative to the redshift thresholds we use to
select subsamples of galaxies. {\it Middle Panel.} Age gradients
as a function of stellar mass. {\it Right Panel.} Metallicity
gradients as a function of stellar mass. The red and blue symbols
with bars are relative to the full galaxy sample for ETGs and LTGs
respectively, as discussed in the text, and the continue and
dashed lines are for subsamples with $z_{spec}>0.02$ and
$z_{spec}>0.03$.} \label{fig: fig_app2}
\end{figure*}

We analyze various systematics in the galaxy sample and in the
selection criteria.

\subsection{Sample systematics}\label{app:appA4}

We quantify the effect of source incompleteness on our
results, in particular on the age and metallicity gradients
as a function of stellar mass.

Our sample is magnitude limited, and misses increasingly fainter
and less massive galaxies at larger redshifts; e.g. at
$z_{spec}=0.02$, the lowest mass galaxies have $\mst \sim
10^{8.2}\, \rm \Msun$, while at $z_{spec}=0.05$ this limit
increases to $\mst \sim 10^{9}\, \rm \Msun$ (see
\citealt{Blanton05a} and the left panel in Fig. \ref{fig:
fig_app2}). We already mentioned that dwarf galaxies can be missed
due to surface brightness selection effects (photometric
incompleteness). Other less obvious sources of incompleteness are
defects in deblending and the absence of reliable redshift
measurements.

By analyzing the surface brightness completeness of this sample of
galaxies, \cite{Blanton05a} found that for $r < -18$ mag ($\mst
\lsim 10^{9} \, \rm \Msun$) the catalog is complete at more than
$90\%$, while at fainter magnitudes the photometric pipeline could
mistakenly deblend galaxies or overestimate the background sky
level. However, down to $\mst \sim 10^{8.5} \, \rm \Msun$ the
overall incompleteness is not dramatic (\citealt{Blanton05a},
\citealt{BGD08}, \citealt{LW09}). Moreover, at $z_{spec}\lsim
0.02$ the distribution of galaxies seems patchy and the fraction
of massive galaxies with $\mst \gsim 10^{9}\, \rm \Msun$ turns out
to be smaller, and not homogeneously distributed as the one at
larger redshifts. In order to account for all possible biases due
to these missing galaxies we have redone the age and metallicity
gradient plots for the subsamples of galaxies having $z_{spec}>
0.02$ and $z_{spec} > 0.03$, in order to avoid the regions with
higher incompleteness. The trends found for the two subsamples are
almost identical to the ones for the full data set, which suggests
that the impact on our results of all sources of incompleteness is
negligible, both at low and high masses.

It has been suggested that the S$\rm \acute{e}$rsic fitting
procedure adopted in B05 contain significant biases. In
particular, at very high luminosities, S$\rm \acute{e}$rsic
indices, effective radii and fluxes can be underestimated by $\sim
0.5$, $\sim 10-15\%$ and $\sim 10\%$, respectively (B05). Similar
considerations hold for systems with high effective radii and
S$\rm \acute{e}$rsic indices. The S$\rm \acute{e}$rsic procedure
used in B05 have been extensively analyzed in \cite{Guo+09}, who
have shown that the main source of systematics is introduced by
the 1D S$\rm \acute{e}$rsic fit and background sky level estimate.
However, these systematics turn out to produce shallower CGs (of
$\sim 0.03$) than our estimates, thus leaving the main trends with
luminosity and mass unaffected. On the other hand, for the systems
with very low S$\rm \acute{e}$rsic indices, the B05 fitting
procedure recovers quite well the parameters, although a slight
underestimate of fluxes and an overestimate of $n$ might still
occur (\citealt{Guo+09}).

\begin{figure}
\centering \psfig{file= 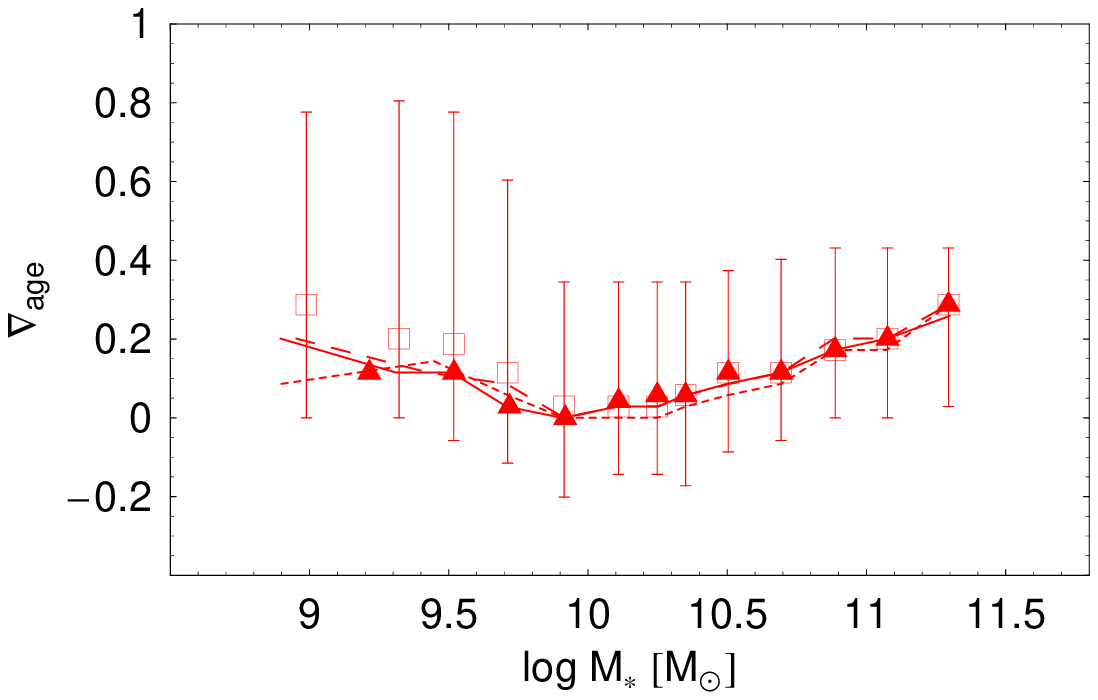, width=0.4\textwidth}
\psfig{file= 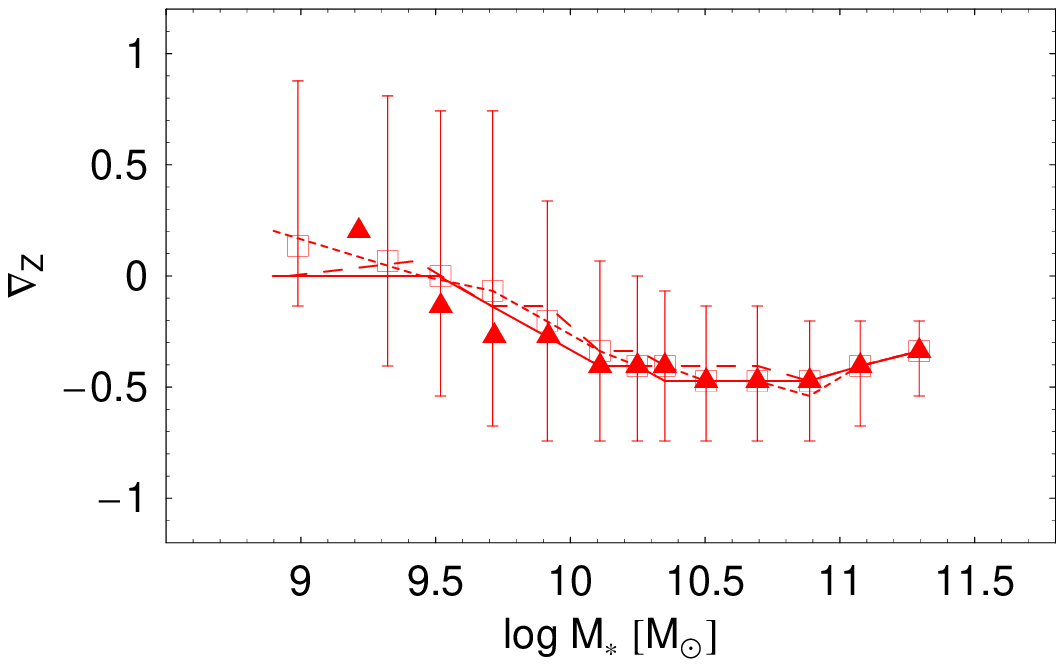, width=0.4\textwidth} \caption{Systematics
in the \gage\ and \gZ\ as a function of stellar mass due to
changes in selection criteria. Boxes and bars are for our
reference results. Continue, long-dashed, and short dashed lines
are for a) $C > 2.6$, b) $C>3$ and c) $2.5< n <5.5$ respectively,
and triangles for our reference criteria with a cut in the total
color ($g-r > 0.55$).} \label{fig: fig_app3}
\end{figure}

\subsection{Morphological separation and contamination}\label{app:appA5}

Seeking for a trustful morphological classification, the eyeball
check or the B/T are the best approaches, but unfortunately they
are not suitable procedure for very large sample of galaxies
(\citealt{Allen+06}, \citealt{Lintott+08}). Indeed, the use of
some objective automated morphological classification is more
efficient. For these reasons, we decided to adopt a classification
with the minimal parameter set to minimize the probability of
erroneously discarding ETGs, even if this would make the LTG
contamination non-negligible. We have ``a posteriori'' checked the
effect of such a contamination on the results. We have verified
that the use of the {\it concentration} and S$\rm \acute{e}$rsic
$n-$index parameters is efficient to this purpose. In fact, the
S$\rm \acute{e}$rsic index is a measure of the steepness of the
surface brightness profile and is historically considered as a
good indicator of galaxy type: LTGs are well fitted by light
profiles with lower $n$, being $n=1$ the prototype of light
profile for galaxy disks, and $n>1$ the ones typical of ETGs.
Recently, the concentration parameter determined from the SDSS
pipeline has been claimed to be an even better proxy of galaxy
morphology (\citealt{Shimasaku+01}, \citealt{Strateva+01},
\citealt{Nakamura+03}, \citealt{weinmann09}). We have checked that
the results on the gradient trends do not strongly depend on the
constraints adopted on the two parameters for the ETG/LTG
separation (see Fig. \ref{fig: fig_app3}) with some significant
change raising for age gradients at $\mst \lsim 10^{9.5}\, \rm
\Msun$. We also analyzed the impact on our results when a cut in
colour is applied.

\begin{figure*}
\psfig{file= 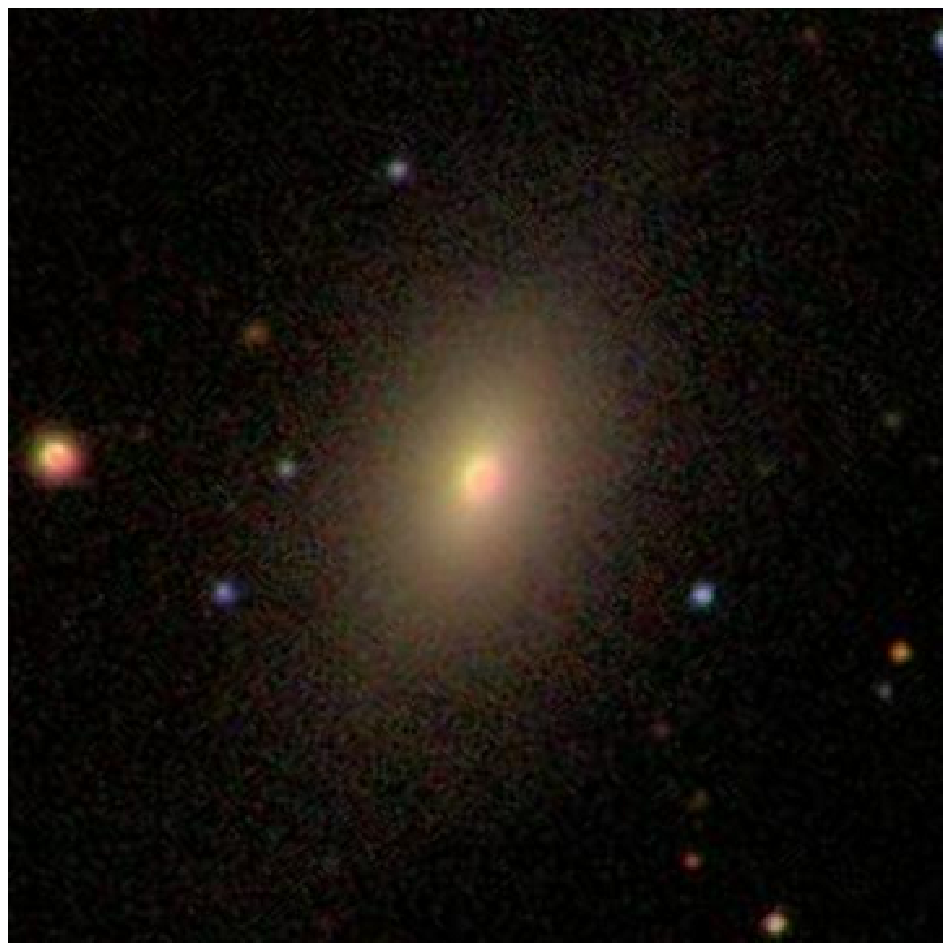, width=0.23\textwidth} \psfig{file=
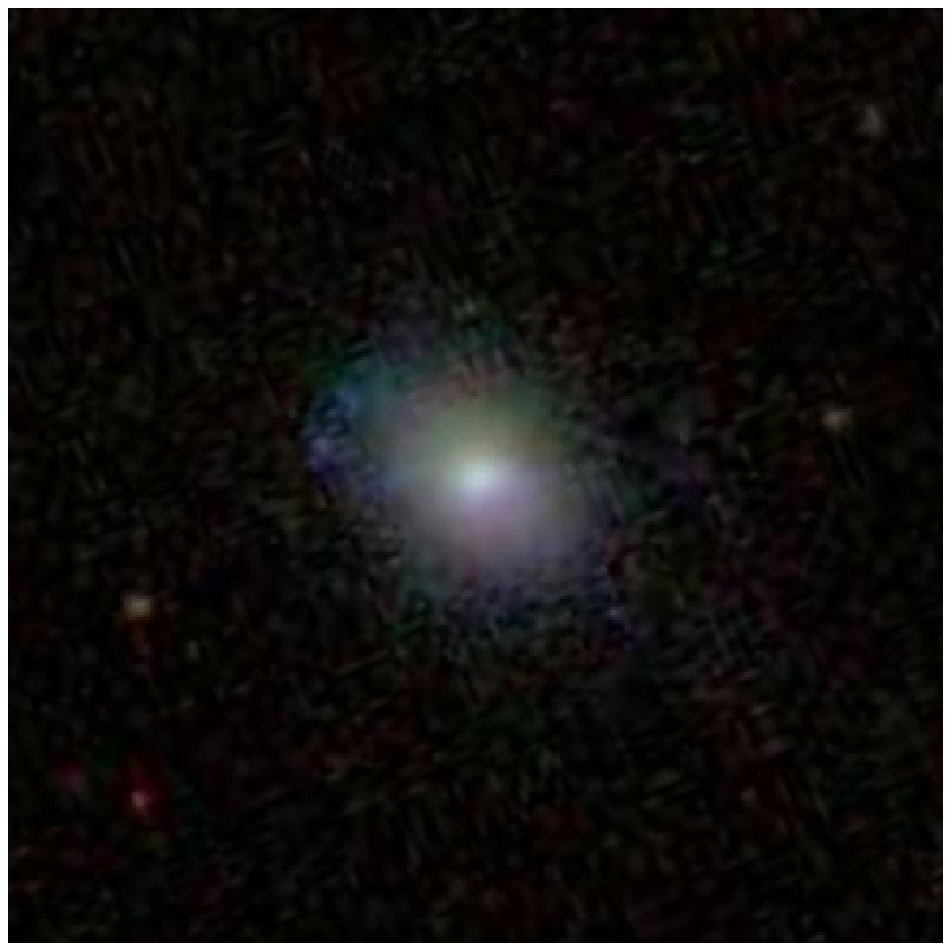, width=0.23\textwidth} \psfig{file= 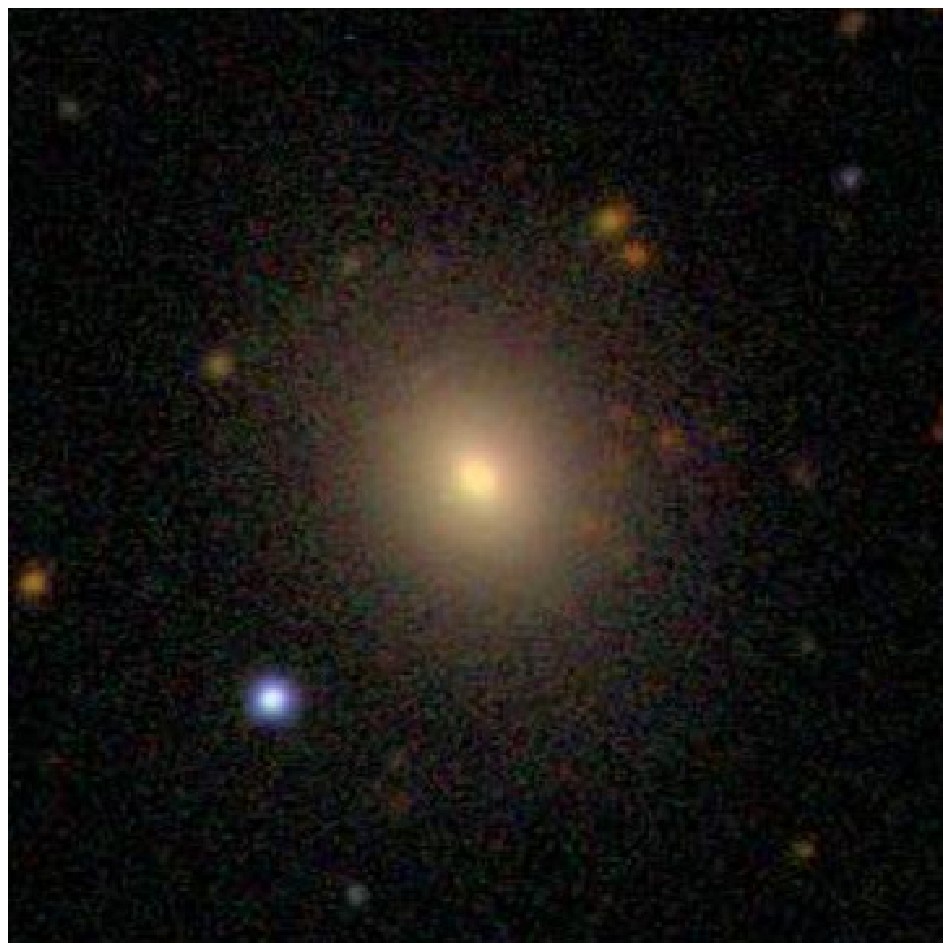,
width=0.23\textwidth} \psfig{file= 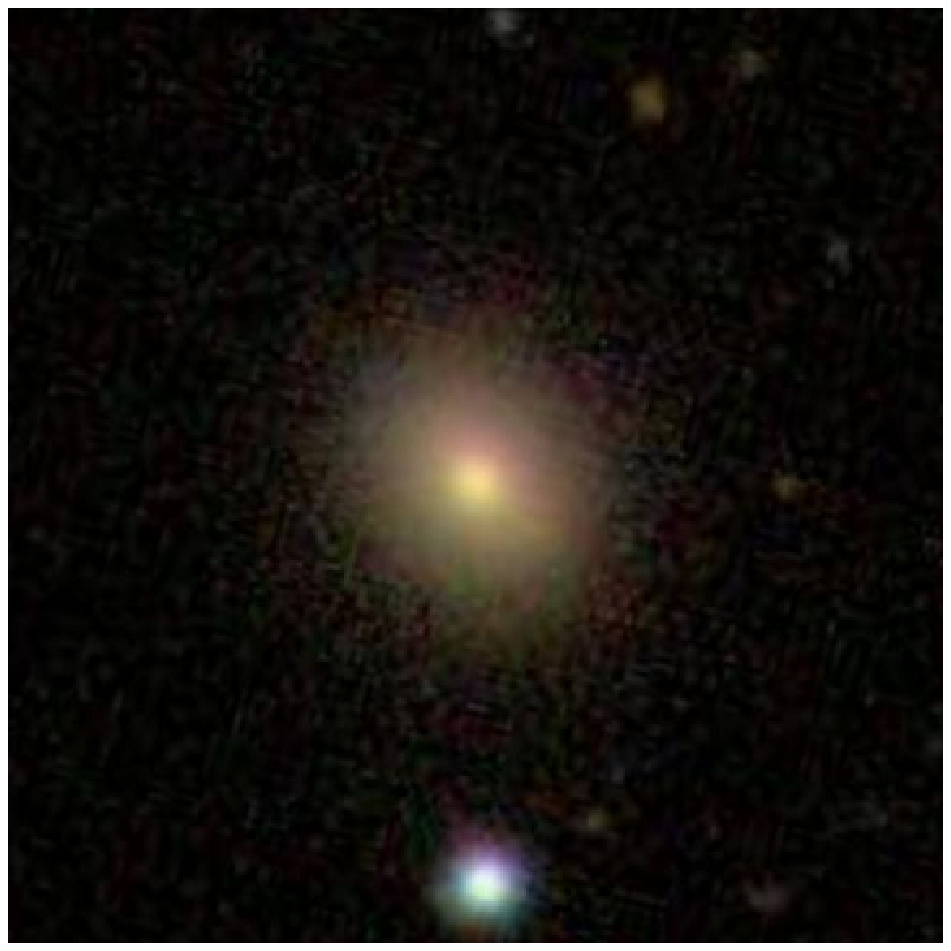,
width=0.23\textwidth}\\ \psfig{file= 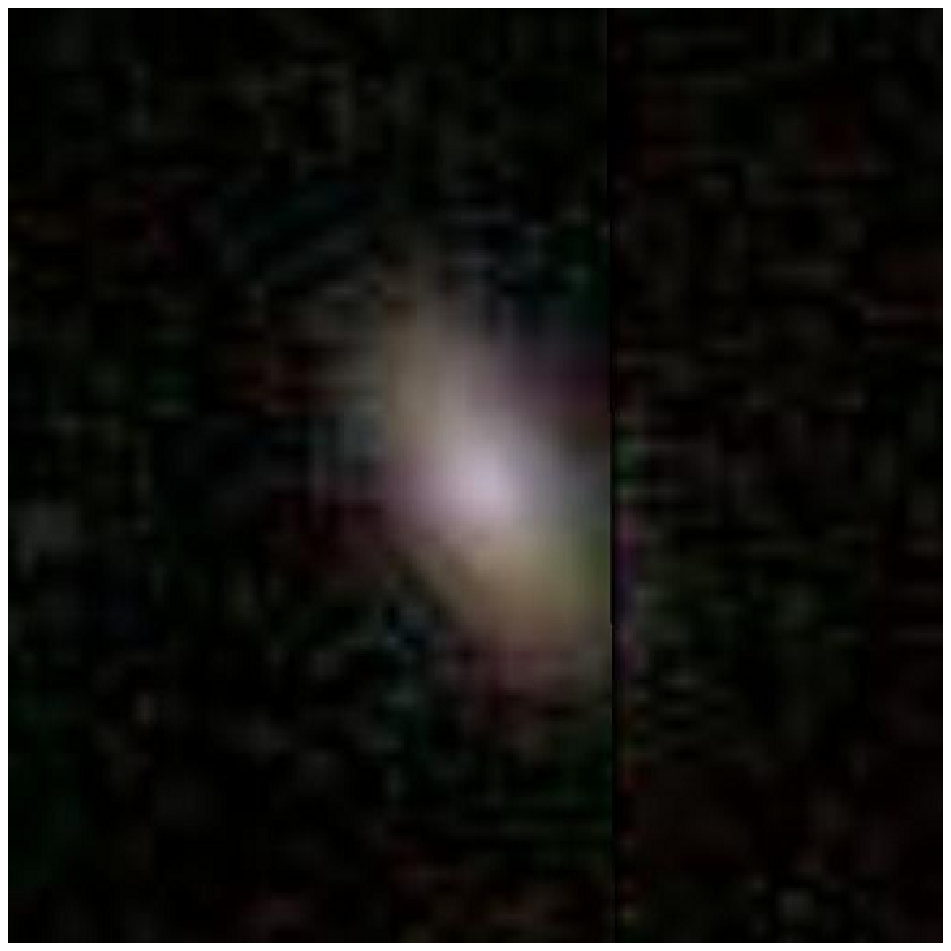,
width=0.23\textwidth} \psfig{file= 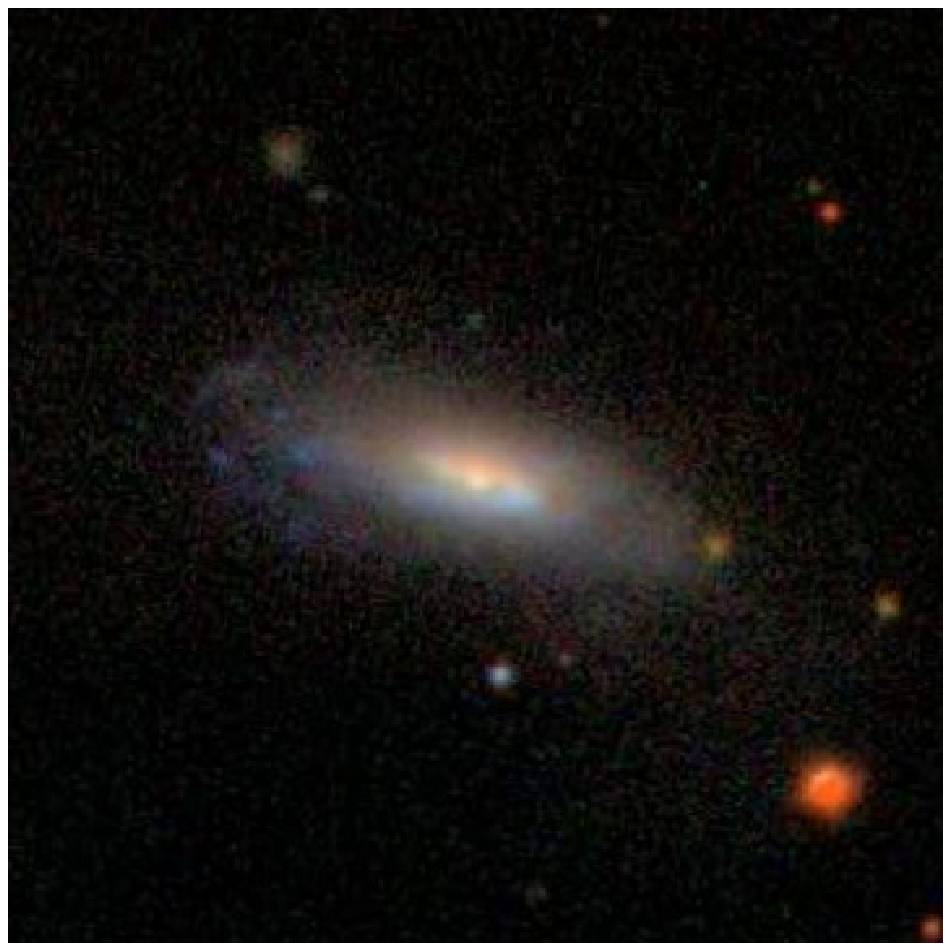,
width=0.23\textwidth} \psfig{file= 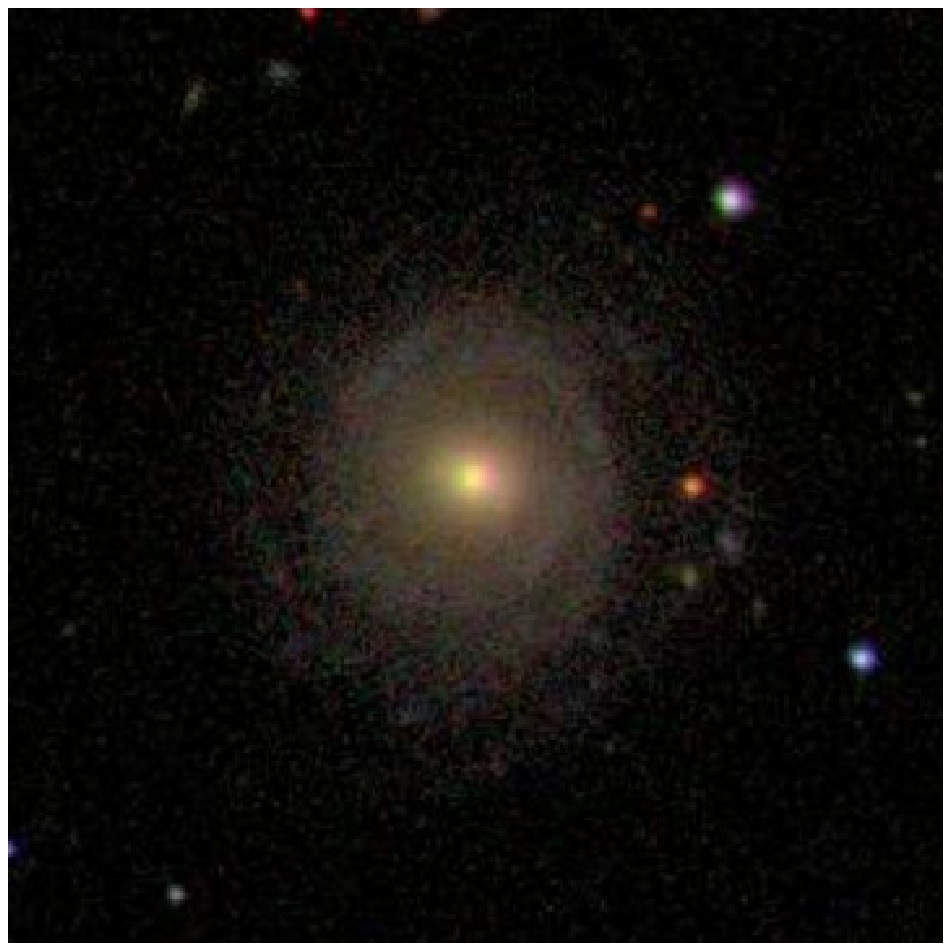,
width=0.23\textwidth} \psfig{file= 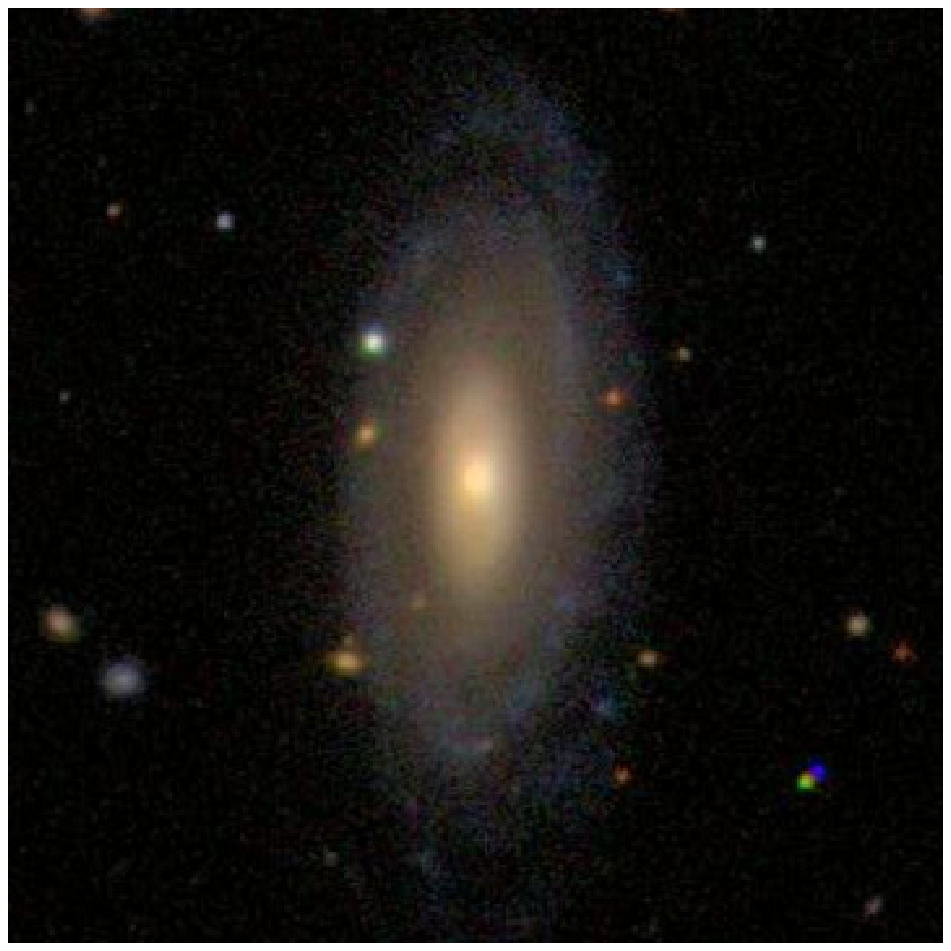,
width=0.23\textwidth}\caption{Images of 8 galaxies classified by
eyeball checking. The galaxies in the top panels are classified as
E/S0, the first one in the bottom panels is uncertain, while the
last three are clear example of spirals.} \label{fig:
images_eyeball}
\end{figure*}

\begin{figure*}
\psfig{file= 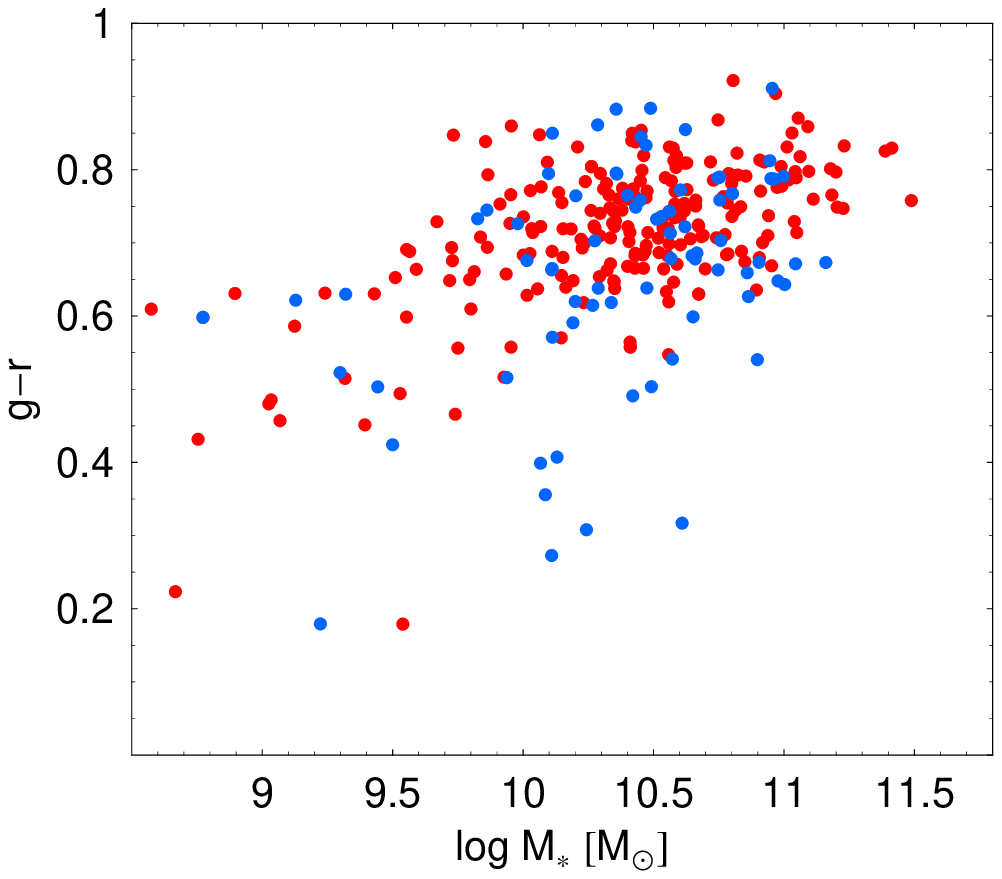, width=0.33\textwidth} \psfig{file=
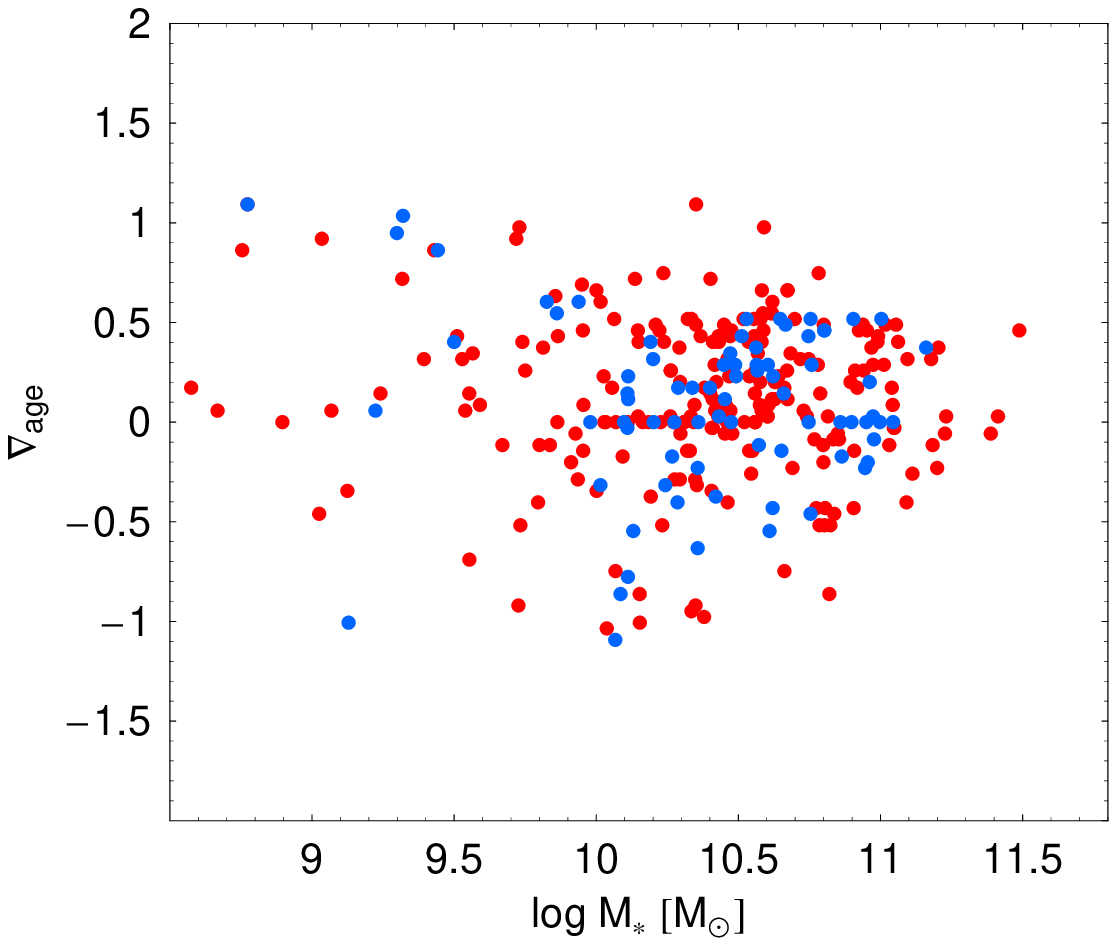, width=0.33\textwidth} \psfig{file= 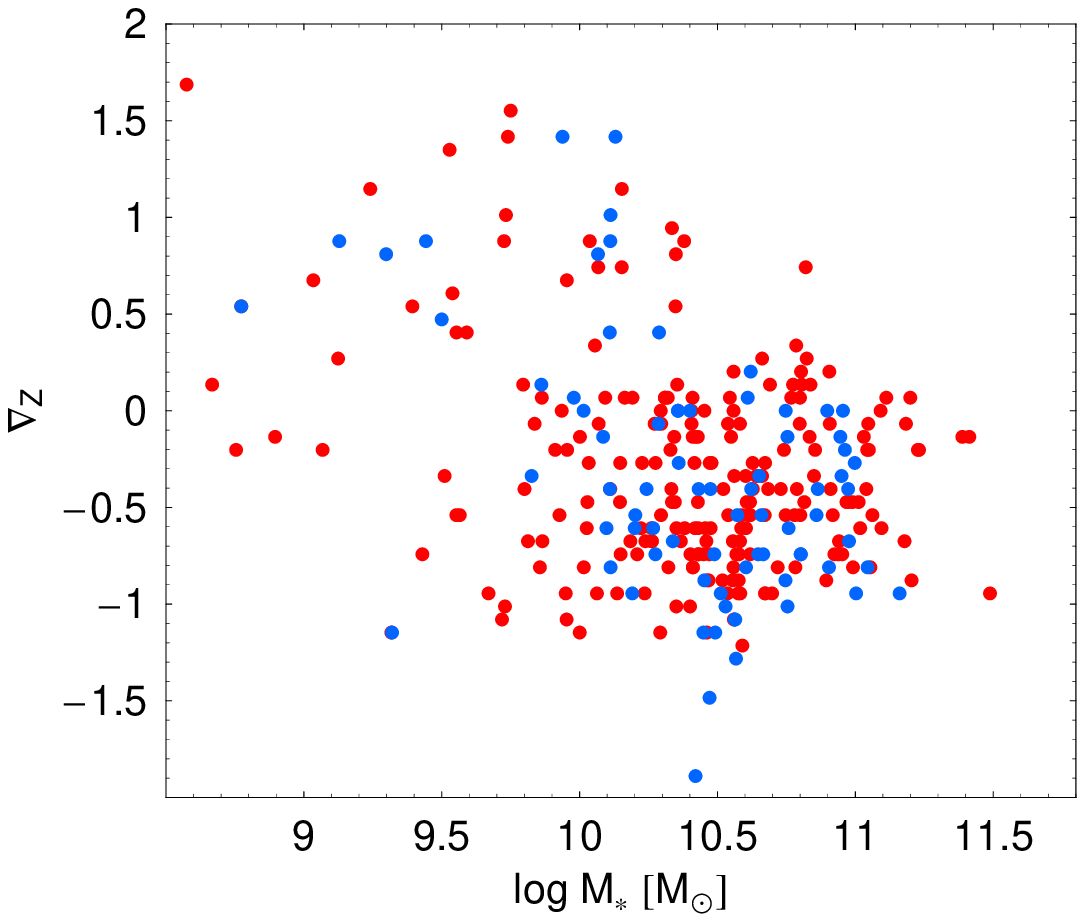,
width=0.33\textwidth} \caption{Results for galaxies extracted for
eyeball check. The red points are for galaxies classified as
ellipticals, while blue ones are for the galaxies with an
uncertain classification or spirals. {\it Left Panel.} Color-mass
diagram. {\it Middle Panel.} \gage\ as a function of stellar mass.
{\it Right Panel.} \gZ\ as a function of stellar mass.}
\label{fig: fig_app4}
\end{figure*}

As anticipated, our selection criteria are not meant to be
contamination free, with ETGs being possibly contaminated by disk
galaxies: e.g., both $n$ and $C$ depend on the inclination and
edge-on disk or spiral galaxies can show parameter values typical
of ETGs. An approach to avoid this kind of contamination can rely
on the axial ratio, but in this case we would loose edge-on S0s
misclassified as LTGs (\citealt{Maller+09}, \citealt{weinmann09}).
In order to evaluate the fraction of contaminants in our ETG
sample we have made an eyeball inspection of a subsample of 300
randomly selected galaxies. This classification may be still prone
to subjectivity, thus four of us have checked the galaxies and
produced an average classification. Eight galaxy images are shown
in Fig. \ref{fig: images_eyeball}. We find that $\sim 70-75\%$ of
galaxies are ETGs, while the remaining galaxies are uncertain
objects or clear spirals. These uncertain or misclassified
galaxies mainly fill the intermediate-mass region and many of them
have red colors $g-i \gsim 0.5$. Although the fraction of such
contaminants is not negligible, we have found that the gradients
are not strongly biased on average, as shown in Fig. \ref{fig:
fig_app4}. The bottom line of this check is that the inclusion of
contaminants leaves the average gradients unaffected. For
instance, on the mass range $10 < \log \mst < 11$, we obtain as
median value $\gage = 0.12$, and $0.03$ for the ETGs, and the
contaminants respectively, as shown in Fig. \ref{fig: fig_app4},
while we find $\gZ = -0.47$ and $-0.54$ for the same systems.
However, when computing the median values for the whole sample,
the result is unchanged from the one of systems correctly
classified as ETGs. No net difference in the central galaxy age is
found, while central metallicity for contaminants is larger than
the one for ETGs, consistently with the steeper metallicity
gradients for galaxies with larger metallicity (see Fig. \ref{fig:
fig5bis}). We have also checked that the presence of contaminants
in ETG sample, in particular, does not affect the dependence on
the central age of \gage\ and \gZ\ as discussed in Fig. \ref{fig:
fig4}. Thus, we can safely assume that the central age is a
genuine parameter which rules the scatter of the relations with
mass and \sigc .

\begin{figure}
\centering \psfig{file= 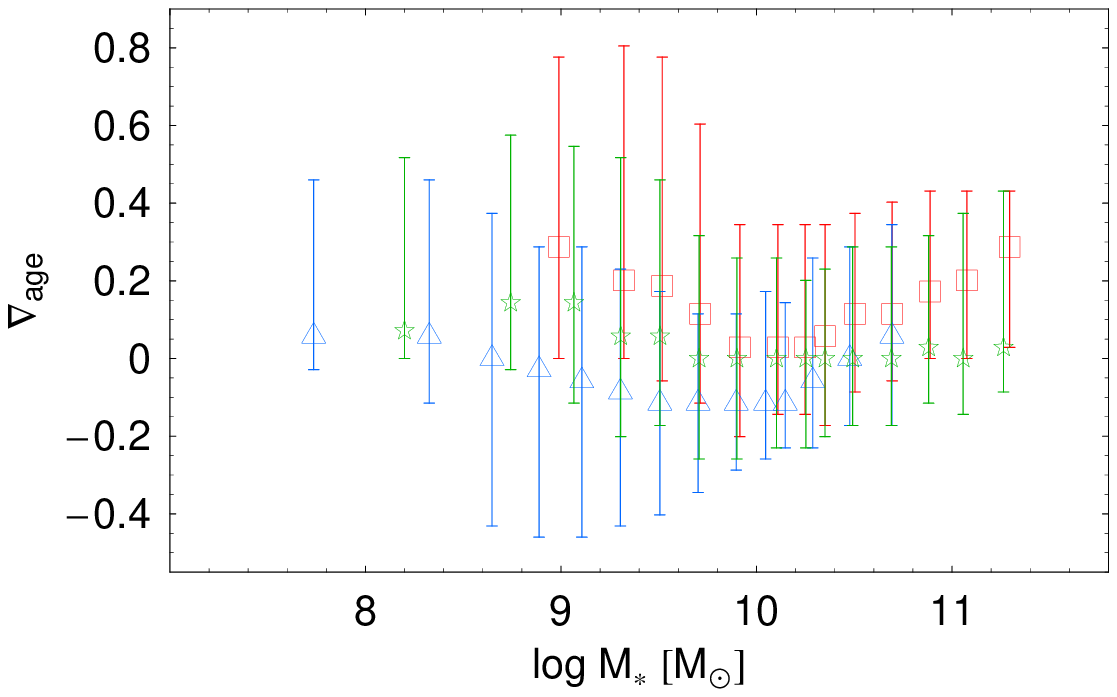, width=0.4\textwidth}
\psfig{file= 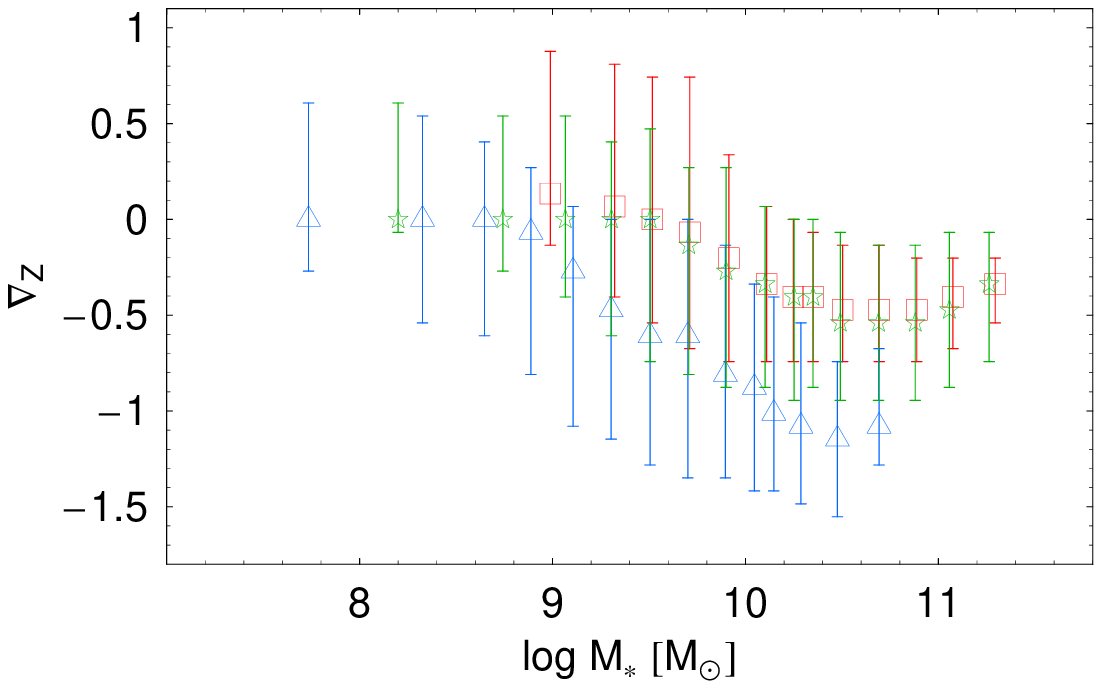, width=0.4\textwidth} \caption{Age (left
panel) and metallicity (right panel) gradients as a function of
stellar mass for ETGs (red symbols), LTGs (blue symbols) and ITGs
(green symbols).} \label{fig: fig_intermediate}
\end{figure}

Finally, the conservative choice of both $n$ and $C$ provided a
quite sharp separation between ETGs and LTGs. For sake of
completeness, the residual galaxy sample with a mixed morphology,
i.e. ITGs, turned out to have also an intermediate distribution of
the age gradients as shown in Fig. \ref{fig: fig_intermediate}. On
the contrary, metallicity gradients of ITGs have a trend which is
more similar to the one of ETGs.

\section{Systematics in stellar fitting}\label{app:app_fit}

We examine the role of systematic uncertainties on the stellar
populations results. First, we address the problem of degeneracy
in synthetic spectral models (\S\ \ref{app:appA1}), then derive
the optical-infrared and infrared CGs from our fitted stellar
populations and check the consistency with results from literature
(\S\ \ref{app:appA2}). Dust extinction is discussed in \S\
\ref{app:appA3}.

\subsection{Parameter degeneracies and wavelength
coverage}\label{app:appA1}

We have analyzed the optical CGs, fitting the internal and
external colors with synthetic spectral models to derive the age
and metallicity of stellar populations. Because of the well known
age-metallicity degeneracy (\citealt{Worthey94}, \citealt{BC03},
\citealt{Gallazzi05}), the stellar parameters and the relative
gradients might be severely biased, in particular by using the
optical colors only. Increasing the wavelength coverage or adding
spectral information would be the best approach to the problem.

\begin{table}
\centering \caption{Results from Montecarlo simulations. We show
the median relative age
$\Delta(age)=(age_{fit}-age_{in})/age_{in}$ and metallicity
$\Delta(Z)=(Z_{fit}-Z_{in})/Z_{in}$, where {\it fit} and {\it in}
are for estimated and input parameters. The results for different
initial perturbations $\delta$ and  without or with IR photometry
are shown.}\label{tab:appA}
\begin{tabular}{lccc} \hline \hline
  $$ & $$ & ugriz & ugrizJHKs   \\
  \hline
$\Delta(age)$ &  $\delta = 0.01$  & $0.01_{-0.10}^{+0.12}$ & $0_{-0.07}^{+0.08}$ \\
$$ &  $\delta = 0.03$  & $0.02_{-0.20}^{+0.27}$ & $0_{-0.14}^{+0.12}$ \\
$$ &  $\delta = 0.05$  & $0.03_{-0.26}^{+0.49}$ & $0_{-0.17}^{+0.25}$ \\

$\Delta(Z)$ &  $\delta = 0.01$  & $-0.01_{-0.13}^{+0.13}$ & $0_{-0.08}^{+0.07}$ \\
$$ &  $\delta = 0.03$  & $0_{-0.27}^{+0.28}$ & $0_{-0.12}^{+0.13}$ \\
$$ &  $\delta = 0.05$  & $-0.02_{-0.40}^{+0.39}$ & $0_{-0.20}^{+0.19}$ \\
\hline
\end{tabular}
\end{table}

\begin{figure}
\psfig{file= 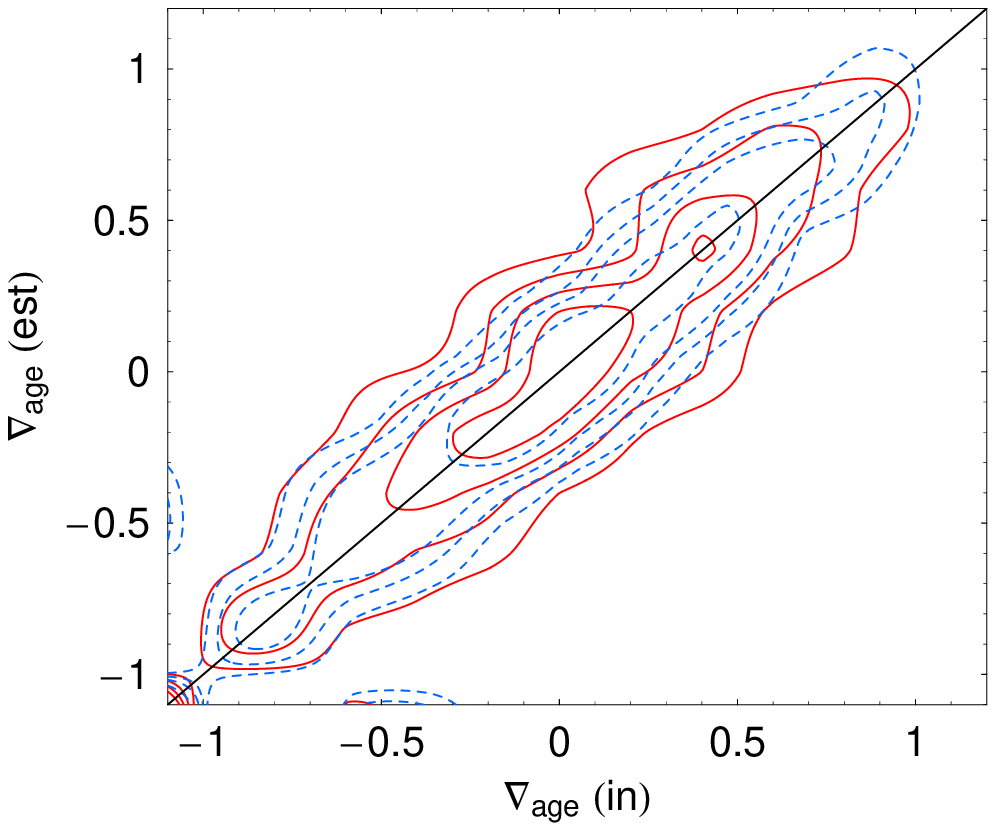, width=0.23\textwidth} \psfig{file=
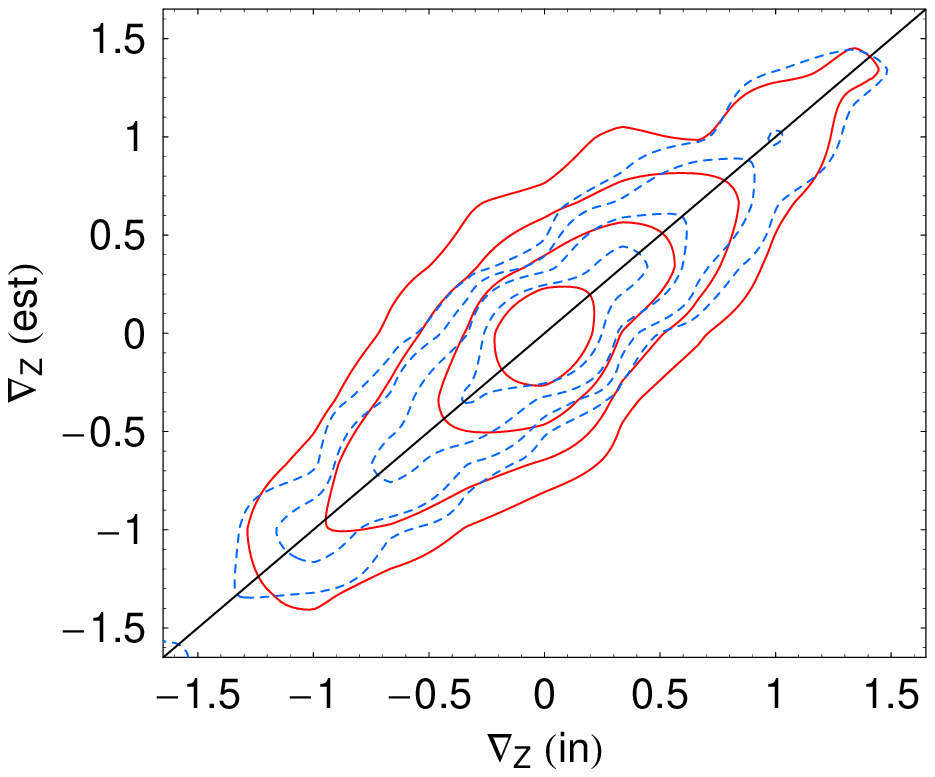, width=0.23\textwidth} \caption{Comparison of age
(left) and metallicity (right) gradients estimated from Montecarlo
repetitions (y-axis), given the initial input values (x-axis).
Here we consider our extreme error budget, $\delta=0.05$. Red and
dashed blue contours are the isodensity corresponding to the
results from the stellar population fit, using only optical and
optical+infrared colors respectively.} \label{fig: fig_app1}
\end{figure}

Near-IR bands are primarily sensitive to the red stellar
populations which represent the main fraction of stars forming an
evolved galaxy, allowing to trace the visible mass of old
galaxies, and are less affected by dust extinction. Their
inclusion is suitable to alleviate the age-metallicity degeneracy
(\citealt{deJ96}, \citealt{Cardiel+03}, \citealt{MacArthur+04},
\citealt{Chang+06}), but is not expected to be definitive.
Different colors are found to have a different sensitivity to age
and metallicity (\citealt{Li+07}). For instance, both optical and
infrared colors are found to be sensitive to metallicity, while
age mostly affects the optical colors (\citealt{Chang+06}).
\cite{Wu+05} showed that the inclusion of near-IR bands allows one
to recover much more accurate stellar parameters from the fit, but
the confidence contours are still affected by the age-metallicity
relation, and the gain with respect the optical bands is minimal.
It is reasonable to argue that the use of a large statistical
sample, as we adopt in this paper, is more effective in reducing
the overall uncertainties of the galaxy properties than the
adoption of a larger filter baseline.

\begin{figure*}
\psfig{file= 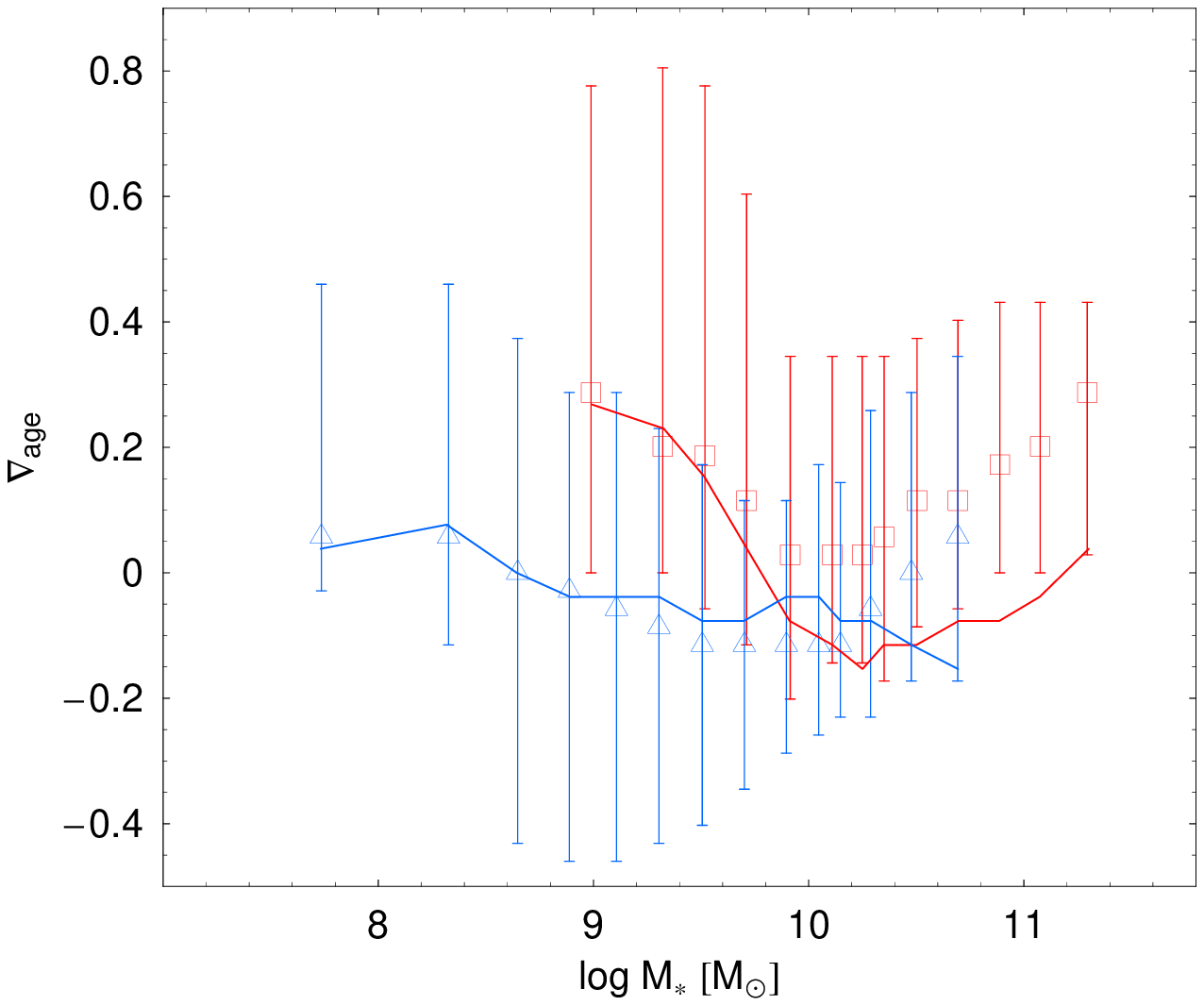, width=0.33\textwidth} \psfig{file=
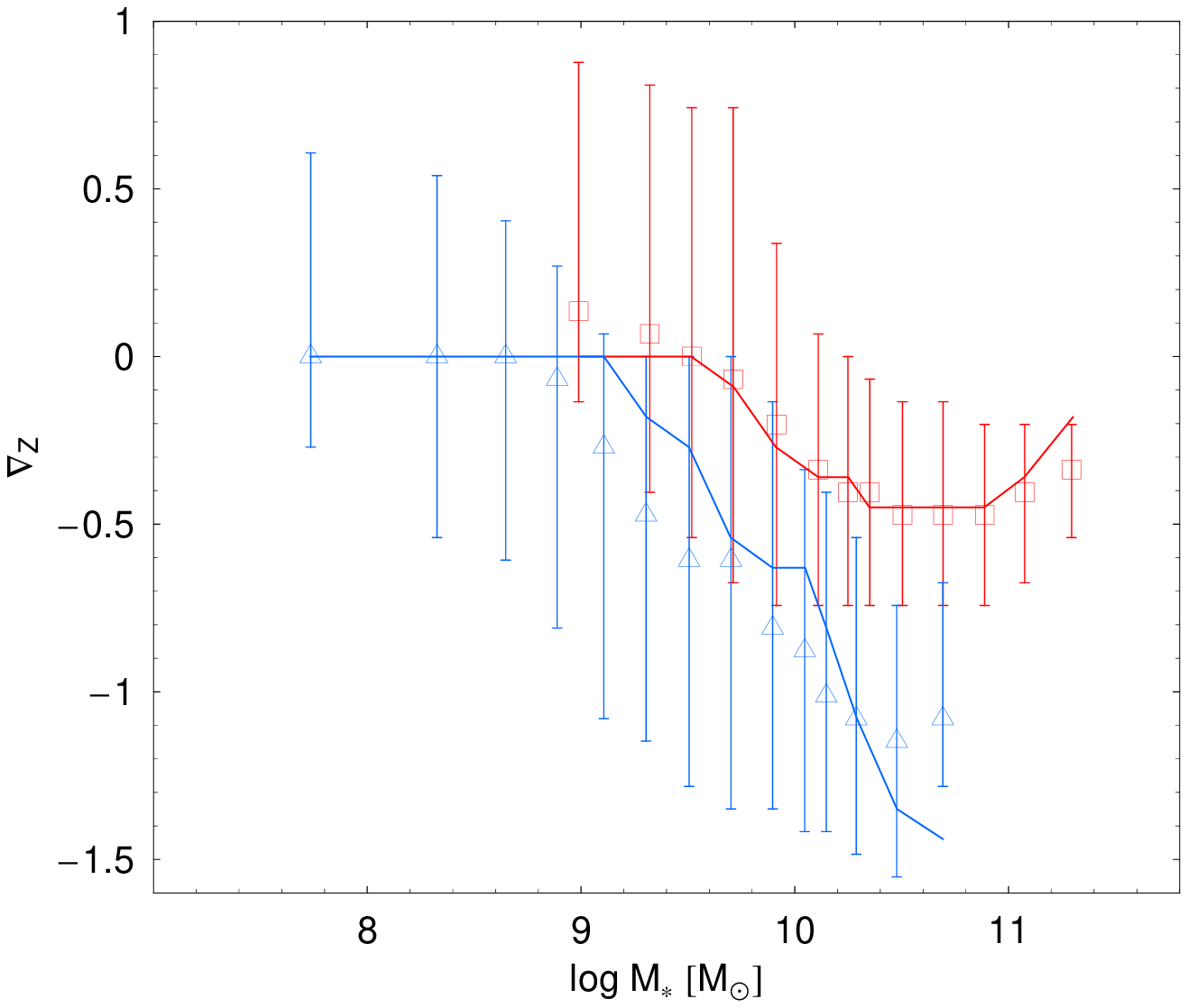, width=0.33\textwidth} \psfig{file= 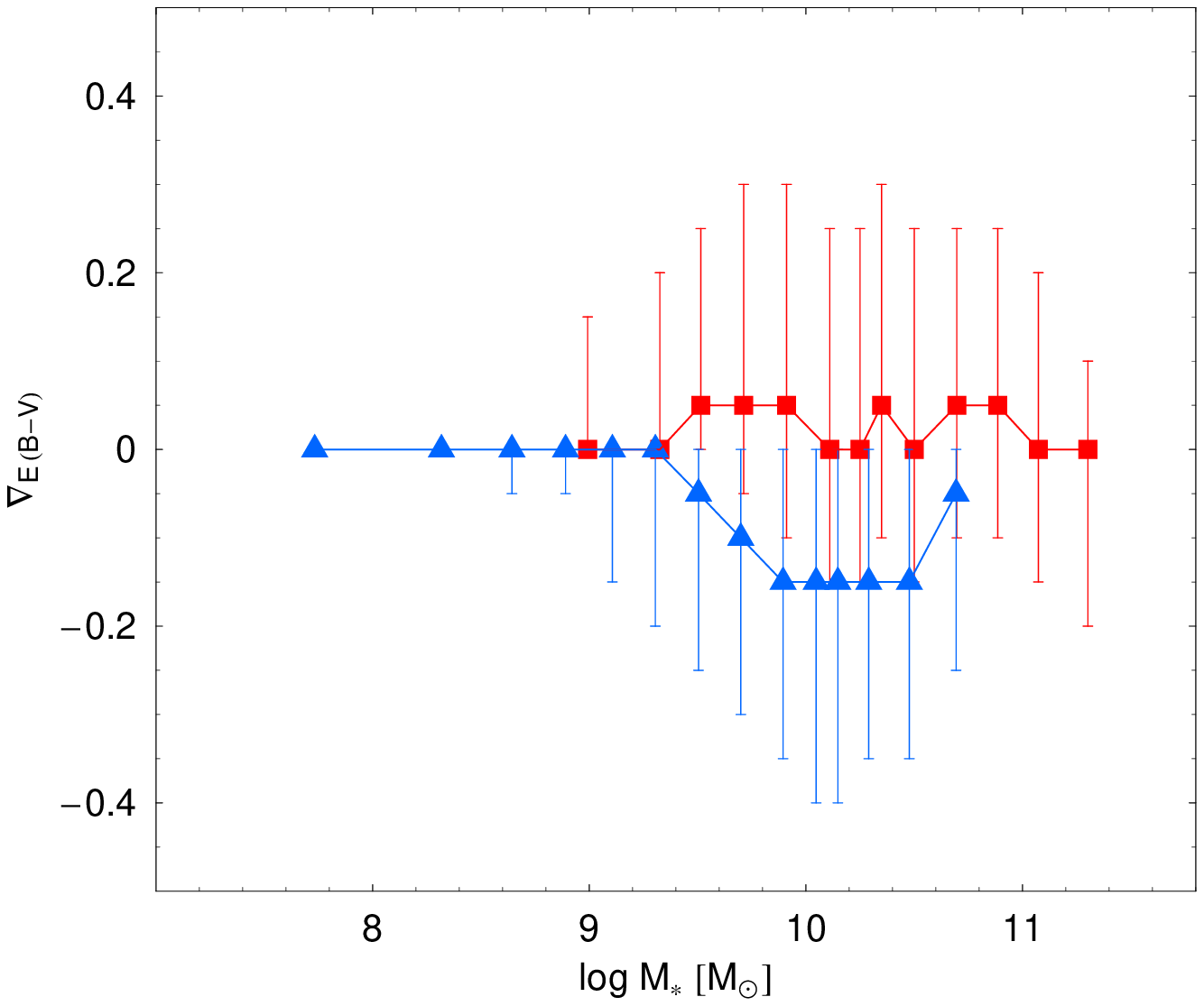,
width=0.33\textwidth} \caption{Gradients for the spectral model
leaving free the galaxy age, metallicity, and dust content. The
open symbols with bars are for our reference fit, while the line
is for the fit with dust extinction free to vary. {\it Left
Panel.} \gage\ as a function of stellar mass. {\it Middle Panel.}
\gZ\ as a function of stellar mass. {\it Right Panel.} Gradient in
dust extinction as a function of stellar mass.} \label{fig:
fig_appdust}
\end{figure*}

Of course, the adoption of spectroscopic line absorption indices
would be the best indicators of stellar population parameters and
gradients, as they permit to break the age-metallicity degeneracy
(e.g., \citealt{Burstein+84}, \citealt{HW99}, \citealt{CDB93},
\citealt{Davies+93}, \citealt{Mehlert+03}). However, spectral
indices of large galaxy sample are still difficult to obtain for
the observational limits imposed by the low starlight fluxes (e.g.
around \Re\ and beyond). Recently, integral-field spectroscopy has
been used to determine gradients in line absorption indices (e.g.
\citealt{Kuntschner06}, \citealt{Rawle+09}) of elliptical and
lenticular galaxies, but this approach is only applicable to
limited samples of systems.

Despite the limited wavelength coverage, but taking advantage of
the large statistical sample we have shown that both spectroscopy
and photometric analysis converge to similar results for stellar
population gradients (see e.g. \S\ref{sec:compar}). Similarly,
\cite{Rawle+09} have shown that optical colors and CGs derived
from spectroscopic age and metallicity are generally consistent
with values directly recovered from photometry. Moreover, the
recovered correlations between metallicity and age gradients are
not spurious and not generated by correlated measurement errors on
these quantities.

However, in the attempt of putting our result further on a more
solid ground, we have tested the reliability of our color
modelling technique and checked for the presence of spuriously
generated correlations between age and Z or gradients $\gage$ and
$\gZ$, running several Monte Carlo simulations. We extracted 1000
simulated galaxy spectra from our BC03 SED libraries with random
$(age_i, Z_i)$ for $i=1,2$ (i.e. with no correlations among these
parameters and with \gZ\ , and \gage\ randomly distributed), and
applied our fitting procedure, comparing the estimated parameters
with the input model values. We perform the fit 1) only using the
optical SDSS bands ugriz and 2) adding the JHKs photometry from
2MASS survey (\citealt{Jarrett+03}), to understand the systematics
that the fit of optical colors can induce. Typical observational
uncertainties for each band are assigned (i.e. $\delta u \sim
0.06$, $\delta g \sim 0.035$, $\delta r \sim 0.035$, $\delta i
\sim 0.035$, $\delta g \sim 0.04$, $\delta J \sim 0.08$, $\delta H
\sim 0.12$, $\delta Ks \sim 0.09$, from matched catalog in B05)
and the input model colors are perturbed adding a random step in
the interval $(-\delta, +\delta)$ with $\delta = 0.01,0.03,0.05$
to account for measurement systematics. Results of this analysis
are collected in Table \ref{tab:appA} and Fig. \ref{fig:
fig_app1}. We found that, on average, both the stellar parameters
and gradients are fairly well recovered from the fitting
procedure, with a scatter increasing with $\delta$, and no
spurious correlations are induced at more than $99\%$ confidence
level (see also \cite{Wu+05} and \cite{Tortora2009} for similar
analysis). We confirm the finding of \cite{Wu+05} that the
inclusion of near-IR bands allows a better recovery of the input
stellar parameters with a scatter in the range $5-25 \%$, while
the scatter is larger (in the range $10-50\%$) when only SDSS
bands are used. Thus, the most remarkable result of this analysis,
is that, notwithstanding the larger scatter, we are able to
successfully recover the stellar population parameters, even
without near-IR photometry.

\subsection{Comparison with IR CGs}\label{app:appA2}

One of the advantages of synthetic spectral modelling is to allow
the recover of synthetic IR colors by using the best fitted
spectral parameters. Therefore, we derive the colors $g-J$, $g-H$
and $g-Ks$ at $0.1 \Re$ and 1$\Re$ and the relative CGs, which
show trends with mass similar to those discussed in this paper. If
we select luminous ($r < -20$ mag) ETGs we find median values of
$\nabla_{g-J} = -0.33$, $\nabla_{g-H} = -0.37$ and $\nabla_{g-Ks}
= -0.40$, with 25-75 percentiles in sample distribution of $0.17$,
$0.18$, and $0.20$, respectively. As already extensively discussed
in the text, to compare these synthetic CGs with the observed
values in \cite{LaBarbera2009}, we select older systems with
$age_{1} > 6 \, \rm Gyr$. In this case we have $\nabla_{g-J} =
-0.26$, $\nabla_{g-H} = -0.29$, and $\nabla_{g-Ks} = -0.31$ (with
scatter of $0.16$, $0.18$, and $0.20$), which are fairly
consistent with the results reported by these authors. A fairly
good agreement is also found with the median $\nabla_{g-Ks} =
-0.29\pm 0.07$ in \cite{Wu+05}.

\subsection{Internal dust extinction}\label{app:appA3}

In our analysis we have assumed no internal extinction. This
assumption is more reasonable for ETGs, where only a negligible
fraction of interstellar matter is in the form of dust (e.g.
\citealt{KD89}, \citealt{WS96}, \citealt{Dorman+03}), but it might
be unappropriate for LTGs, as these systems have a larger amount
of internal dust (e.g. \citealt{Brammer+09}). Usually no dust
extinction is included in the analysis of ETGs (\citealt{T+00},
\citealt{TO2000}, \citealt{TO2003}, \citealt{Wu+05}), but dust
gradients would have effects in CGs (e.g. \citealt{GdeJ95}).
However, for ETGs it has been shown that metal absorption lines
gradients, being less affected by dust extinction than colors
(e.g. \citealt{MacArthur05}), are unlikely to be produced by pure
dust extinction gradients to mimic the population gradients
inferred by broadband colors (\citealt{Peletier+90a},
\citealt{Davies+93}, \citealt{CDB93}, \citealt{KoAr99},
\citealt{Rawle+09}). Thus they are not dominant but could still
contribute to a fraction of CGs. Similarly, in LTGs the dust seems
to have a negligible effect in shaping CGs (e.g.
\citealt{MacArthur+04}), even thought there are still open
questions (\citealt{MacArthur+09}) which might lead to a different
conclusion. To analyze the impact of dust extinction on our
results, we fit the colors at $0.1\Re$ and $1\Re$ using a library
of spectra with age, metallicity and dust content free to change.
The extinction curve of \cite{CCM89} is used to account for the
extinction contribution in different wavebands and the color
excess $E(B-V)$ in the range $0-0.5$ is used as a free parameter.
The results are shown in Fig. \ref{fig: fig_appdust}, where they
are compared with our reference results with no dust. The main
trends are unaffected by the dust, although we note some
differences in the \gage\ at $\mst \gsim 10^{9.5}\, \rm \Msun$. On
the other hand, only slight differences are observed for
metallicity gradients, and the qualitative trends are unaffected
both in ETGs and LTGs. Thus this analysis has shown that, although
dust extinction is not negligible in both ETGs and LTGs, the
trends and the amplitude of gradients are still dominated by
changes in stellar population parameters. We show the dust
extinction gradients, defined as $\nabla_{E(B-V)} = E(B-V)_{2} -
E(B-V)_{1}$ in the right panel of Fig. \ref{fig: fig_appdust}. On
average, the ETGs exhibit null or slightly positive dust gradients
(with a tail towards negative values, too), while LTGs have
opposite behaviour with a larger dust content in the inner
regions, showing steeper dust gradients at intermediate high
masses (e.g., \citealt{DB02}, \citealt{MacArthur+04}). Finally, we
have checked the absence of any correlation between age and
metallicity gradient with both $\nabla_{E(B-V)}$  and the central
extinction for LTGs which has suggested that the results of these
systems are not affected by the presence of dust. We remark that
the use of only optical bands and the addition of IR photometry is
not sufficient to break the degeneracy between stellar population
parameters and dust, for this reason we are tempted to consider
these results more as a caveat for this particular data-set rather
than a conclusive check. If the correction of the age gradients
for the dust extinction is right, the net effect is that the age
gradient might be less affecting the color gradient for massive
systems, and only metallicity gradient should matter.

\end{document}